\documentclass[usenatbib]{mn2e}
\usepackage{graphicx}
\usepackage{longtable}
\input psfig.sty

\title[The Influence of Binary Interactions in IR bands of populations]
{The Influence of Binary Interactions in Infrared passbands of populations}

\author[F. Zhang, L. Li, Z. Han]
{Fenghui~Zhang\thanks{E-mail: gssephd@public.km.yn.cn;
zhang\_fh@hotmail.com}, Lifang~Li and Zhanwen~Han\\
National Astronomical Observatories/Yunnan Observatory, Chinese
Academy of Sciences, Kunming, 650011, China \\}

\begin{document}

\date{\today}

\pagerange{\pageref{firstpage}--\pageref{lastpage}}

\pubyear{2008}

\maketitle

\label{firstpage}

\begin{abstract}
In our evolutionary population synthesis models, the samples of binaries are reproduced by the '{\sl patched}' Monte Carlo simulation and the stellar masses, integrated $J, H, K, L, L2$ and $M$ magnitudes, mass-to-light ratios and broad colours involving infrared bands are presented, for an extensive set of instantaneous-burst binary stellar populations. In addition, the fluctuations in the integrated colours, which have been given by \citet{zha05a}, are reduced.

By comparing the results for binary stellar populations with (Model A) and without (Model B) binary interactions we show that the inclusion of binary interactions makes the stellar mass of a binary stellar population smaller ($\sim$\,3.6-4.5\,\% during the past 15\,Gyr); magnitudes greater (except $U$, $\sim$\,0.18\,mag at the most); colours smaller ($\sim$\,0.15\,mag for $V-K$ at the most); the mass-to-light ratios greater ($\sim$\,0.06 for $K$-band) except those in the $U$ and $B$ passbands at higher metallicities. And, Binary interactions make the $V$ magnitude less sensitive to age, $R$ and $I$ magnitudes more sensitive to metallicity.

Given an age, the absolute values of the differences in the stellar mass, magnitudes, mass-to-light ratios (except those in the $U$ and $B$ bands) between Models A and B reach the maximum at $Z=0.0001$, i.e., the effects of binary interactions on these parameters reach the maximum, while the differences in some colours reach the maximum at $Z \sim$\,0.01-0.0004. On the contrary, the absolute value of the difference in the stellar mass is minimal at $Z=0.03$, those in the $U,B,V$ magnitudes and the mass-to-light ratios in the $U$ and $B$ bands reach the minimum at $Z \sim$\,0.01-0.004.
\end{abstract}


\section{Introduction}

In previous papers \citep{zha05a,zha05b,zha06} we took into account binary interactions (BIs) in evolutionary population synthesis (EPS) models, presented the integrated $U-B$, $B-V$, $V-R$ and $R-I$ colours, the integrated spectral energy distributions (ISEDs) and Lick/IDS spectral absorption feature indices for binary stellar populations (BSPs) with and without BIs, while did not give the infrared (IR) magnitudes and colours involving IR passbands (such as, $V-K$, $J-H$ and $J-K$ colours) because larger fluctuations caused by Monte Carlo simulation exist.

For populations of age $\tau \ga 1\,$Gyr, $J$, $H$ and $K$ magnitudes are mainly dominated by cooler stars, i.e., by giant branch (GB) and asymptotic giant branch (AGB) stars, while the lifetime of these stars is relatively short, this causes that the number of stars on these evolutionary stages produced by Monte Carlo simulation is relatively small, and the fluctuations in the results involving IR passbands are large.
However, the results in IR bands are very important in EPS  models. Firstly, IR light can reflect the metallicity of populations because that the metallicity effect on GB and AGB stages is very significant, the colours involving IR and visible bands ($U$, $B$ and $V$ magnitudes are better candidates determining the age of populations), for example, $V-K$, are the candidates of breaking the degeneration between age and metallicity. Moreover, the IR results can affect the determination of the galaxy mass derived from the correlation between colours and stellar mass-to-light ratios, further affect the studies of galaxy evolution. \citet{ret06} have ever used multiband photometry to derive the stellar masses of early-type galaxies, then compared them with their dynamical masses and obtained the relevance of the dark matter component of early-type galaxies as a function of the total mass.

In this paper we adopt the so-called {\sl 'patched'} Monte Carlo simulation to produce the samples of binaries in BSPs (see Sect. 2) and present the stellar masses, integrated IR magnitudes, mass-to-light ratios and colours for BSPs. The results presented in this paper are very similar to those calculated from the BSPs composed of $2.5 \times 10^7$ binaries, the fluctuations in our previous results, which are calculated from the samples of $1 \times 10^6$ binaries, are largely reduced. Using the {\sl 'patched'} Monte Carlo simulation the number of binaries calculated is largely decreased, and the calculation efficiency is largely increased.

Our EPS models use Hurley's single stellar evolution (SSE) and binary stellar evolution (BSE) codes \citep{hur00,hur02}, which include the thermally pulsing AGB (TPAGB) stars but it is in a simple approach. Recently, several EPS models (\citealt[hereafter B07]{bru07}; \citealt[hereafter M05]{mar05}) also included TPAGB stars because M05 and \citet{mar07a} found that
AGB stars are important contributors to the integrated bolometric and near-IR luminosities with a maximum located at an age of $\sim$\,1\,Gyr, accounting for $\sim$\,40-80\% of the total clulster's luminosity, especially, B07 found that TPAGB stars contribute to $\sim$\,70\% of the $K$-light.
The different treatment of the TPAGB stars is a source of major
discrepancy in the determination of the spectroscopic age and mass
of high-z ($1.4<z<2.7$) galaxies \citep{mar06}.
For those galaxies observed by {\it Spitzer}, their mid-UV spectra
indicate that their ages are in the range of 0.2-2\,Gyr, at which the
contribution of TPAGB stars in the rest-frame near-IR
is expected to be at maximum (B07), therefore,
the inclusion of TPAGB stars also
plays a key role in the interpretation of the {\it Spitzer} data.
In this work we also compare our model results with theirs.

In our EPS models we assume that all stars are born in binaries and born at the same time, i.e., an instantaneous BSP. To investigate the effect of BIs on the results, we perform two sets of calculations: one includes all BIs, we call Model A; another neglects BIs, we call Model B. Models A and B have the same binary sample. A full model description and algorithm are given in \citet{zha05a}, we refer the interested reader to part 2 for them.

The outline of the paper is as follows: we describe the method used to reduce the fluctuations in EPS models, and compare the new results with the old ones in Section 2; our results and the comparisons with literature and observations are presented in Section 3; and then finally, in Section 4, we give our conclusions.

\section{Reduction of the fluctuations}
\begin{figure}
\centering{
\includegraphics[bb=135 122 530 625,height=3.8cm,width=4.6cm,clip,angle=-90]{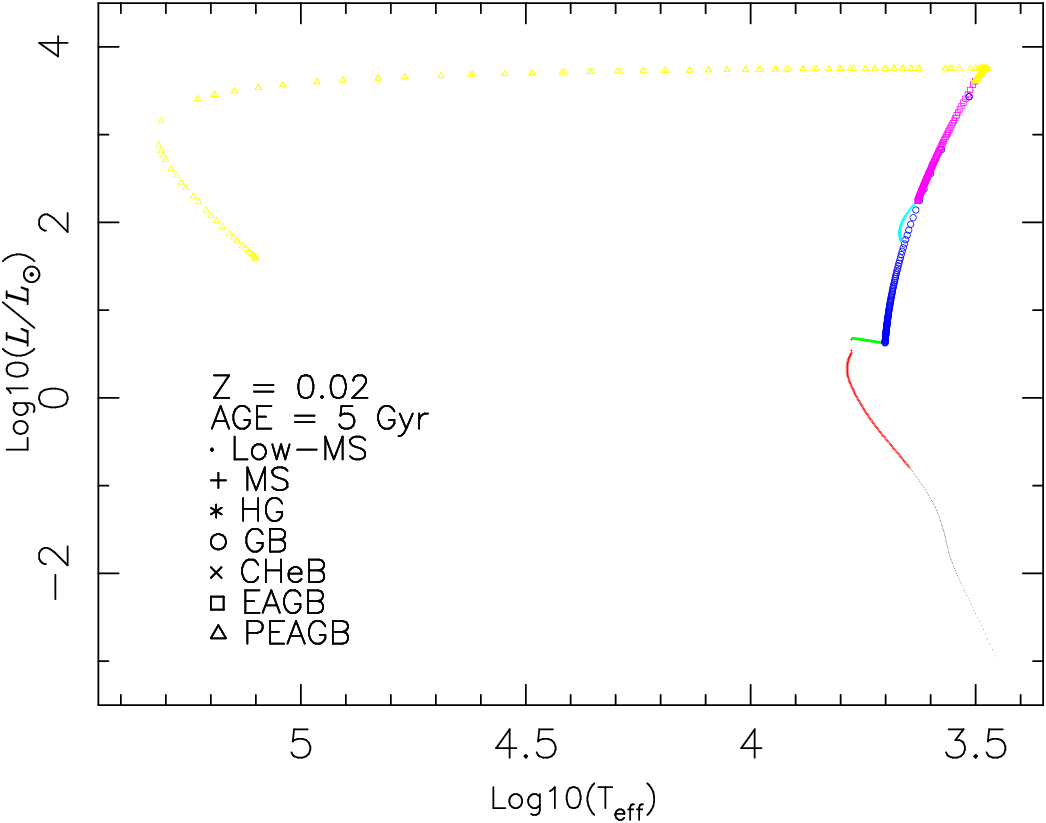}
\includegraphics[bb=80 109 591 629,height=4.0cm,width=4.5cm,clip,angle=-90]{5Gp00-K-contri.ps}
}
\caption{Isochrone (Left panel) and the contribution of stars along the isochrone to $K$-band light (Right panel) for solar-metallicity $\tau=1$\,Gyr BSPs without BIs (Model B). The abbreviations are explained in the text. In right panel from right to left $i$ corresponds to the $i-$bin along the isochrone from the end to the beginning of the corresponding evolutionary stage. Solid rectangles correspond to the GB stars on the tip and those with ${\rm log}(L/L_{\sun}) \simeq 2.0$.}
\label{Fig:kcontri}
\end{figure}

\begin{figure}
\centering{
\includegraphics[bb=131 105 551 624,height=4.0cm,width=4.5cm,clip,angle=-90]{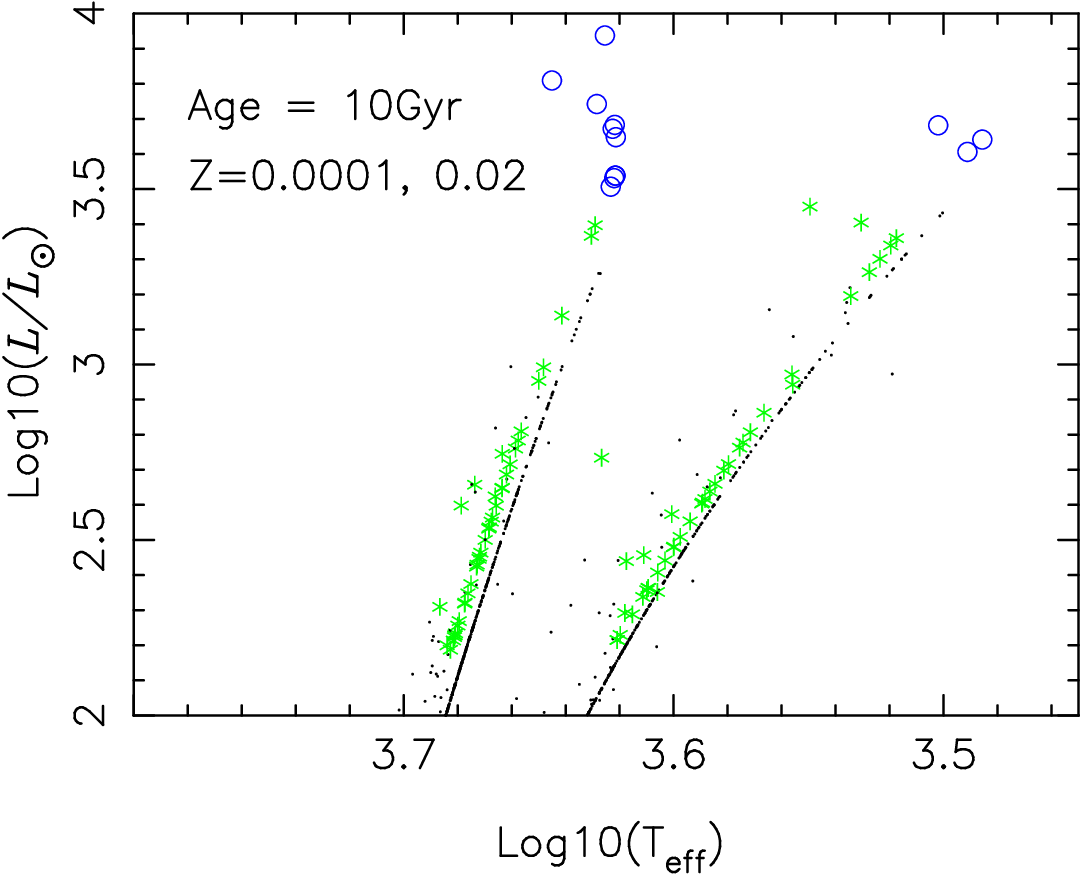}
\includegraphics[bb=131 105 551 624,height=4.0cm,width=4.5cm,clip,angle=-90]{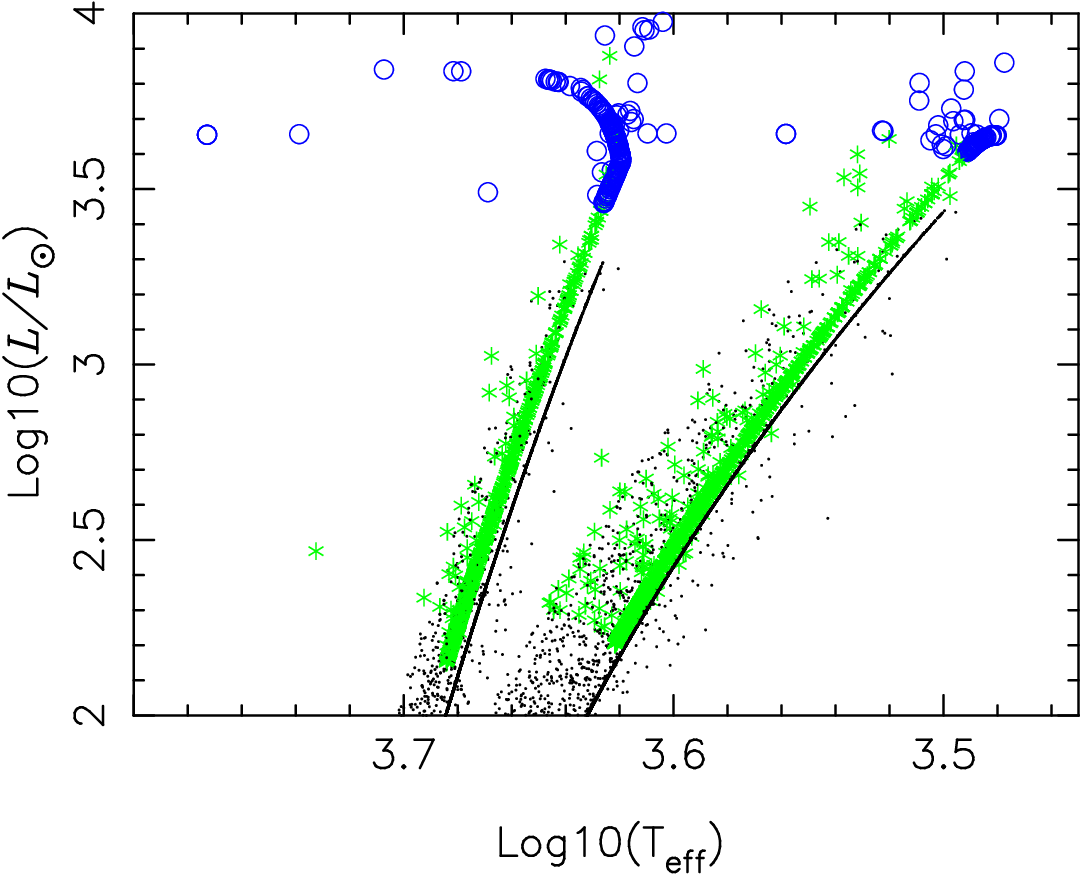}}
\caption{The theoretical distribution of GB (dots),EAGB (stars) and PEAGB (circles) stars with ${\rm log} (L/L_{\sun}) \ga 2.0$ in HR diagram for Model A. The age $\tau=10$\,Gyr and metallicity $Z=0.0001$ and 0.02. Left and right panels are for the old and new versions, respectively.}
\label{Fig:iso-adp}
\end{figure}

\begin{figure}
\centering{
\includegraphics[bb=108 46 503 684,height=4.0cm,width=3.0cm,clip,angle=-90]{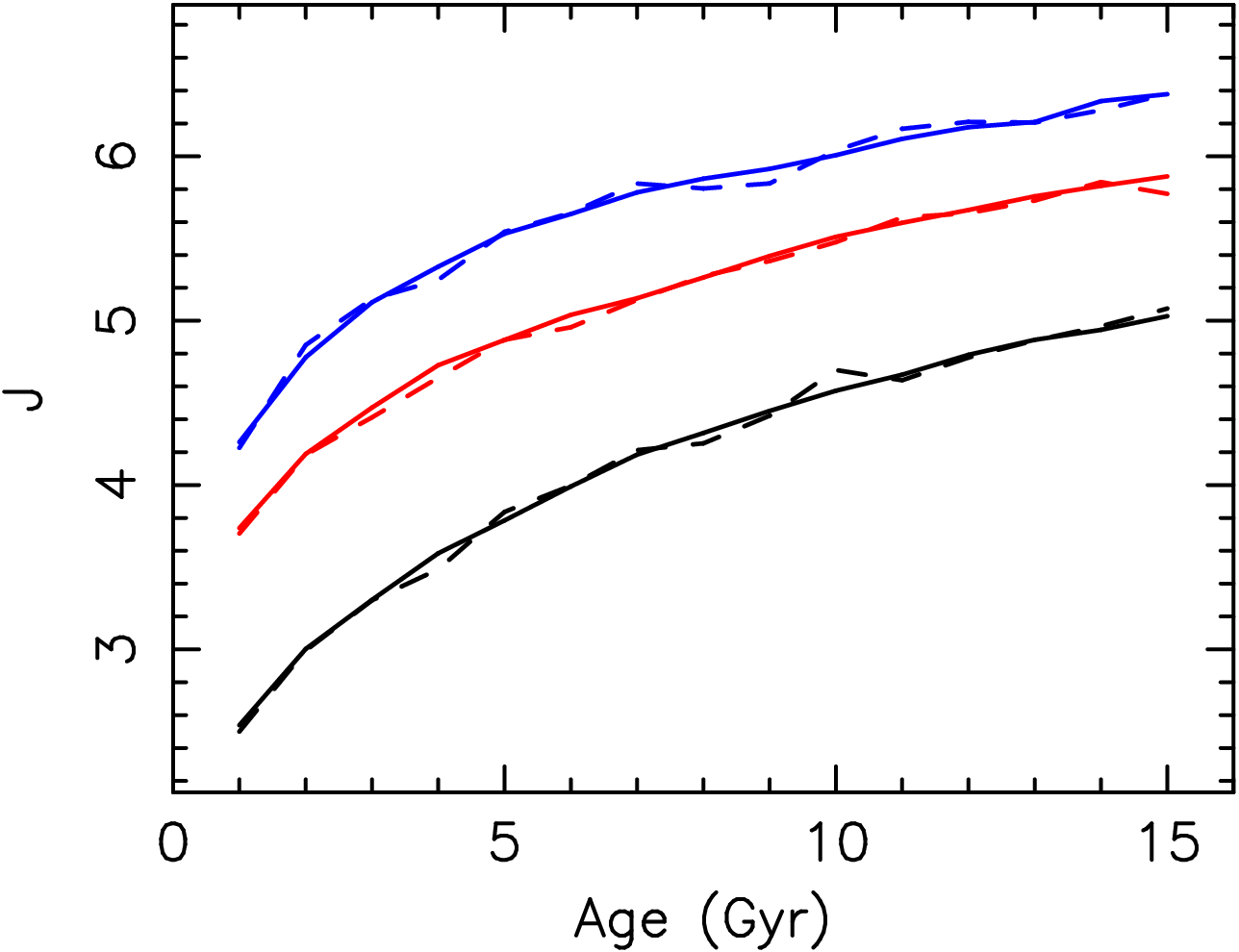}
\includegraphics[bb=108 46 503 684,height=4.0cm,width=3.0cm,clip,angle=-90]{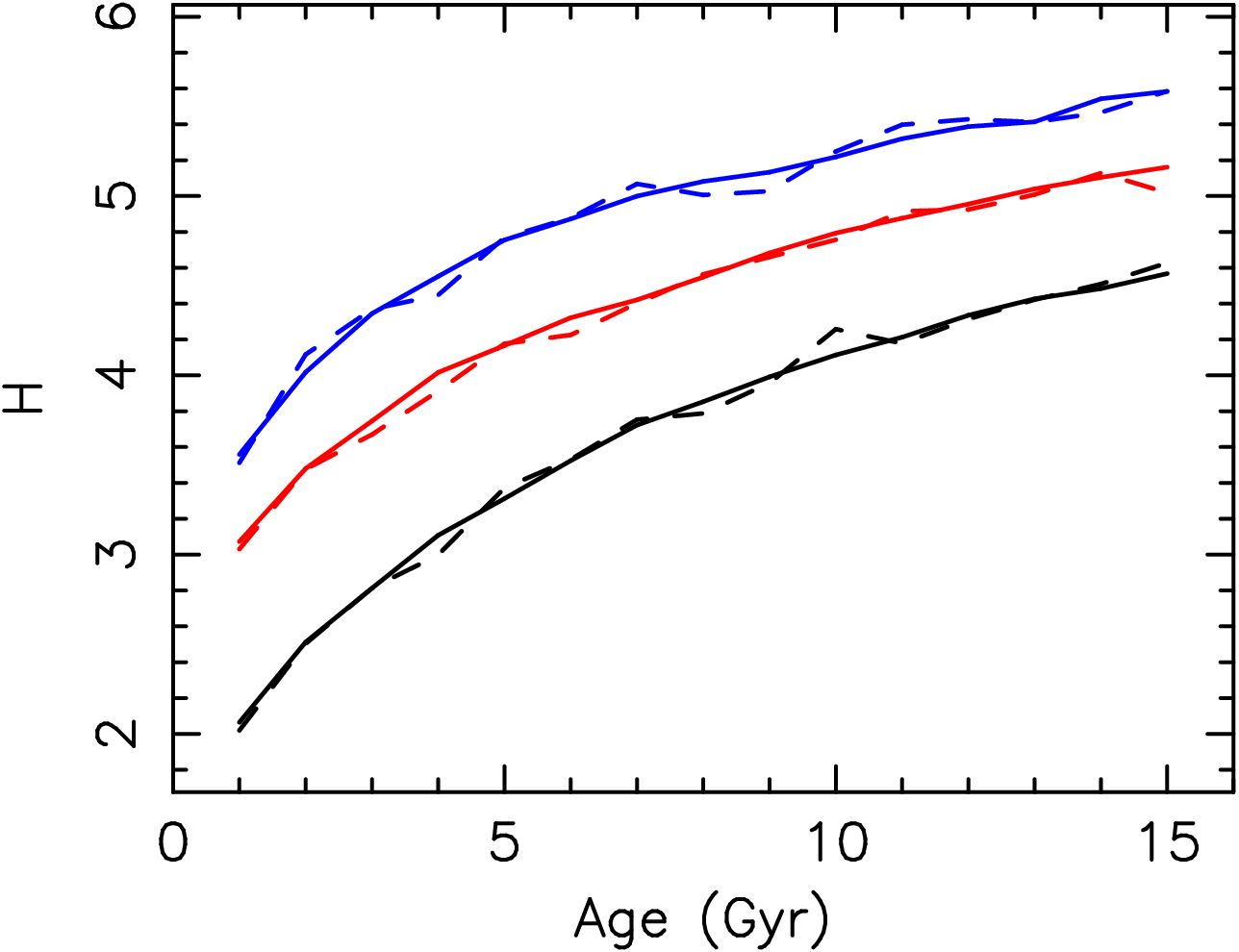}
}\\
\centering{
\includegraphics[bb=108 46 503 685,height=4.0cm,width=3.0cm,clip,angle=-90]{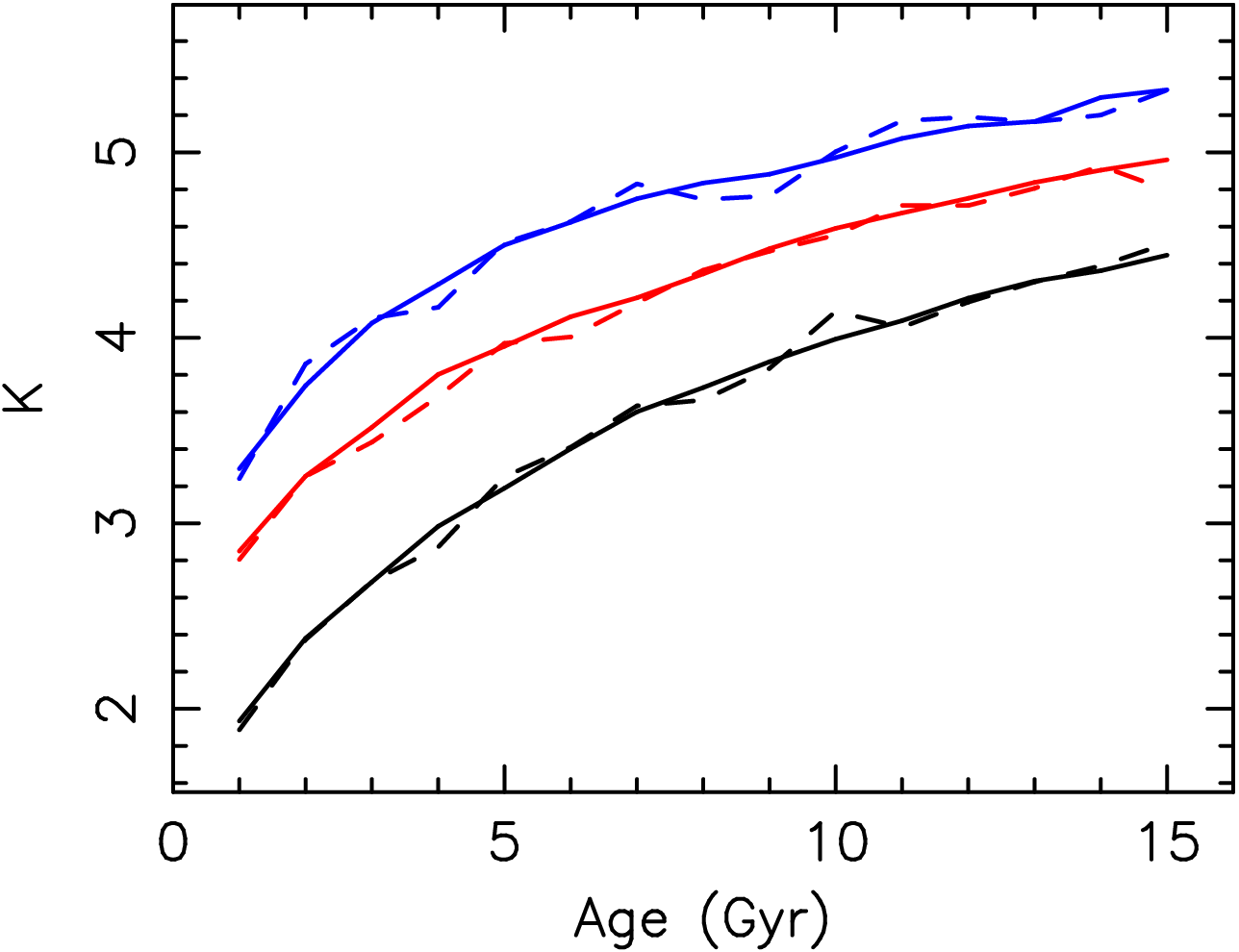}
\includegraphics[bb=108 46 503 685,height=4.0cm,width=3.0cm,clip,angle=-90]{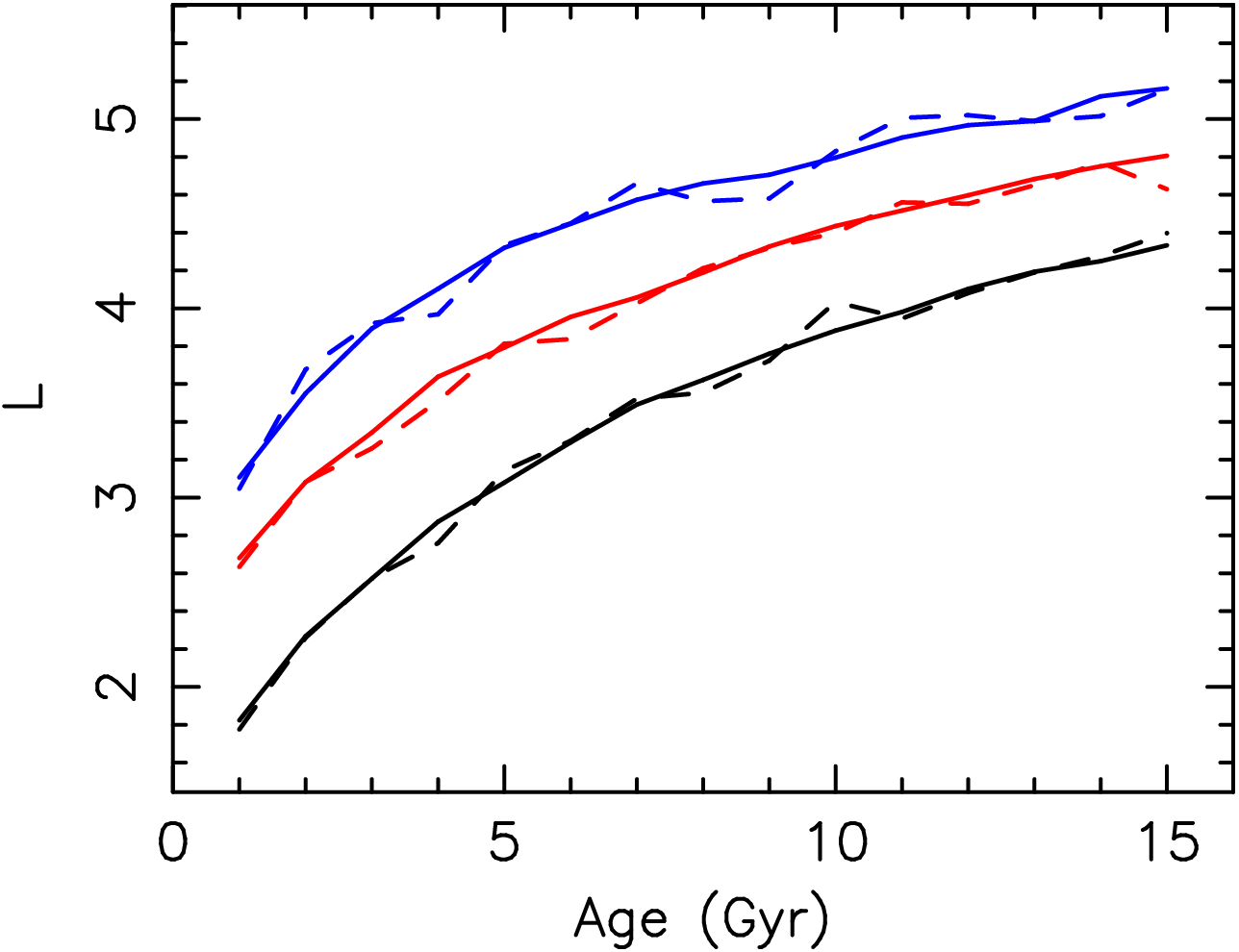}
}\\
\centering{
\includegraphics[bb=108 36 582 677,height=4.0cm,width=3.6cm,clip,angle=-90]{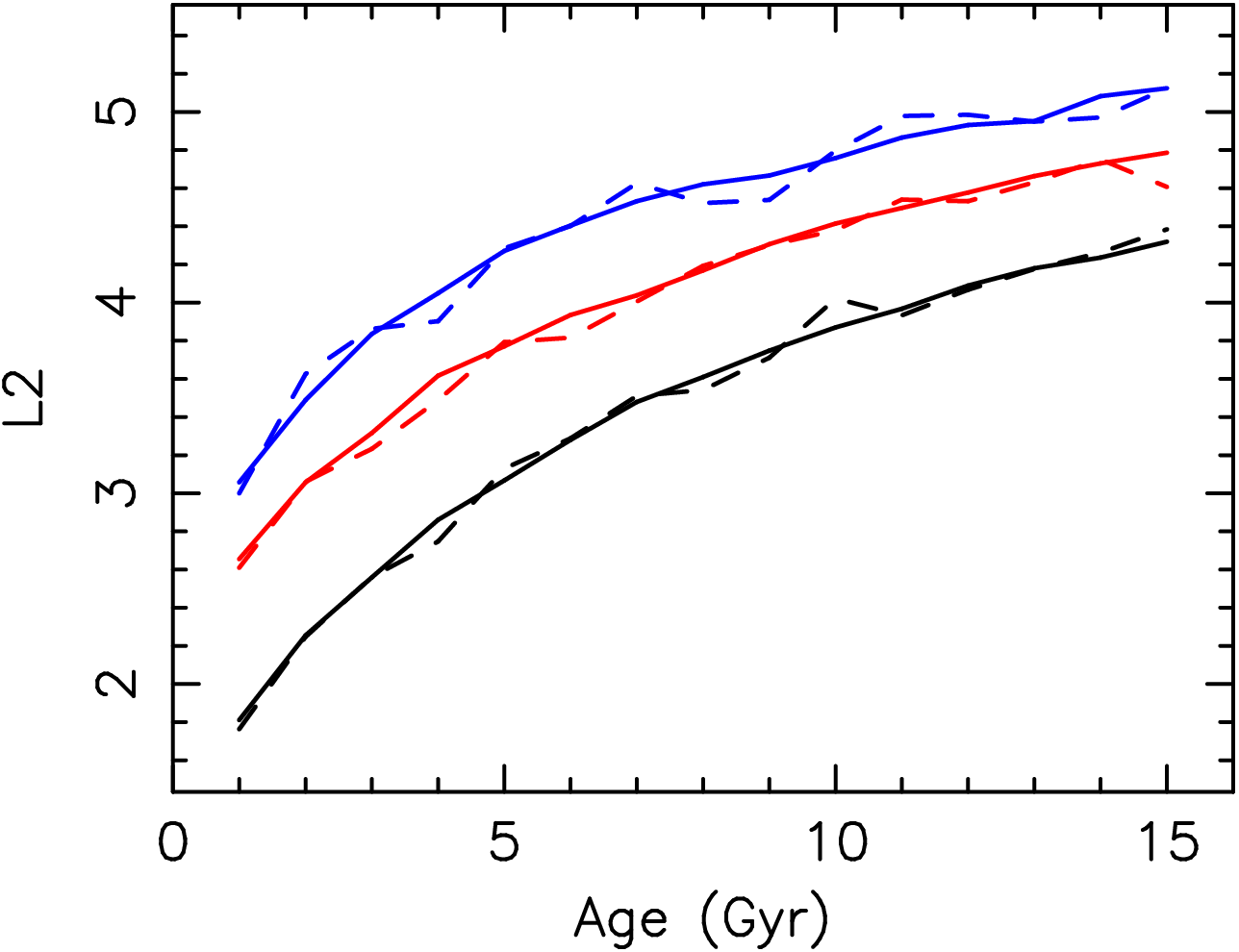}
\includegraphics[bb=108 38 582 677,height=4.0cm,width=3.6cm,clip,angle=-90]{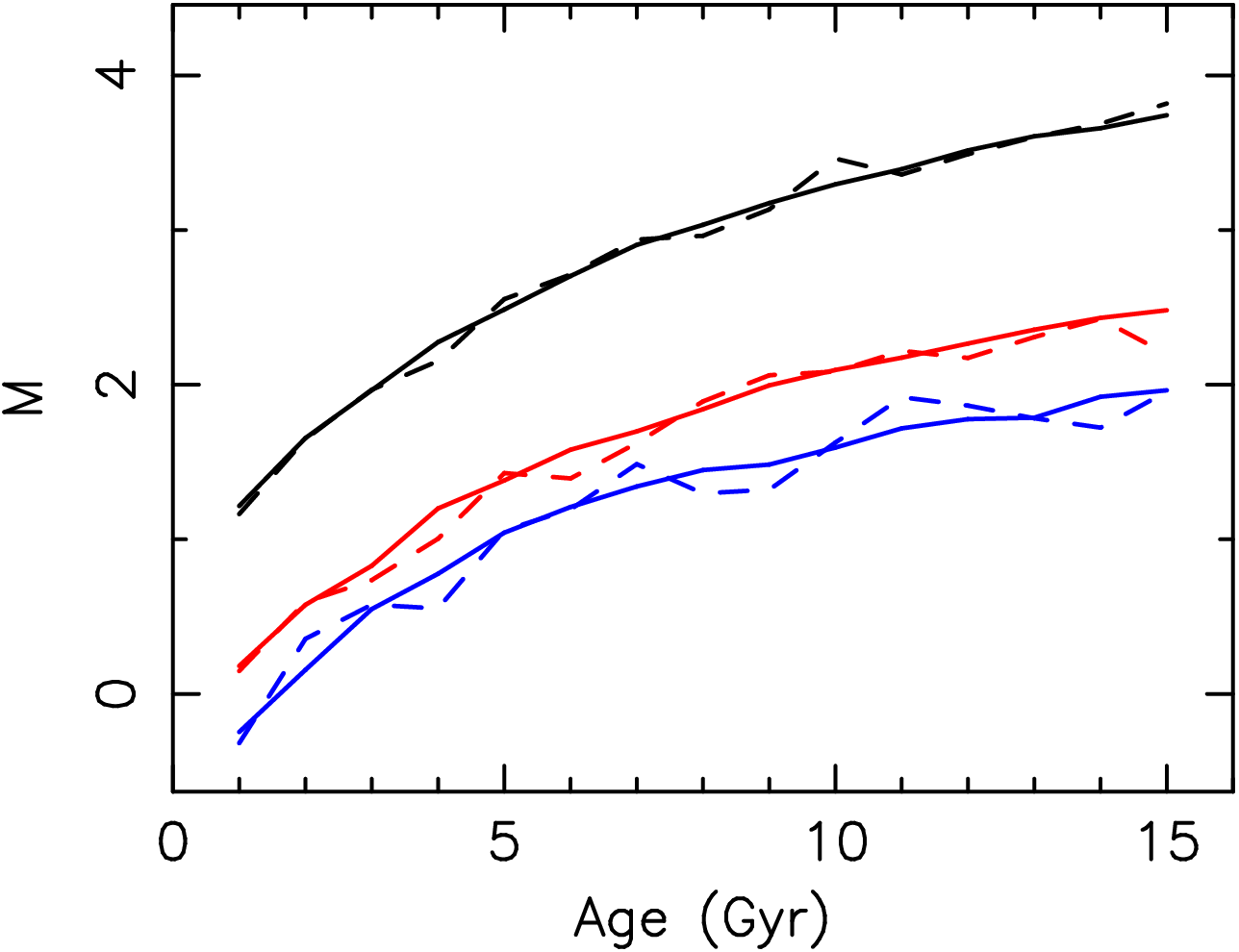}
}
\caption{The new (full line) and old (dashed line) $J, H, K, L, L2, M$ magnitudes for Model A. From top to bottom $Z=0.03, 0.01$ and 0.0001 (opposite for $M$ magnitude). The curves of $Z=0.01$ and 0.03 are shifted upwards (downwards for $M$ magnitude) by difference amount for different magnitude.}
\label{Fig:mag-com}
\end{figure}

\begin{figure}
\centering{
\includegraphics[bb=101 46 582 677,height=4.0cm,width=3.6cm,clip,angle=-90]{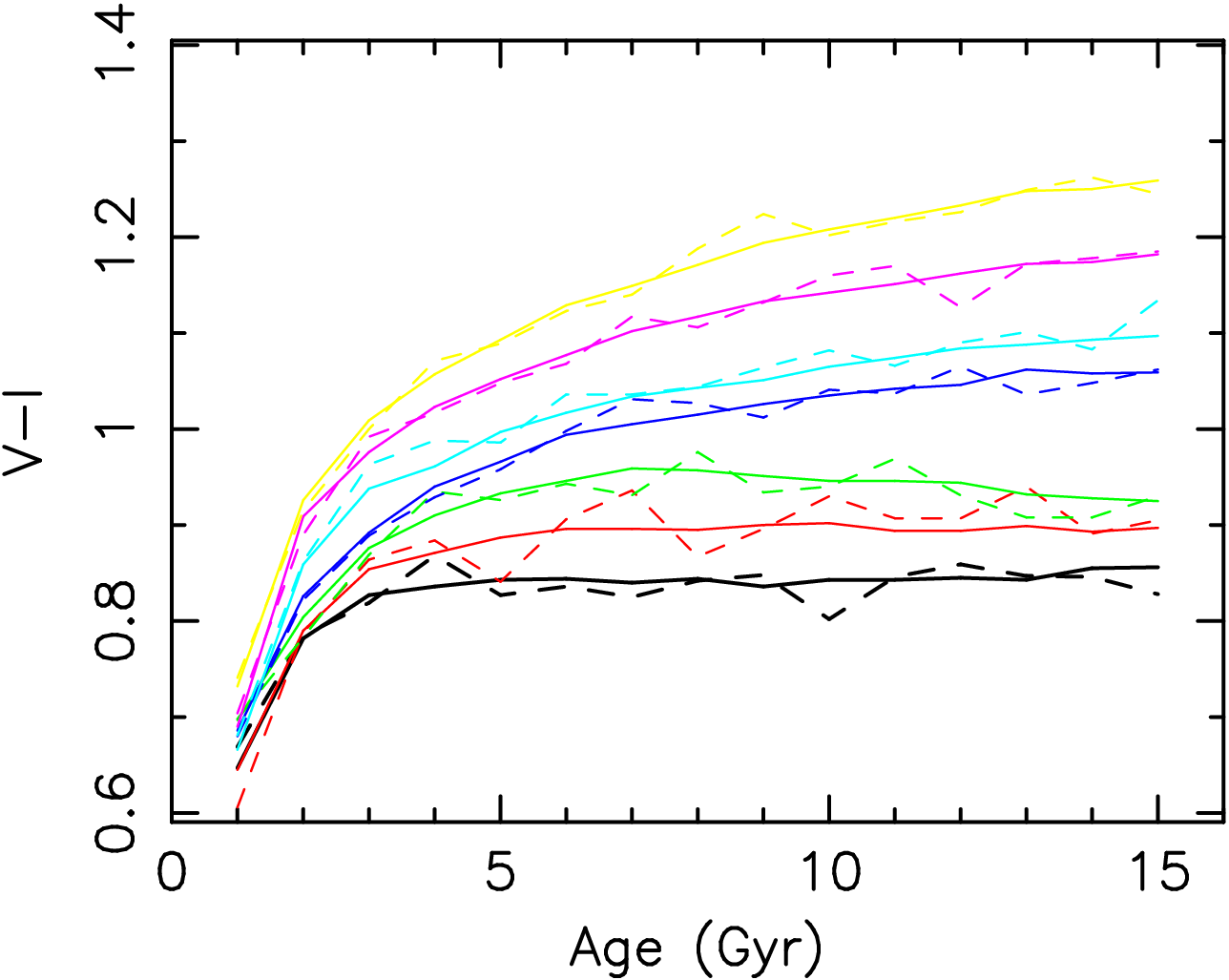}
\includegraphics[bb=101 46 582 677,height=4.0cm,width=3.6cm,clip,angle=-90]{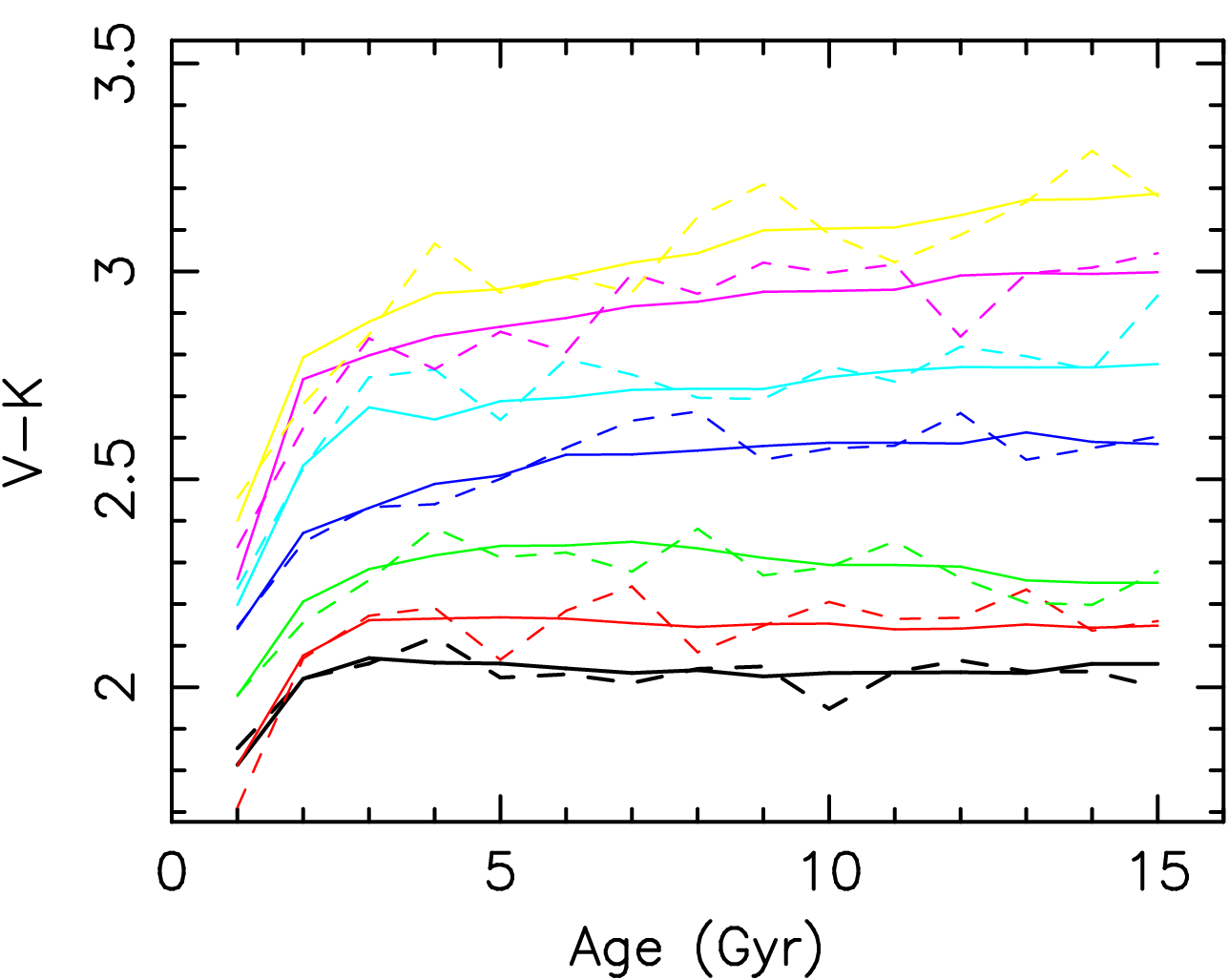}}
\caption{The new (full line) and old (dashed line) integrated $V-I$ and $V-K$ colours for Model A. From top to bottom $Z$=0.03, 0.02, 0.01, 0.004, 0.001, 0.0003 and 0.0001}
\label{Fig:col-com}
\end{figure}

For the sake of clarity, we call the results calculated from the samples of $1 \times 10^6$ binaries in BSPs old one, the results presented in this paper new one.

\subsection{Method of reducing the fluctuations}
To reduce the fluctuations in the results in IR passbands, we need to
increase the number of binary systems in BSPs by Monte Carlo simulation.
One method is to increase directly the number of binary systems, this will
increase the consumption of calculation-time. Another method is to only
increase the number of some binary systems on the basis of the original samples, for which the contribution to
IR light is larger and the number is smaller. Because these binary systems
always locate within the specific ranges of initial input parameters (e.g.,
the primary mass $m_1$, the secondary mass $m_2$ and orbital separation
$a$), so it is called {\sl 'patched'} Monte Carlo simulation method.

In this paper we adopt the second method.
In order that this method can be understood easily, we take the integrated $K$-luminosity ($L_{\rm K}$) as example to explain what is the {\sl 'patched'} Monte Carlo simulation method and how to get the results:
\begin{itemize}
\item First, we use Monte Carlo Simulation method to produce $1 \times 10^6$ binaries, then, evolve them and obtain the integrated $K$-luminosity ($L_{\rm K,0} = \sum_{i=1}^{1 \times 10^6} L_{{\rm K},i}$, i.e., the old result) by adding the $K$-luminosity of all stars in this sample (the same procedure as in our original papers);
\item Second, determine which stars have larger contribution to $K$-light and the number is smaller, these factors lead to the fluctuations in the results involving $K$-band, from the sample of $1 \times 10^6$ binaries;
\item Third, according to the results of binary evolution, determine the initial input parameter ranges ('{\sl patched}' regions) spanned by these binaries (determined from the second step). For Models A and B the {\sl 'patched'} ranges in the initial input parameter spaces are different;
\item Forth, obtain the integrated $K$-luminosity for those stars in the {\sl 'patched'} regions ($L_{\rm K,p,0}$) from the sample of $1 \times 10^6$ binaries;
\item Fifth, obtain the {\sl 'patched'} sample of binaries, for which initial input parameters locate within the {\sl 'patched'} regions, from the sample of $N/10^6$ binaries, then evolve them and obtain their integrated $K$-luminosity ($L_{\rm K,p,2}$);
\item At last, obtain the new $K$-luminosity by using the following formula:
\begin{equation}
L_{\rm K}= L_{\rm K,0} - L_{\rm K,p,0} + (L_{\rm K,p,0} +L_{\rm K,p,2})/(1+N)
\end{equation}
\end{itemize}
The {\sl 'patched'} Monte Carlo simulation method is the sum of the first and fifth steps.

As an illustration of the second step in the above procedure,
in Fig.~\ref{Fig:kcontri} we give the isochrone
and the contribution of stars along the isochrone to $K$-band light
(note the real contribution should to be $10^{(-0.4 \cdot K)}$) for
solar-metallicity 1-Gyr BSPs without BIs (Model B). In left panel
various abbreviations are used to denote the evolutionary phases. They
are as follows: 'MS' stands for main-sequence stars, which are divided
into two phases to distinguish deeply or fully convective low-mass
stars  ($m \la 0.7 {\rm M_{\sun}}$, therefore Low-MS) and stars of higher
mass with little or no convective envelope ($m \ga 0.7 {\rm M_{\sun}}$);
'HG' stands for Hertzsprung gap; 'CHeB' stands for core helium burning;
'EAGB' stands for early AGB; 'PEAGB' (post EAGB) refers to those phases
beyond the 'EAGB'-including thermally pulsing giant branch/protoplanetary
nebula/planetary nebula(TPAGB/PPN/PN).
In right panel of Fig.~\ref{Fig:kcontri} two solid rectangles correspond
to the GB stars with $ {\rm log} (L/L_{\sun}) \simeq 2.0$ and the GB
stars on the tip, respectively. From it we see that $K$-band light is
dominated by cooler and luminous GB and AGB stars. Also is true for Model
A.
%

\subsection{Comparison with the old results}

Because that GB, EAGB and PEAGB stars with ${\rm log} (L/L_{\sun}) \ga 2.0$ are those stars for which the number are needed to increase to deduce the fluctuations in the IR bands (see above). In Fig.~\ref{Fig:iso-adp} we give the comparison of the distribution of these stars in Hertzprung-Russel (HR) diagram between old and new versions for Model A. The age $\tau=10$\,Gyr and metallicity $Z=0.0001$ and 0.02. From it we see the number of these stars increases significantly. This will lead to the reduction of the fluctuations in the results in IR bands.

In Fig.~\ref{Fig:mag-com} we present the old and new evolutionary curves of $J,H,K,L,L2$ and $M$ magnitudes for Model A. The metallicity $Z=0.03, 0.01$ and 0.0001. It shows that the fluctuations caused by Monte Carlo simulation are significantly reduced. Moreover, $U,B,V,R$ and $I$ magnitudes have been slightly improved because only slight fluctuations exist in the old results (only at $Z=0.0001$ and late ages).

Moreover, those colours involving $J,H,K,L,L2$ and $M$ magnitudes, such as $V-K$, also are improved. In Fig.~\ref{Fig:col-com} we compare the new $V-I$ and $V-K$ colours with the old ones for Model A. The metallicity $Z=$ 0.03, 0.02, 0.01, 0.004, 0.001, 0.0003 and 0.0001.

Other needed to mention is the results of Model B have been improved.

\section{Results and Comparisons with literature and observations}
In this paper we present the new stellar masses, bolometric magnitude, $U, B, V, R, I,$ $J, H, K, L, L2, M$ magnitudes, the stellar mass-to-light ratios and broad-band colours for Models A and B. The ages of BSPs are in the range of $1-15$\,Gyr and metallicity $Z=0.0001, 0.0003, 0.001, 0.004, 0.01, 0.02$ and 0.03. The data of Model A are given in Appendix.
For completeness, we also obtain the results of BSPs at a logarithmic age interval of 0.05 in the age range $5 \le {\rm log(}\tau{\rm /yr)} \le 9$. This part of data is calculated from the samples of $2 \times 10^7$ binaries, and can be obtained by ftp from our website or require from the first author.

\subsection{Stellar mass}
\begin{figure}
\centering{
\includegraphics[bb=81 34 531 705,height=6.70cm,width=3.5cm,clip,angle=-90]{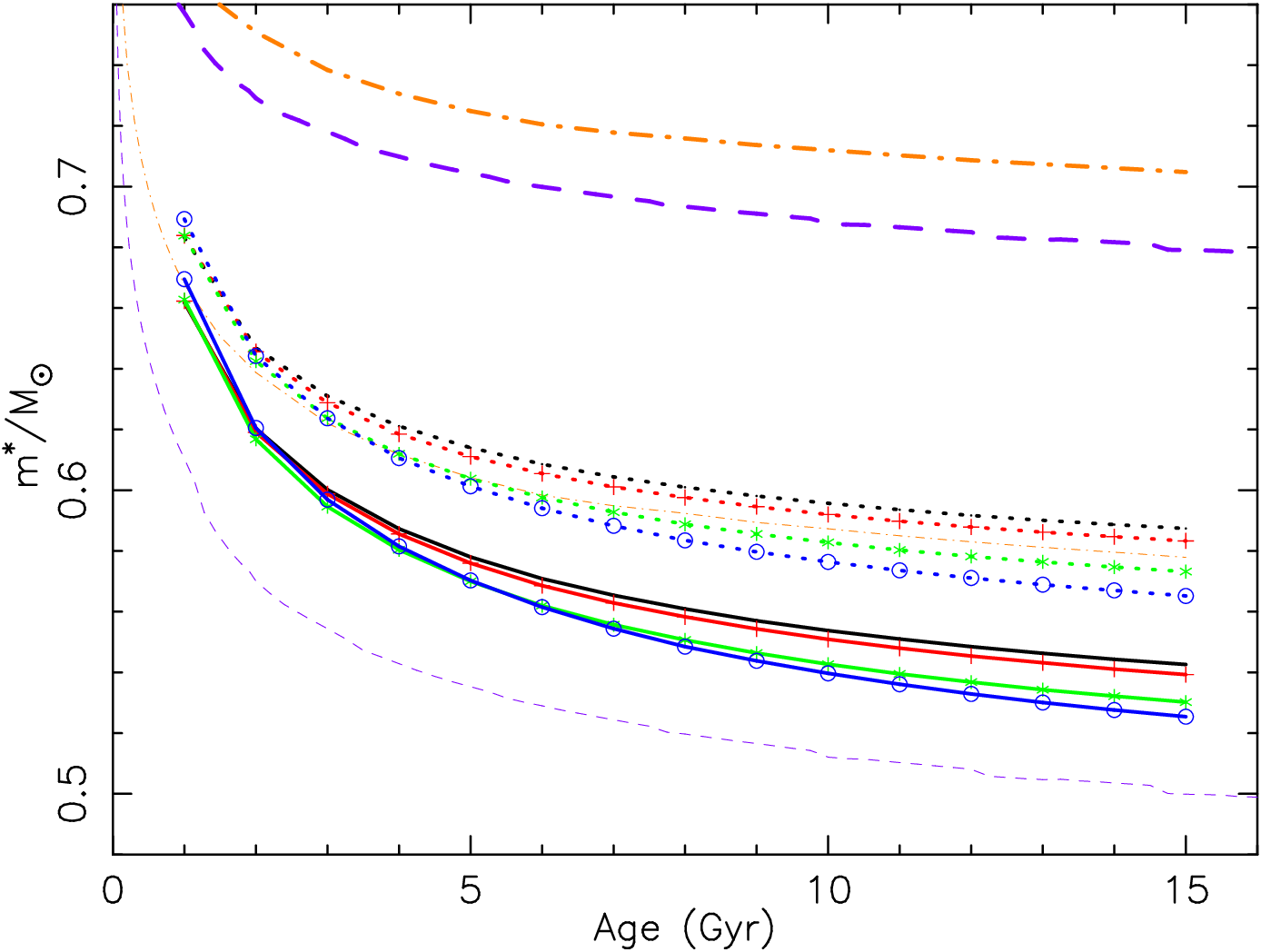}\\
\includegraphics[bb=81 30 583 701,height=6.70cm,width=3.85cm,clip,angle=-90]{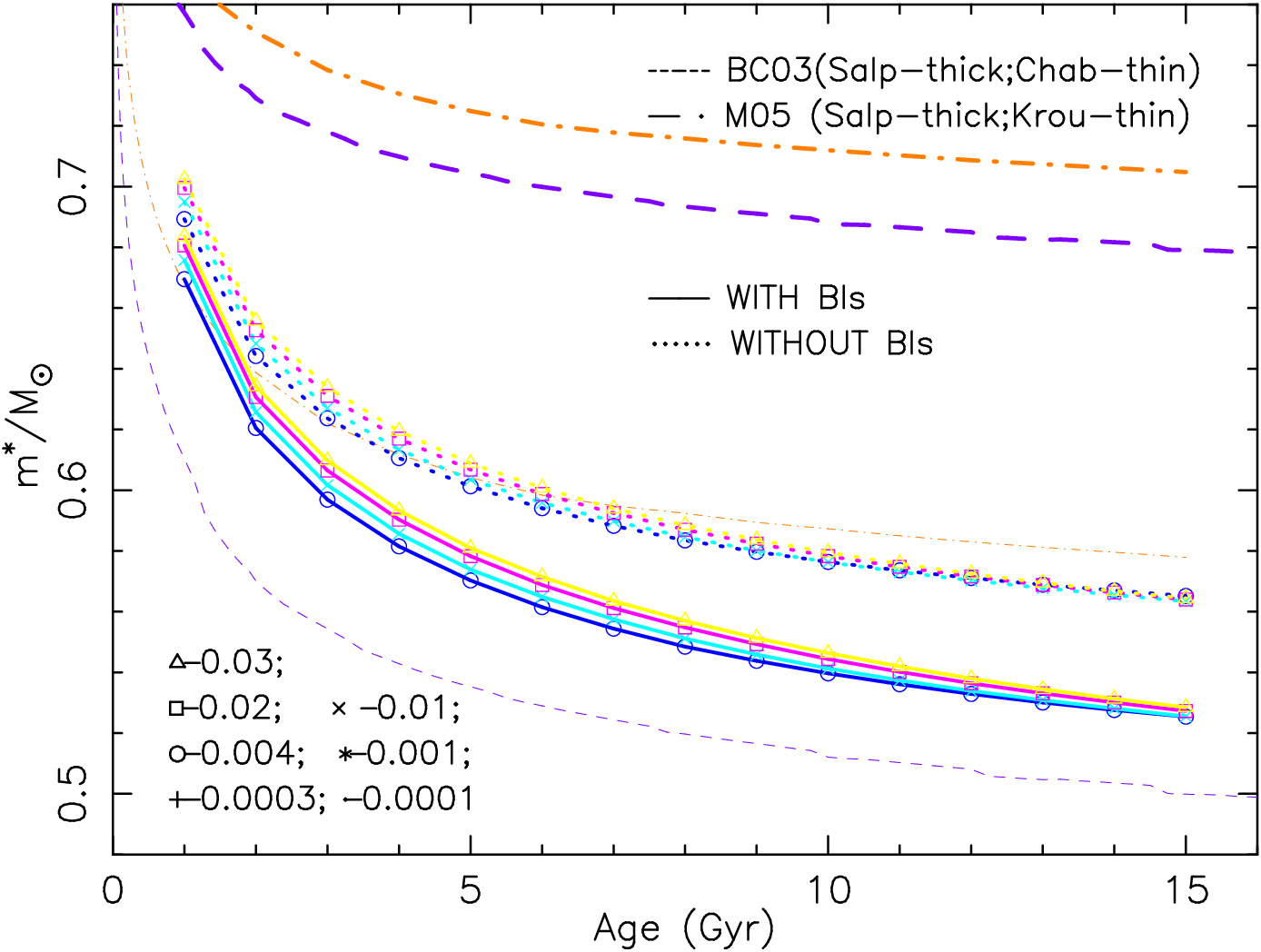}
}
\caption{The stellar mass (normalized to 1\,$\rm M_{\sun}$) of an evolving BSP for Models A (full) and B (dotted) at 7 metallicities. In top and bottom panels $Z=0.0001, 0.0003, 0.001, 0.004$ and $Z=0.03, 0.02, 0.01, 0.004$ from top to bottom, respectively. In both panels the curve of $Z=0.004$ and the results of BC03-S, BC03-C, M05-S and M05-K at solar metallicity are plotted.}
\label{Fig:mt-t}
\end{figure}

\begin{figure}
\centering{
\includegraphics[bb=81 30 583 701,height=6.70cm,width=6.70cm,clip,angle=-90]{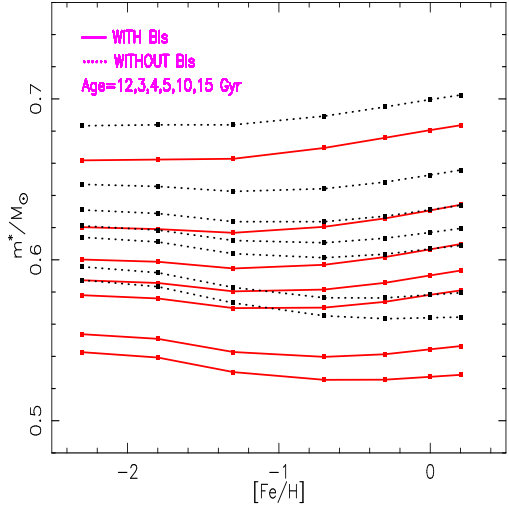}}
\caption{The stellar mass of an evolving BSP as a function of metallicity for Models A (full) and B (dotted). From top to bottom, the ages $\tau = 1, 2,3,4,5,10$ and 10\,Gyr.}
\label{Fig:mt-feh}
\end{figure}

\begin{figure}
\centering{
\includegraphics[bb=81 34 531 705,height=6.70cm,width=3.5cm,clip,angle=-90]{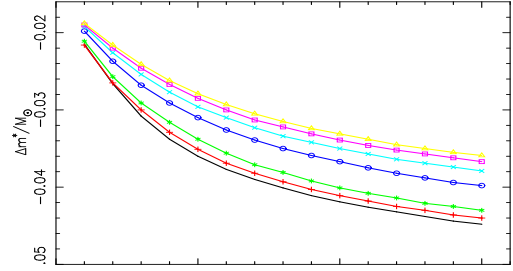}
}
\caption{Difference in the stellar mass of an evolving BSP between Models A and B ($\Delta m^{\star} \equiv m^{\star}_{\rm A} - m^{\star}_{\rm B}$) for 7 metallicites. From bottom to top $Z=0.0001, 0.0003, 0.001, 0.004, 0.01, 0.02$ and 0.03. The symbols have the same meanings as in Fig.~\ref{Fig:mt-t}.}
\label{Fig:dmt-t}
\end{figure}

\begin{figure}
\centering{
\includegraphics[bb=132 128 525 625,height=5.5cm,width=5.5cm,clip,angle=-90]{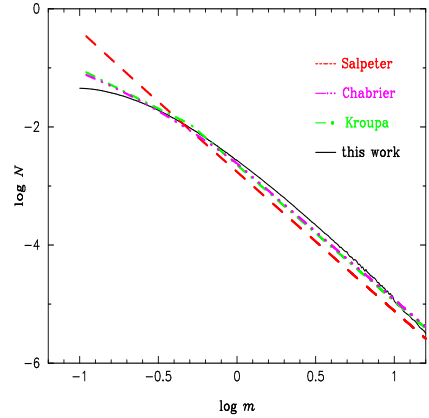}}
\caption{The IMF given by the Salpeter with $\alpha = 2.35$ (dashed), Chabrier (dash-dotted), Kroupa (dashed-dot-dot) IMFs, and the distribution used in this work (solid) at solar metallicity.}
\label{Fig:imf-com}
\end{figure}

Fig.~\ref{Fig:mt-t} shows the stellar mass $m^{\star}$ of an evolving BSP with an initial mass 1\,${\rm M_{\sun}}$ at $Z=0.0001, 0.0003, 0.001, 0.004, 0.01, 0.02$ and 0.03 for Models A and B.
From it we see that the stellar masses of Models A and B decrease with age. During the first Gyr Models A and B lose $\sim$\,33 and 31 per cent of their masses, respectively; then lose $\sim$\,13-15 and $\sim$\,10-13.7 per cent of their masses during the following $1 \le \tau \le 15$\,Gyr.
The correlation between $m^{\star}$ and metallicity $Z$ is not monotonous: at $Z \le 0.01$  $m^{\star}$ decreases with $Z$ (see top panel of Fig.~\ref{Fig:mt-t}), while $Z \ge 0.01$ it increases with $Z$ (see bottom panel of Fig.~\ref{Fig:mt-t}). The correlation between $m^{\star}$ and $Z$ also can refer to Fig.~\ref{Fig:mt-feh}, in it we give the stellar mass as a function of metallicity for Models A and B at ages $\tau = 1,2,3,4,5,10$ and 15\,Gyr. The discrepancy in $m^{\star}$ caused by the difference in metallicity reaches $\sim$\,2$\%$ and $\sim$\,3$\%$ at an age of 15\,Gyr for Models A and B.

Comparing with Model B, Model A has the lower $m^{\star}$ at all metallcities and ages. The absolute value of $\Delta m^{\star}$ ($\equiv m^{\star}_{\rm A} - m^{\star}_{\rm B}$), the difference in the $m^{\star}$ between Models A and B, decreases with $Z$, at $Z=0.0001$ it reaches $\sim$\,4.5 per cent at an age of $\tau=15$\,Gyr (see Fig.~\ref{Fig:dmt-t}). Comparing with the difference caused by metallicity ($\sim$\,3\% at the most), the discrepancy caused by BIs is even greater. In our work we constrain Model A holds the same binary samples as Model B, so the effect of the diffenence in the samples is rejected.
The lower $m^{\star}$ of Model A is caused mainly {by more mass loss ejected to inter-stellar material (ISM) from binary systems via type Ia supernovae explosion (leave nothing) and common-envelope ejection. Also, BIs can make the lifetime of some components in binary systems shorter, so they earlier evolve to remnants.}

In both panels of Fig.~\ref{Fig:mt-t} we also give the values of BC03 with the \citet{sal55} and \citet{cha03} IMFs (therefore BC03-S \& BC03-C) and M05 with the \citet{sal55} and \citet{kro01} IMFs (therefore M05-S \& M05-K) at solar metallicity. From the comparison we see that at solar metallicity the stellar masses of Models A and B are lower than those of BC03-S and M05-S, while greater than that of BC03-C at all ages. For M05-K, it is greater than that of Model A at all ages, while greater than that of Model B only at age $\tau \ge 6$\,Gyr (within the amount of 2\%). This is mainly caused by the differences in the adoption of the mass loss in stellar evolutionary models, initial mass function (IMF) and the differences in stellar evolutionary parameters (such as, lifetime).
The stellar evolutionary model used by each EPS model adopts different mass loss.
In our work the IMF of the primary is chosen from the approximation to the IMF of \citet{mil79} as given by \citet{egg89}, the initial mass ratio distribution takes an uniform form; BC03 use the Salpeter  ($\phi ({\rm log} m) = {\rm C} \cdot m^{1-\alpha}, \alpha=2.35$) and Chabrier IMFs
$$ 
\phi ({\rm log} m) = \Biggl\{
\matrix{
{\rm C_1} \cdot {\rm exp}[-\frac{({\rm log}m-{\rm log}m_{\rm c})^2}{2\sigma^2}],  m \le 1\,{\rm M_{\sun}}  \cr
 \cr
{\rm C_2} \cdot m^{-1.3}, \ \ \ \ \ \ \ \ \ \ \ \ \ \ \ \ \  m > 1\,{\rm M_{\sun}} \cr
}
$$
in which $m_{\rm c} = 0.08$, $\sigma= 0.69$, ${\rm C_1} = 0.851$ and ${\rm C_2} = 0.238$; M05 use the Salpeter and Kroupa IMFs, which is described as:
$$
\phi ({\rm log} m) = \Biggl\{
\matrix{
{\rm C_{1,K}} \cdot m^{-0.3}, \ \ \ \ \ \ \ \ \  0.1 \le m \le 0.5\,{\rm M_{\sun}} \cr
 \cr
{\rm C_{2,K}} \cdot m^{-1.3}, \ \ \ \ \ \ \ \ \ \ \ \ \ \ \ \ \  m > 0.5\,{\rm M_{\sun}} \cr
}
$$
in which ${\rm C_{1,K}}= 0.449$ and ${\rm C_{2,K}}= 0.224 $. In Fig.~\ref{Fig:imf-com} we plot four of the IMFs, from it we see that the IMF used in our work has a relatively high fraction of massive stars than the Salpeter, Chabrier and Kroupa IMFs.

\subsection{Magnitudes}

\begin{figure*}
\leftline{
\includegraphics[bb=75 37 530 700,height=4.00cm,width=4.00cm,clip,angle=-90]{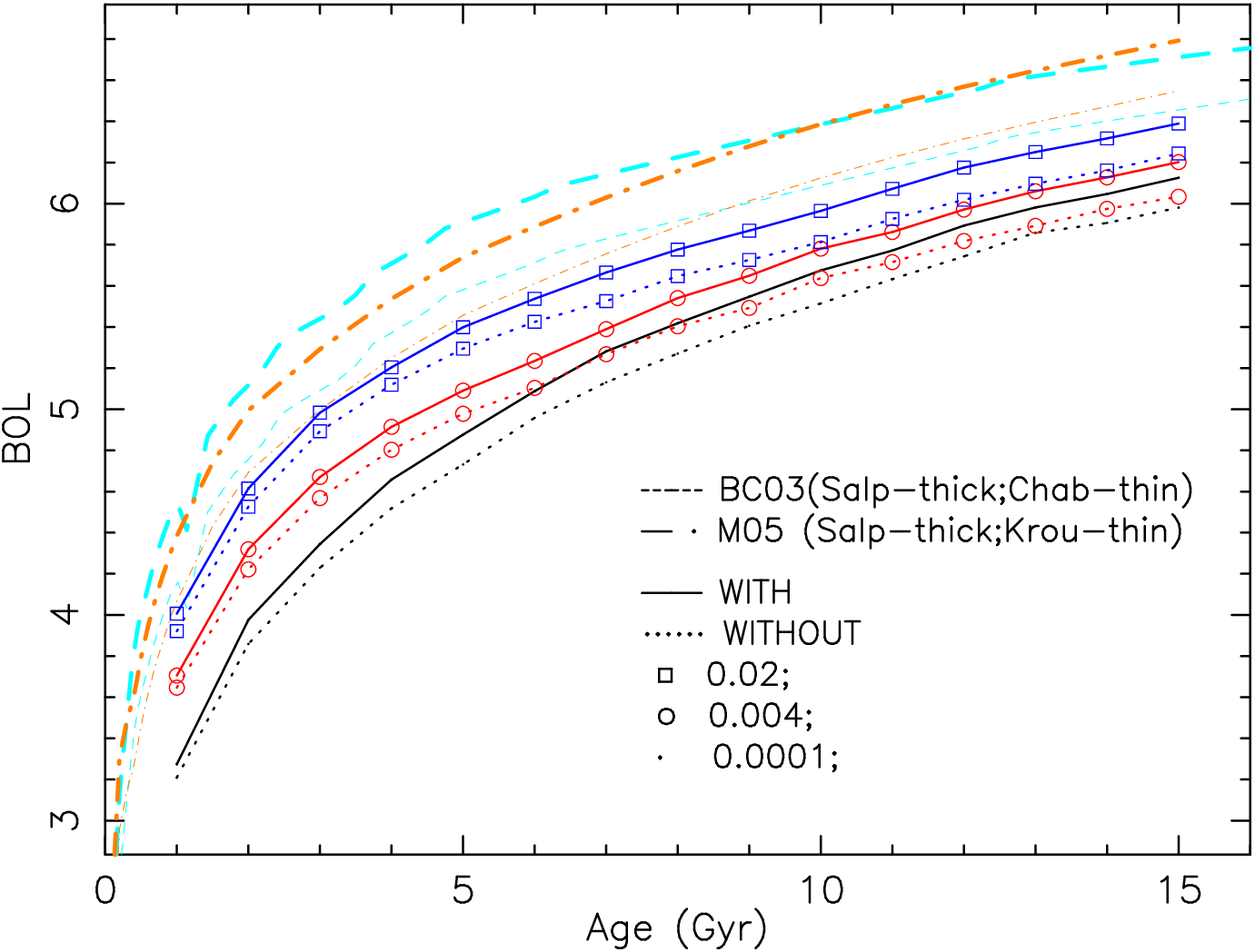}
\includegraphics[bb=75 37 530 700,height=4.00cm,width=4.00cm,clip,angle=-90]{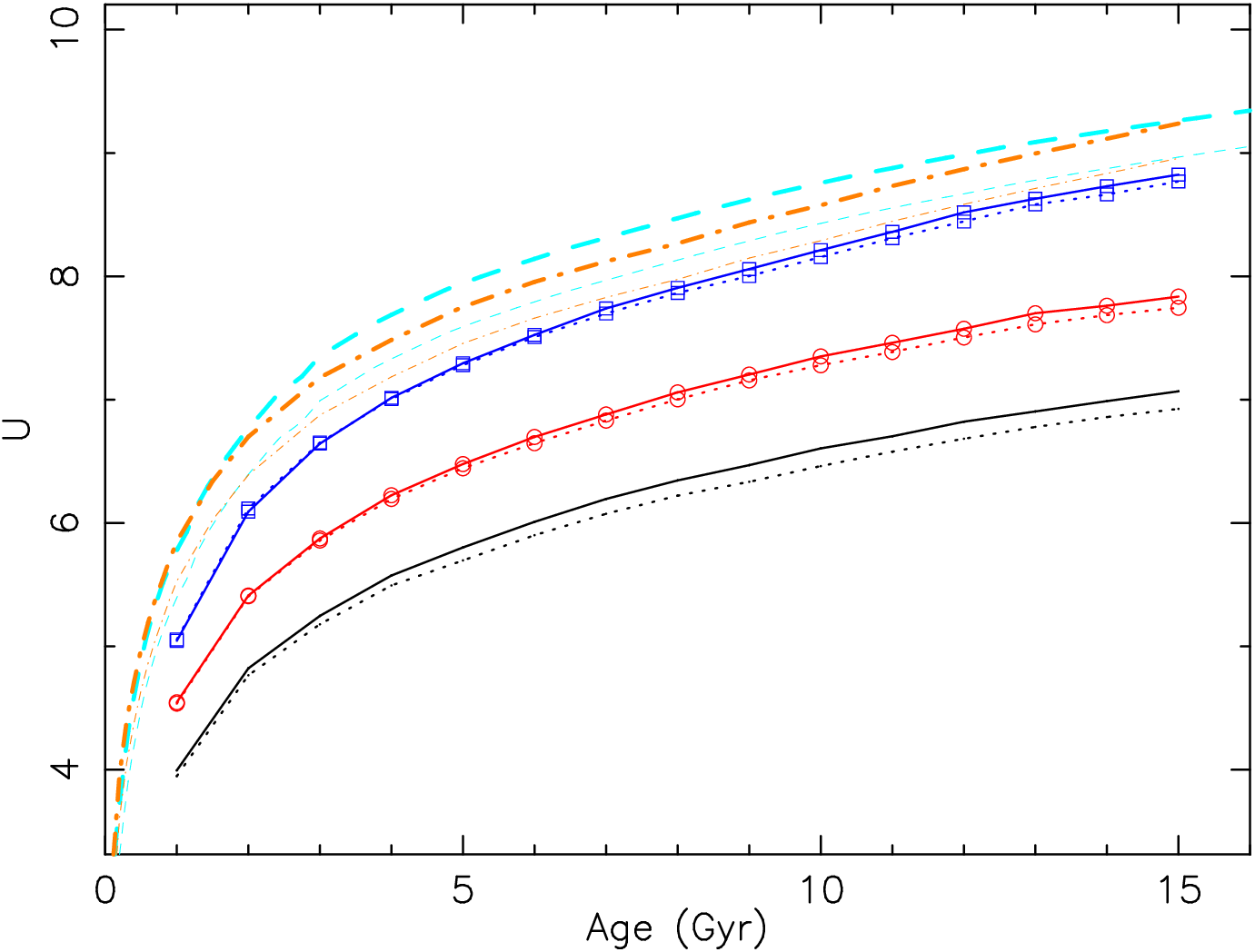}
\includegraphics[bb=75 37 530 700,height=4.00cm,width=4.00cm,clip,angle=-90]{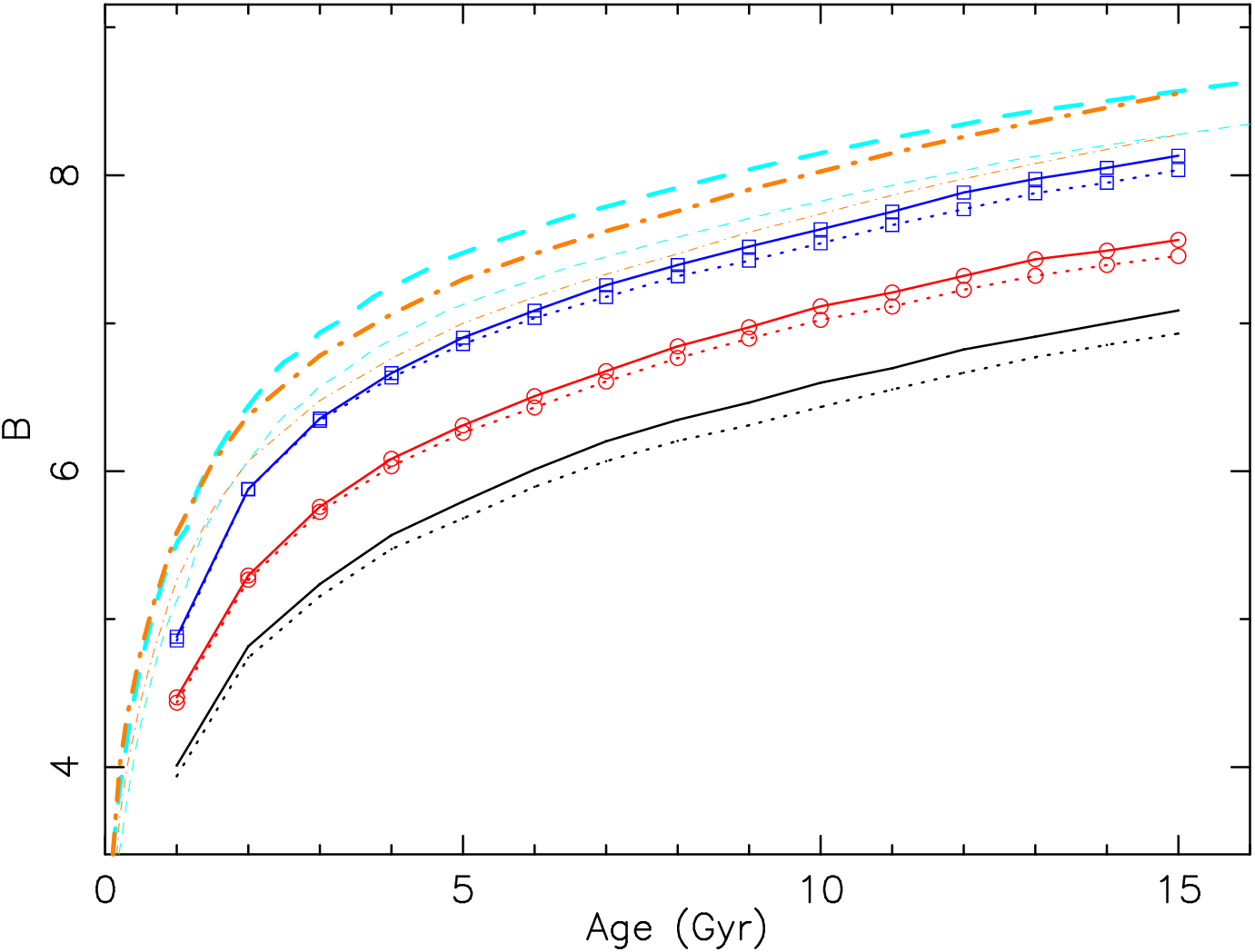}
\includegraphics[bb=75 37 531 700,height=4.00cm,width=4.00cm,clip,angle=-90]{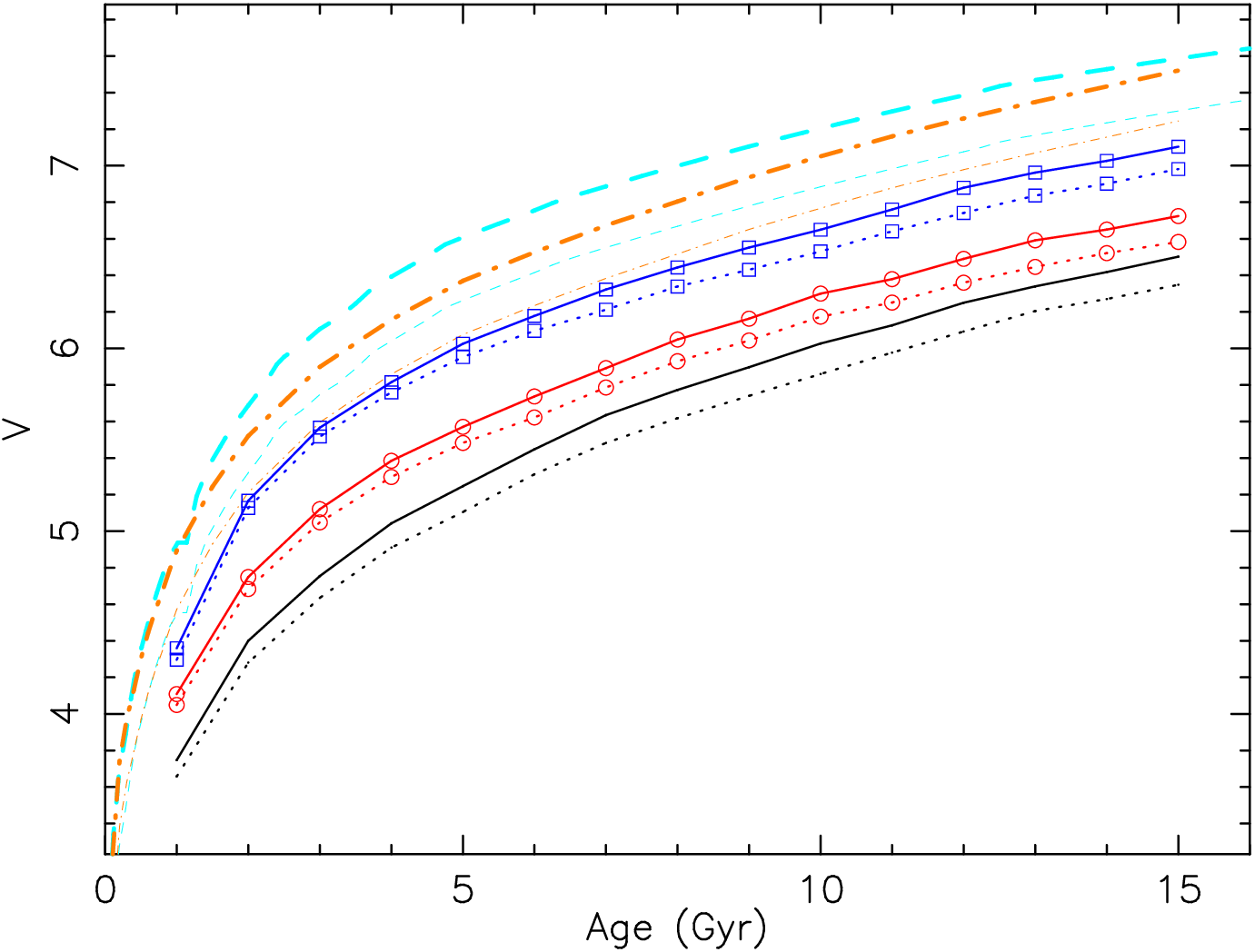}
}
\leftline{
\includegraphics[bb=92 37 580 700,height=4.00cm,width=4.50cm,clip,angle=-90]{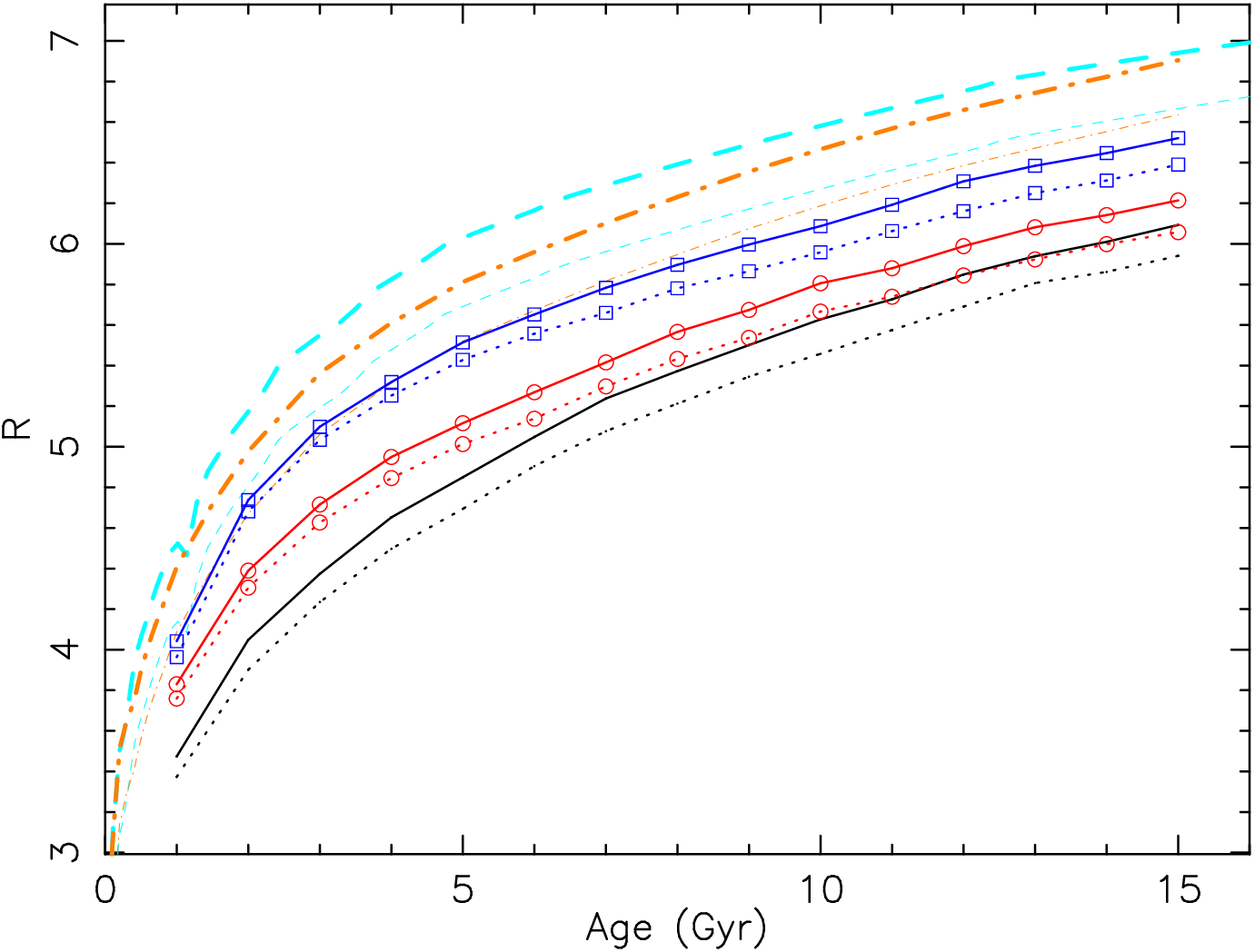}
\includegraphics[bb=92 37 580 700,height=4.00cm,width=4.50cm,clip,angle=-90]{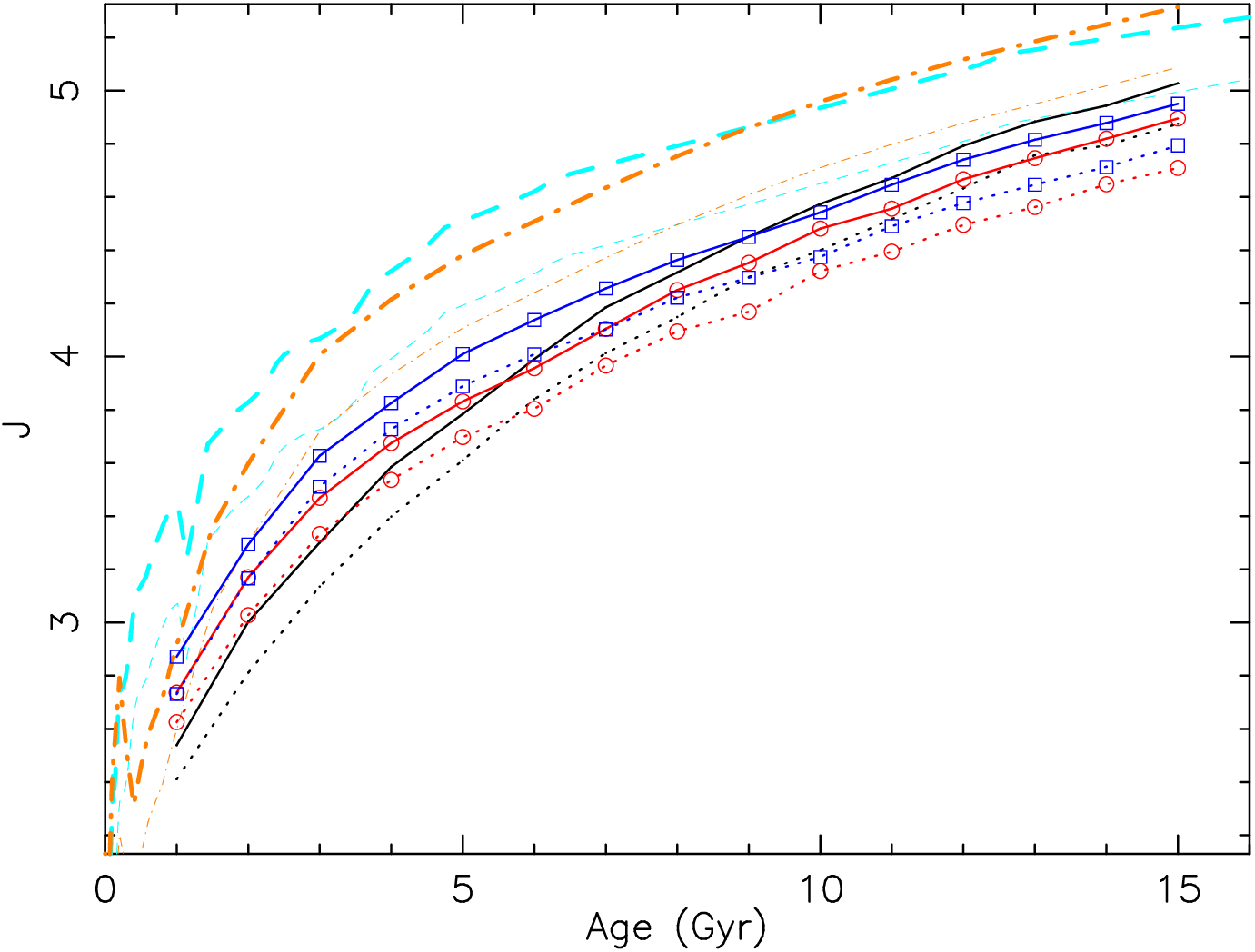}
\includegraphics[bb=92 37 580 700,height=4.00cm,width=4.50cm,clip,angle=-90]{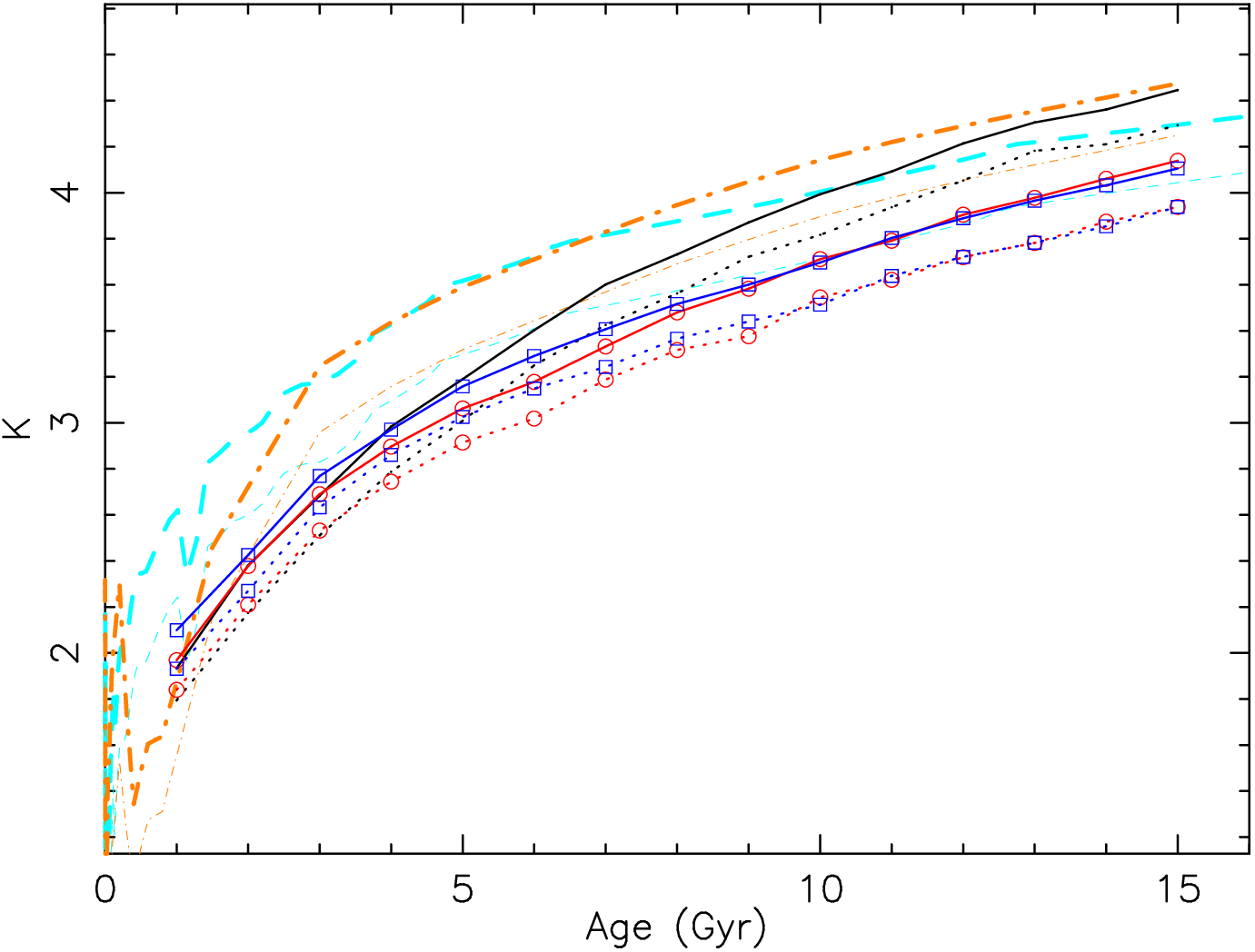}
\includegraphics[bb=92 37 580 700,height=4.00cm,width=4.50cm,clip,angle=-90]{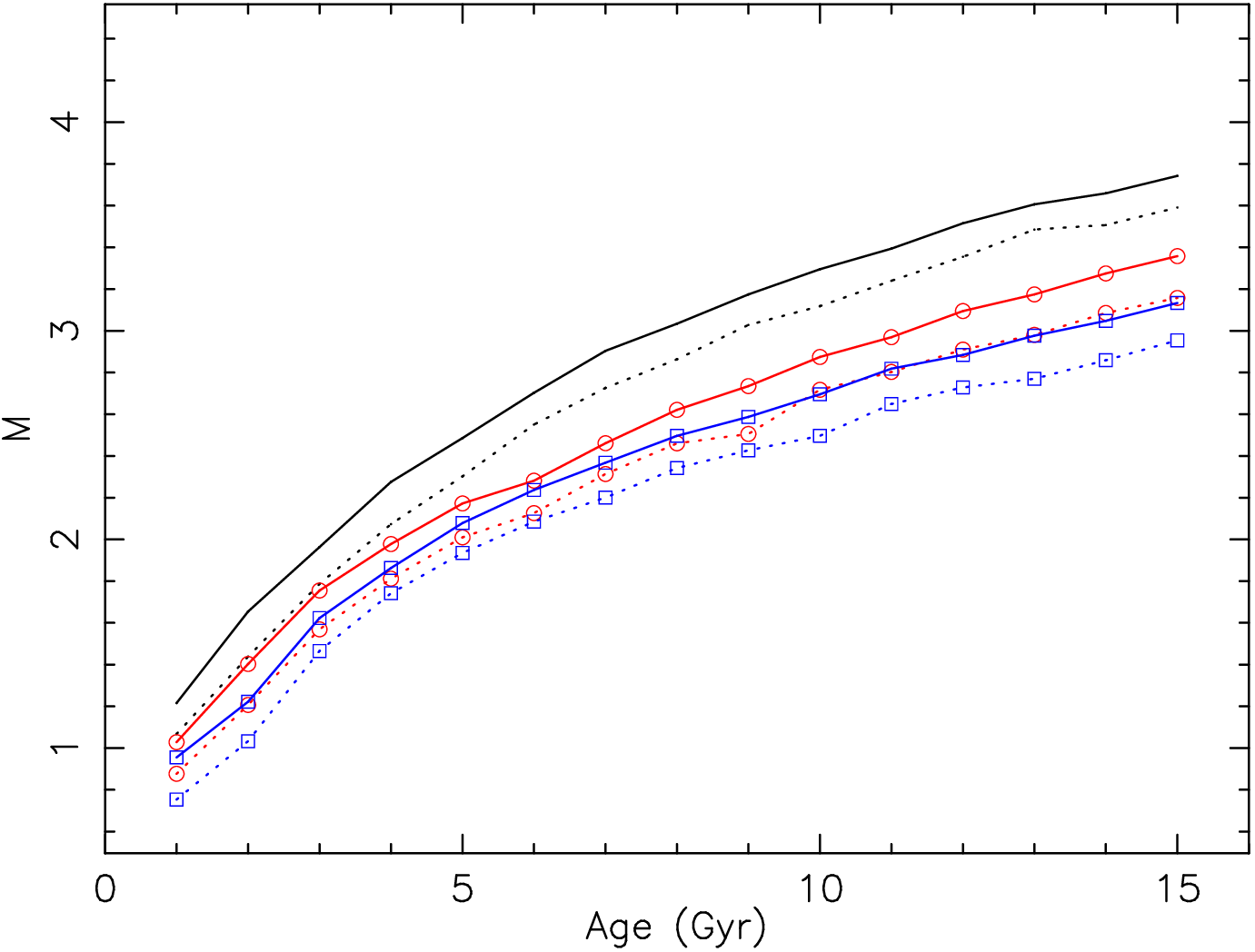}
}
\caption{The bolometric magnitude (${BOL}$), $U, B, V, R, J, K$ and $M$ magnitudes for Models A and B only at $Z=0.02, 0.04$ and 0.0001. Also shown are the results of BC03-S (dashed, thick), BC03-C (dashed, thin), M05-S (dot-dash, thick) and M05-K (dot-dash, thin) except $M$ magnitude at solar metallicity.
$\star$ Note that $K$ magnitude of BC03 is obtained by $J-(J-K)$.
}
\label{Fig:mags-all}
\end{figure*}

\begin{figure}
\leftline{
\includegraphics[bb=122 97 475 624,height=4.00cm,width=4.00cm,clip,angle=-90]{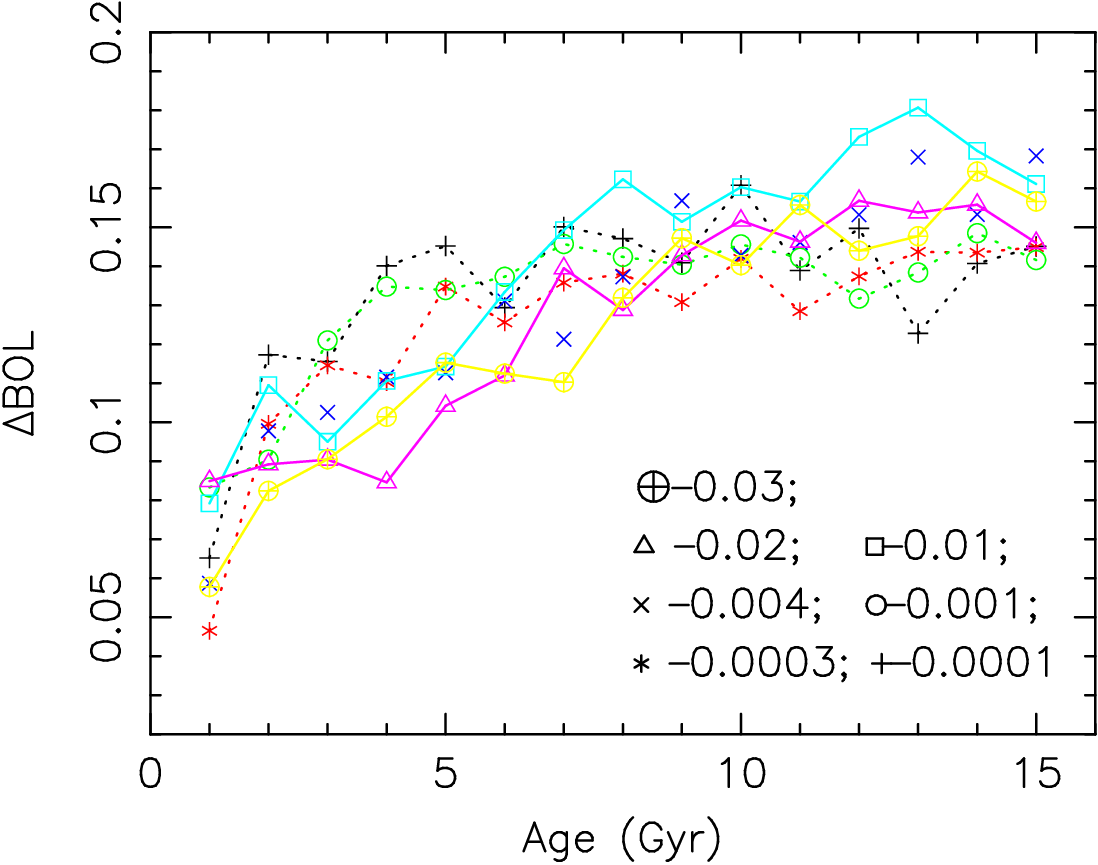}
\includegraphics[bb=122 97 475 624,height=4.00cm,width=4.00cm,clip,angle=-90]{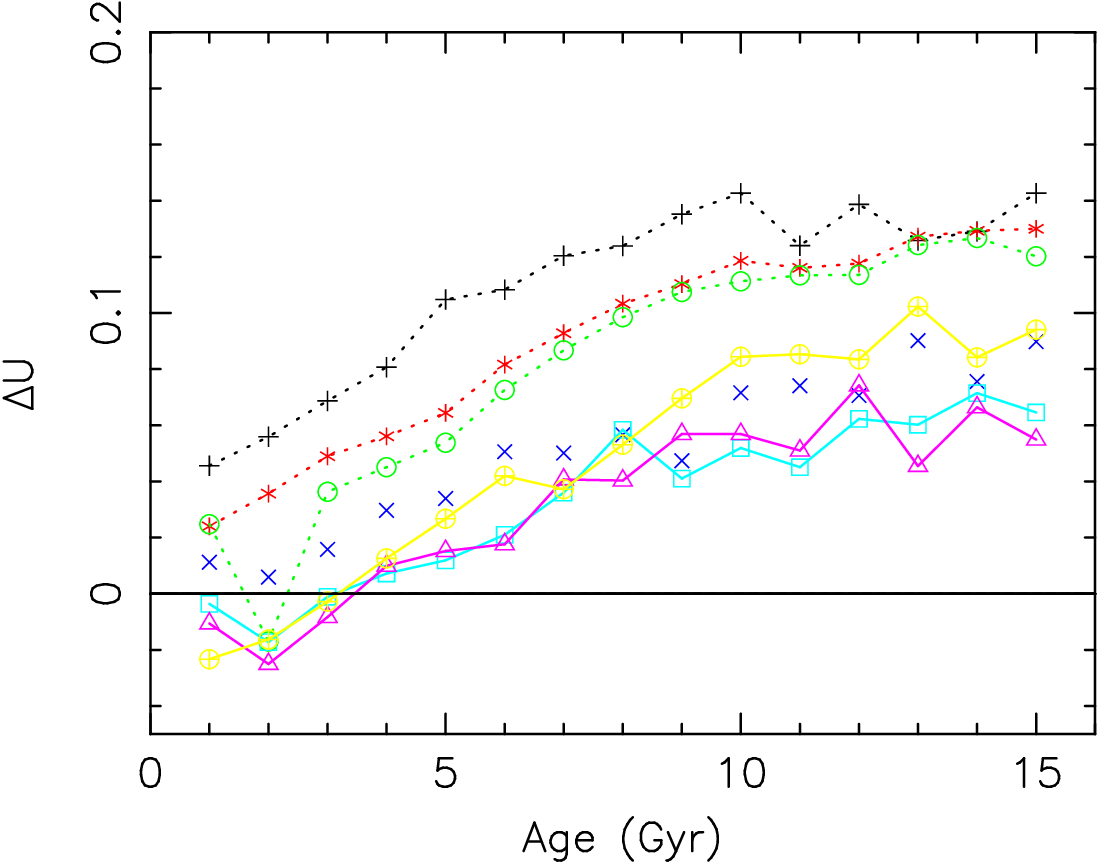}
}
\leftline{
\includegraphics[bb=156 97 475 624,height=4.00cm,width=3.80cm,clip,angle=-90]{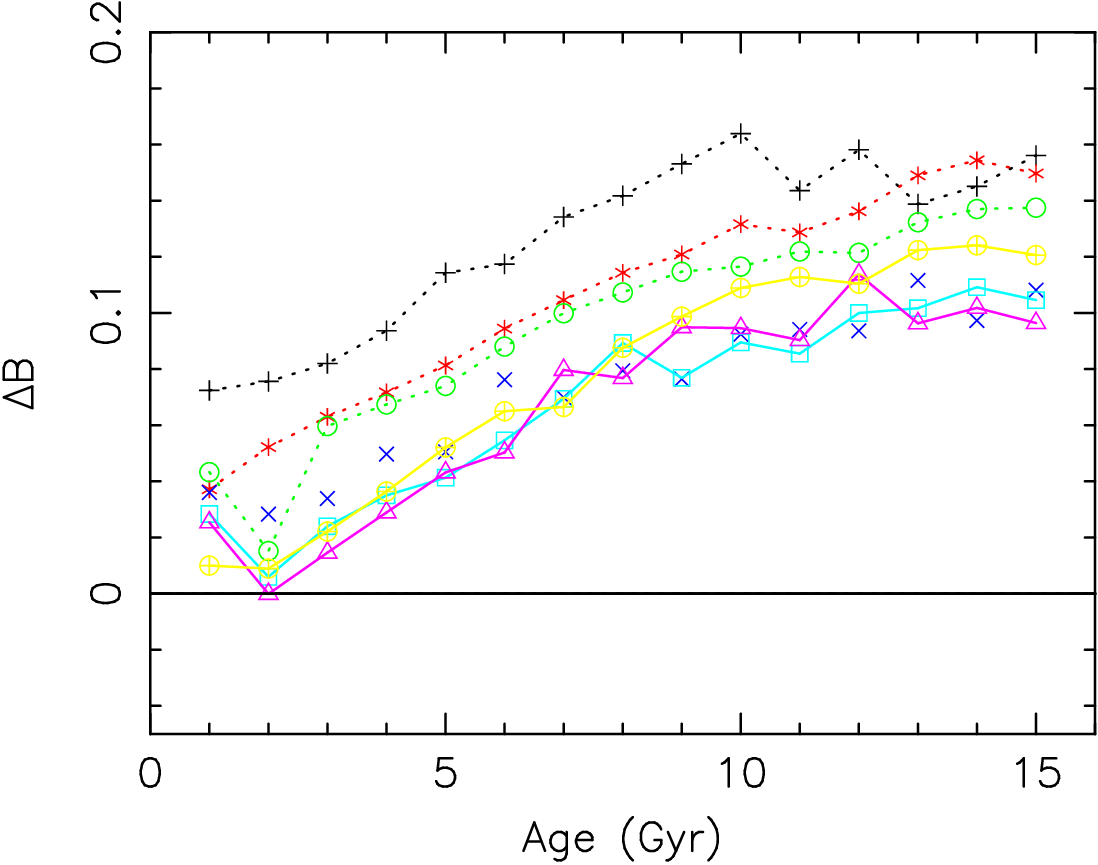}
\includegraphics[bb=156 97 475 624,height=4.00cm,width=3.80cm,clip,angle=-90]{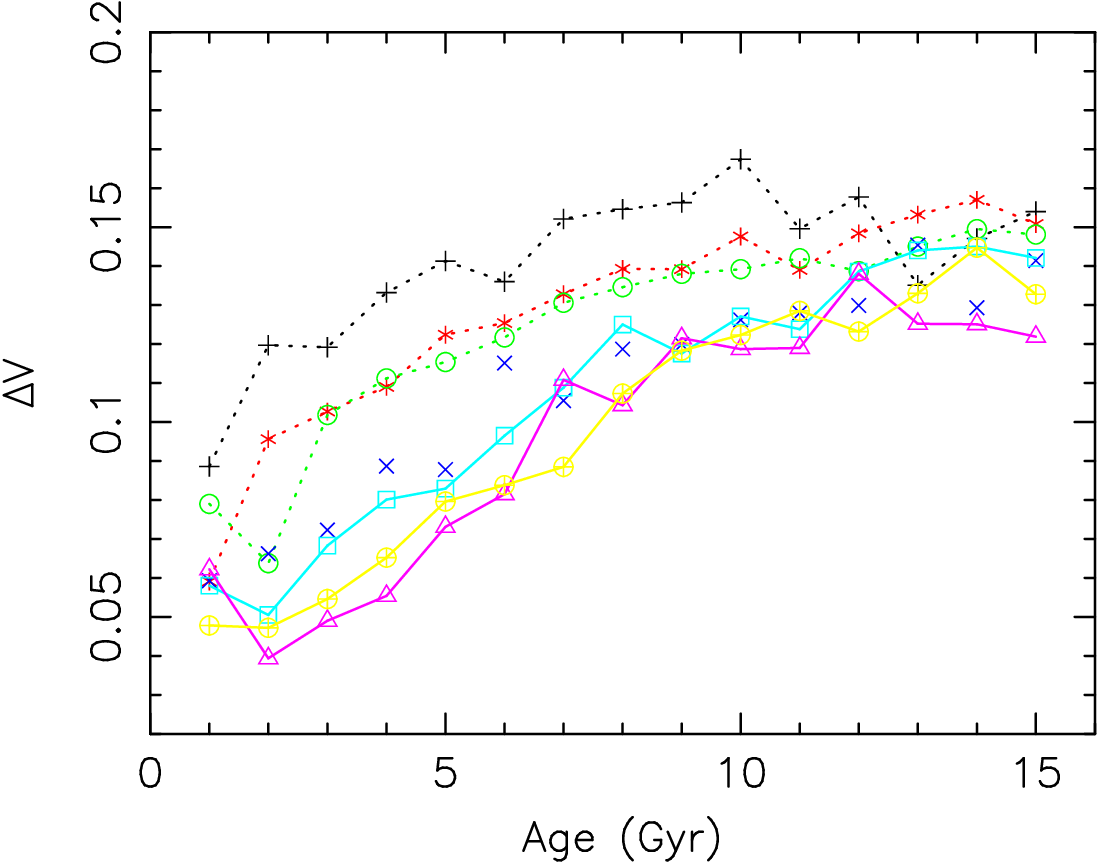}
}
\leftline{
\includegraphics[bb=156 97 537 624,height=4.00cm,width=4.50cm,clip,angle=-90]{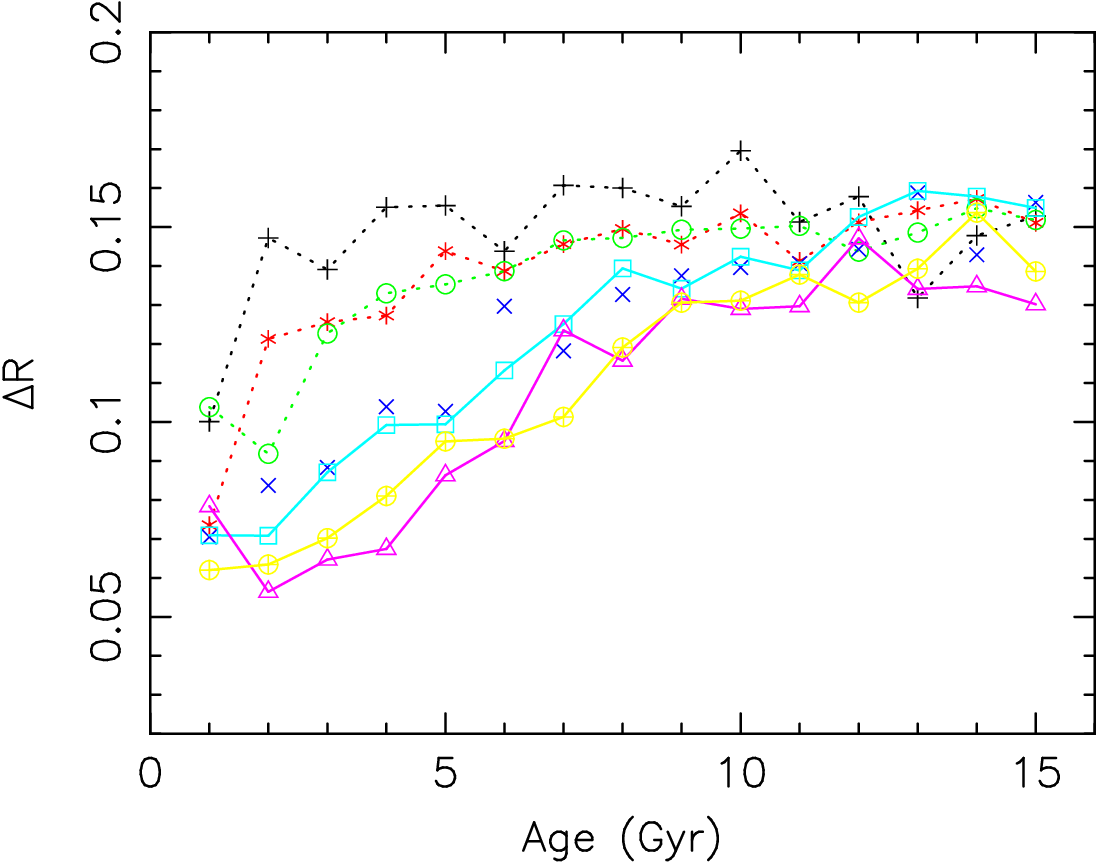}
\includegraphics[bb=156 97 537 624,height=4.00cm,width=4.50cm,clip,angle=-90]{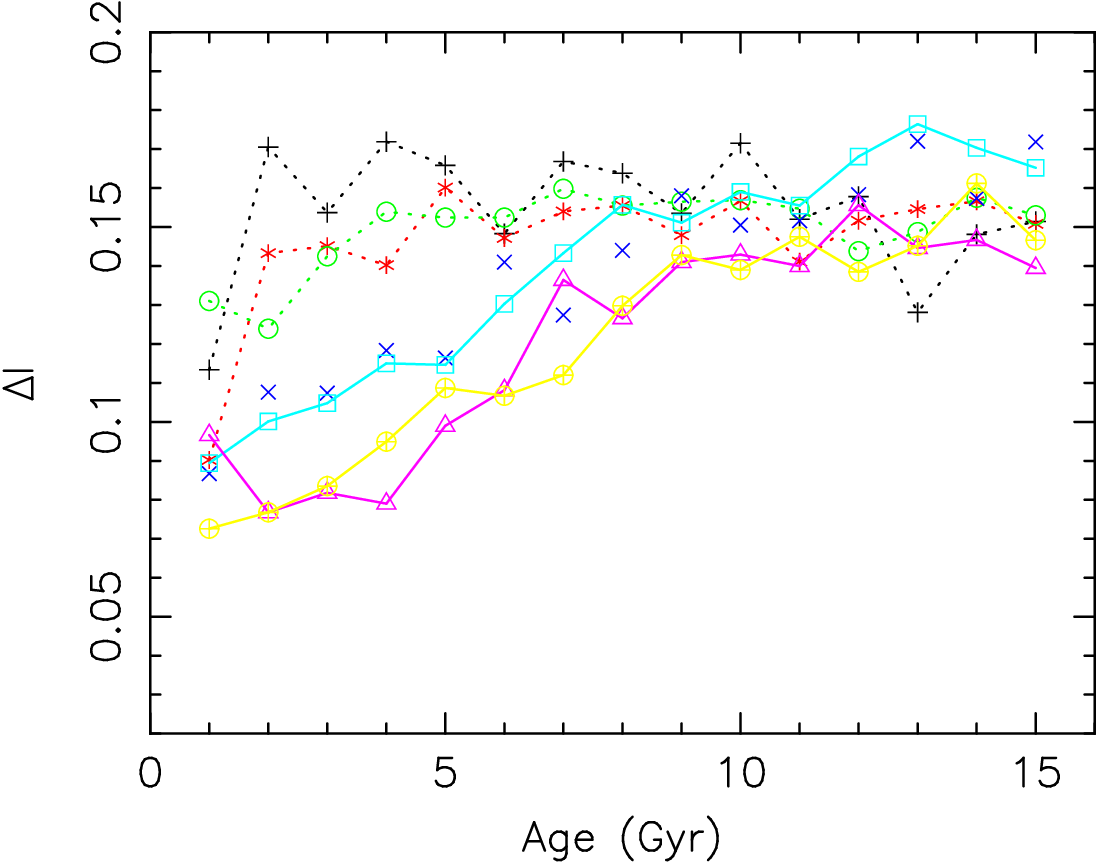}
}
\caption{Differences in the $BOL, U, B, V, R$ and $I$ magnitudes between Models A and B for 7 metallicities. The curves of $Z=0.0001, 0.003, 0.001$ are connected by dotted lines; those of $Z=0.01, 0.02, 0.03$ are by full lines; no line for $Z=0.004$.}
\label{Fig:dmag-t-z}
\end{figure}

\begin{figure}
\centering{
\includegraphics[bb=136 125 527 624,height=6.70cm,width=7.00cm,clip,angle=-90]{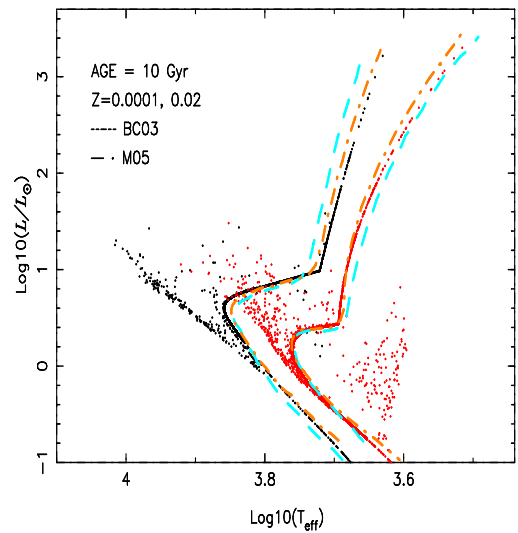}}
\caption{Comparison of isochrones among ours (dots), BC03 (dashed line) and M05 (dot-dash line). The age $\tau=10$\,Gyr and metallicity $Z=0.0001$ and 0.02.}
\label{Fig:iso-com}
\end{figure}

\subsubsection{Bolometric magnitude}
The bolometric magnitude ${BOL}$ increases with age and metallicity for both Models A and B because the luminosity of populations decreases with age and metallicity. At age $\tau \ge 5$\,Gyr the discrepancy in $BOL$ among $Z=0.0001, 0.0003$ and 0.001 is smaller, at age $\tau < 5$\,Gyr the discrepancy between $Z=0.0001$ and 0.0003 becomes to be insignificant. In top left panel of Fig.~\ref{Fig:mags-all} we only present the bolometric magnitude $BOL$ at $Z=0.02, 0.004$ and 0.0001 for Models A and B for the aim of clarity.

Comparing with Model B, Model A has larger $BOL$ at all metallicities and ages, therefore lower bolometric luminosity $L_{\rm BOL}$. The discrepancy in $BOL$ between Models A and B, $\Delta BOL$ ($\equiv BOL_{\rm A} - BOL_{\rm B}$), increases with age and is independent of metallicity (top left panel of Fig.~\ref{Fig:dmag-t-z}). That is to say, the ratio of the bolometric flux of Model A to B ($F_{\rm BOL,A}/F_{\rm BOL,B} \equiv 10^{-0.4 \Delta BOL}$), decreases with age and independent of metallicity. The difference $\Delta BOL$ ranges from $\sim$\,0.07 to 0.16 mag, and the ratio of the bolometric flux ($F_{\rm BOL,A}/F_{\rm BOL,B}$) ranges from $\sim$\,0.93 to 0.87 on average. Lower luminosity of Model A and the increase of $\Delta BOL$ with age are partly caused by that BIs make some components in binary systems fainter or evolve earlier to remnants.

Comparing with BC03 (see top left panel of Fig.~\ref{Fig:mags-all}), Models A and B have lower $BOL$ and larger $L_{\rm BOL}$ at solar metallicity. This is partly due to the warmer GBs than those of the pavoda models which used by BC03 (Fig.~\ref{Fig:iso-com}), other magnitudes are similar to the corresponding ones at solar metallicity.

Comparing with M05 (see top left panel of Fig.~\ref{Fig:mags-all}), Models A and B also have lower $BOL$ at solar metallicity. From Fig.~\ref{Fig:iso-com} we see that GB stars in M05 models have higher temperature than those of us at solar metallicity, but the number of these stars is smaller than ours. For populations of age $1 \le \tau \le 15$\,Gyr, the mass of stars on GB stage ranges from $\sim$\,2.125 to 0.949\,${\rm M_{\sun}}$ at solar metallicity, from Fig.~\ref{Fig:imf-com} we we see the number of these stars in M05 models is smaller than that in our models.

\subsubsection{Other magnitudes}
In general, $U,B,V,R,I,J,H,K,L,L2$ and $M$ magnitudes increase with age. Their variation with metallicity accords to the following rule: in the blue end $U, B, V$ magnitudes increase significantly with metallicity ({\bf called U-type}), $R$ and $I$ magnitudes also increase with metallicity, but insignificant for $Z<0.0003$ at all ages, $\tau > 10$\,Gyr for $R$ and $\tau > 6$\,Gyr for $I$ magnitudes at $Z<0.001$ ({\bf called R-type}); in the red end $M$ magnitude decreases with metallicity ({\bf called M-type}); in the $J,H,K,L,L2$ passbands, the magnitude curves transit gradually from R-type to M-type with increasing wavelength ({\bf called J-type}).
For the sake of it size and clarity, in Fig.~\ref{Fig:mags-all} we only give $U, B, V, R, J, K$ and $M$ magnitudes except for $BOL$ for Models A and B only at $Z=0.02, 0.004$ and 0.0001.

The difference in $X (=U,B,...,M)$ magnitude between Models A and B, $\Delta X$ ($=X_{\rm A} - X_{\rm B}$), is positive at all metallicities and ages except that in $U$ magnitude at age $\tau \le 3$\,Gyr and $Z \ge$ 0.01, this means that the ratio of the integrated flux in $X$ passband ($f_{\rm X,A}/f_{\rm X,B} = 10^{-0.4 \cdot \Delta X}$) is less than 1, while in the $U$ passband it is greater than 1 at age $\tau \le 3$\,Gyr and $Z \ge 0.01$.
For the sake of its size, in Fig.~\ref{Fig:dmag-t-z} we only give $\Delta U, \Delta B, \Delta V, \Delta R$ and $\Delta I$ except for $\Delta BOL$ at $Z=0.0001, 0.003, 0.001, 0.004, 0.01, 0.02$ and 0.03.
From Fig.~\ref{Fig:dmag-t-z} we see that $\Delta U$,  $\Delta B$ and  $\Delta V$ increase with age, at $Z \le 0.004$ decrease with $Z$ while at $Z > 0.004$ increase with $Z$ ({\bf similar to the variation of $m^{\star}$ with metallicity}), i.e., the ratios of the $U$-, $B$- and $V$-flux of Model A to B ($f_{\rm X,A}/f_{\rm X,B}$) decrease with age, increase with $Z$ at low metallicities while decrease with $Z$ at high metallicities. The lower flux ratio at late ages and low metallicities is because $U$-, $B$-, $V$-light is mainly dominated by MS stars, from Fig.~\ref{Fig:iso-com} we see that at low metallicities and late ages the number of luminous MS stars trends to decrease.
From Fig.~\ref{Fig:dmag-t-z} we also see that $\Delta R$ and $\Delta I$ increase with age (only significant at high metallicities) and decrease with $Z$ at age $\tau \le 10$\,Gyr, this is because that $R$- and $I$-light is partly contributed by GB stars, at low metallicities and late ages the temperature range spanned by GB stars is narrow (see Fig.~\ref{Fig:iso-adp}), the difference in the distribution of GB stars in HR diagram caused by the differences in age and metallicity is small, so the discrepancies in $R$ and $I$ magnitudes between Models A and B are insignificant.
The differences $\Delta J$, $\Delta H$, $\Delta K$,  $\Delta L$ and $\Delta M$ are independent of age and metallicity (range from $\sim$\,0.1 to 0.2 mag, at high metallicity they increase with age).
If sensitive parameter $\frac {\Delta X/({\rm d}\tau/\tau)}{\Delta X/({\rm d}Z/Z)} (X=U,B,...,M)$ is greater than 1, $X$ magnitude is sensitive to the metallicity of populations \citep{wor94}. From Fig.~\ref{Fig:dmag-t-z} we see that BIs make $V$ magnitude less sensitive to age, $R$ and $I$ magnitudes more sensitive to metallicity.

In Fig.~\ref{Fig:mags-all} we also give the $U, B, V, R, J$ and $K$ magnitudes of BC03-S, BC03-C, M05-S and M05-K at solar metallicity. Note that BC03 used Busers' system for $U, B (B3)$ and $V$; Cousins' system for $R$ \citep{bes83} and 2MASS system for $J$ and $K$ magnitudes. By comparison we find that $U, B, V, R, J$ and $K$ magnitudes of BC03-S, BC03-C, M05-S and M05-K are greater than the corresponding ones of Models A and B at solar metallicity. The results of M05-K are the closest to ours.

\subsection{Mass-to-light ratios}

\begin{figure}
\leftline{
\includegraphics[bb=81 37 530 700,height=4.00cm,width=4.00cm,clip,angle=-90]{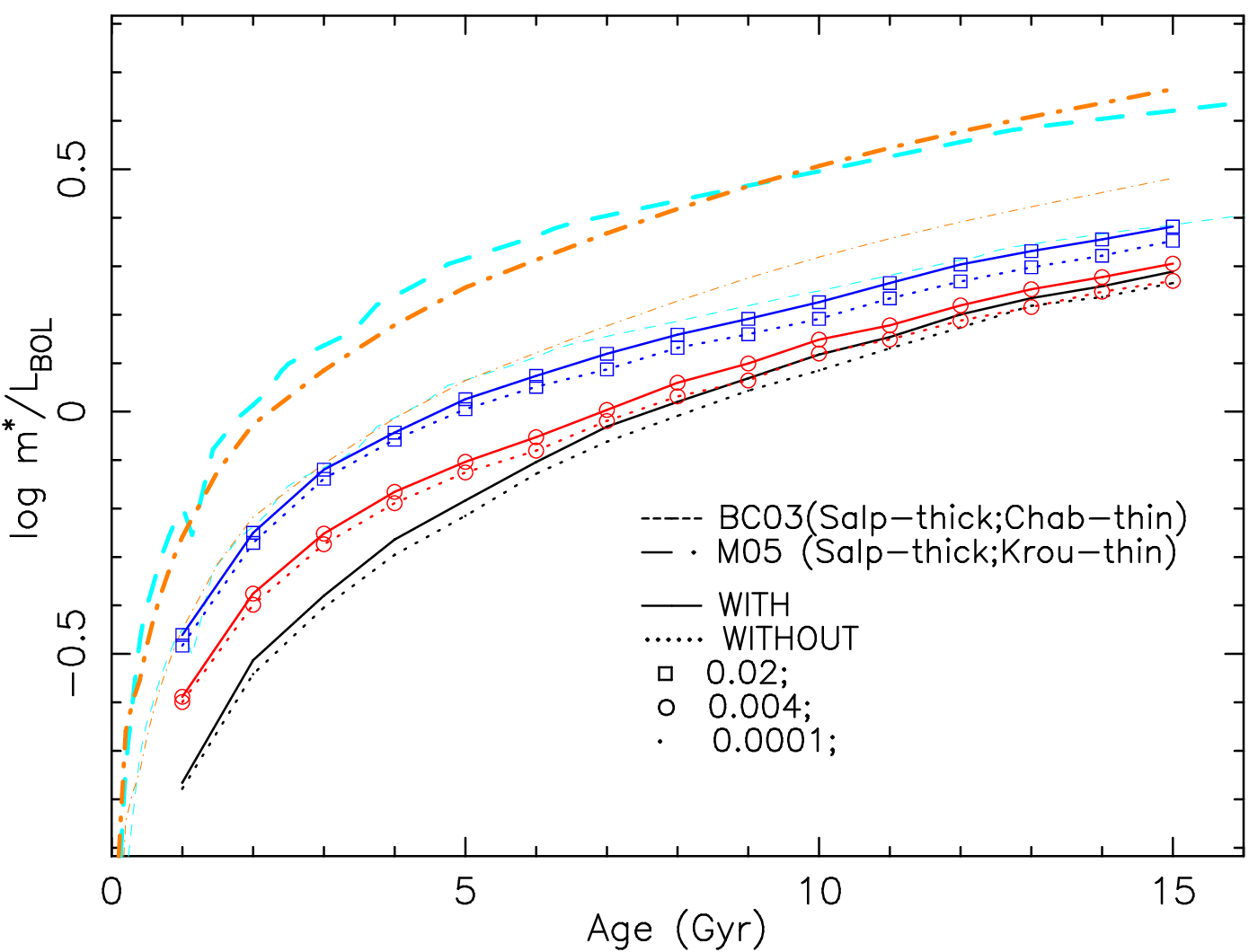}
\includegraphics[bb=81 37 530 700,height=4.00cm,width=4.00cm,clip,angle=-90]{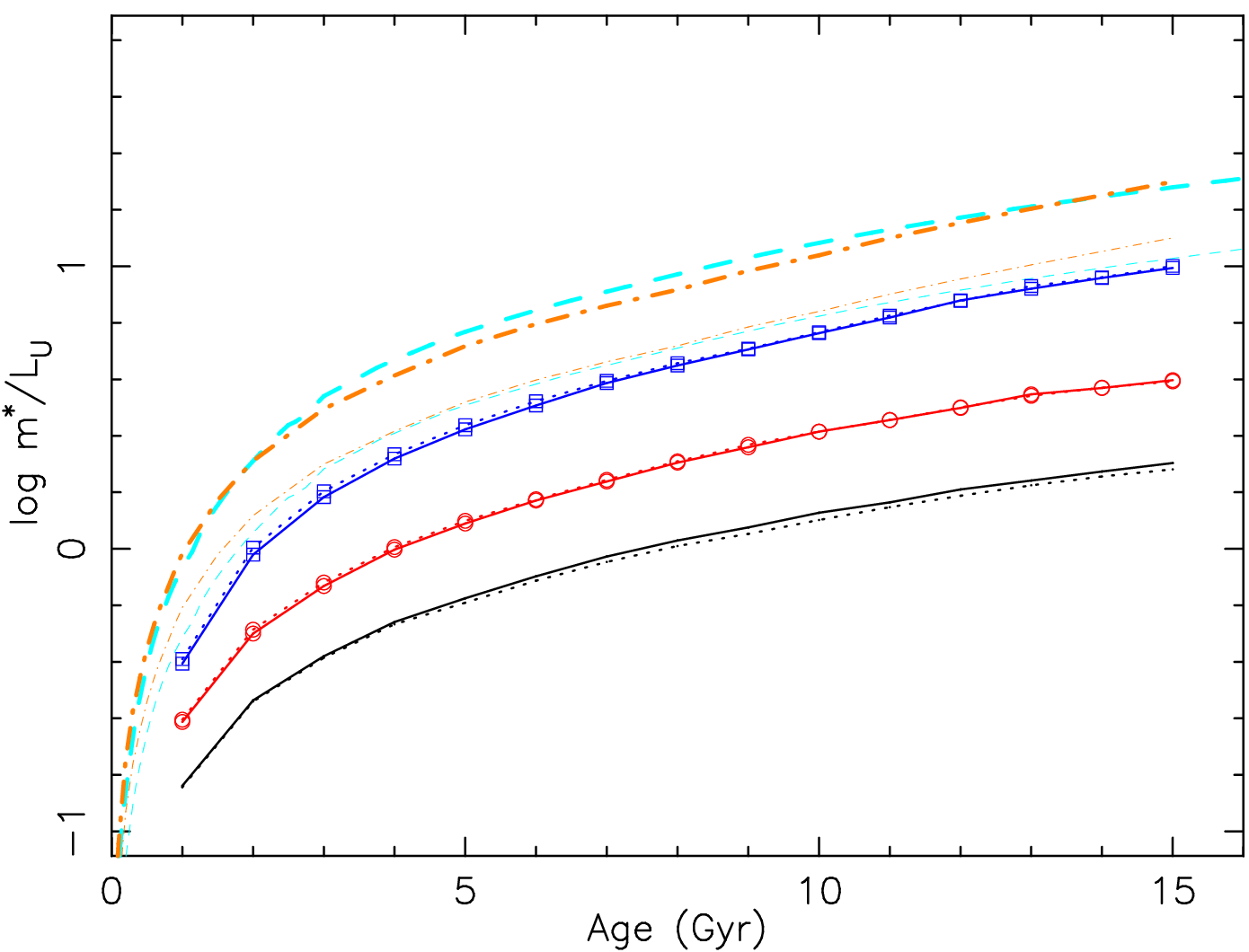}
}
\leftline{
\includegraphics[bb=81 37 531 700,height=4.00cm,width=4.00cm,clip,angle=-90]{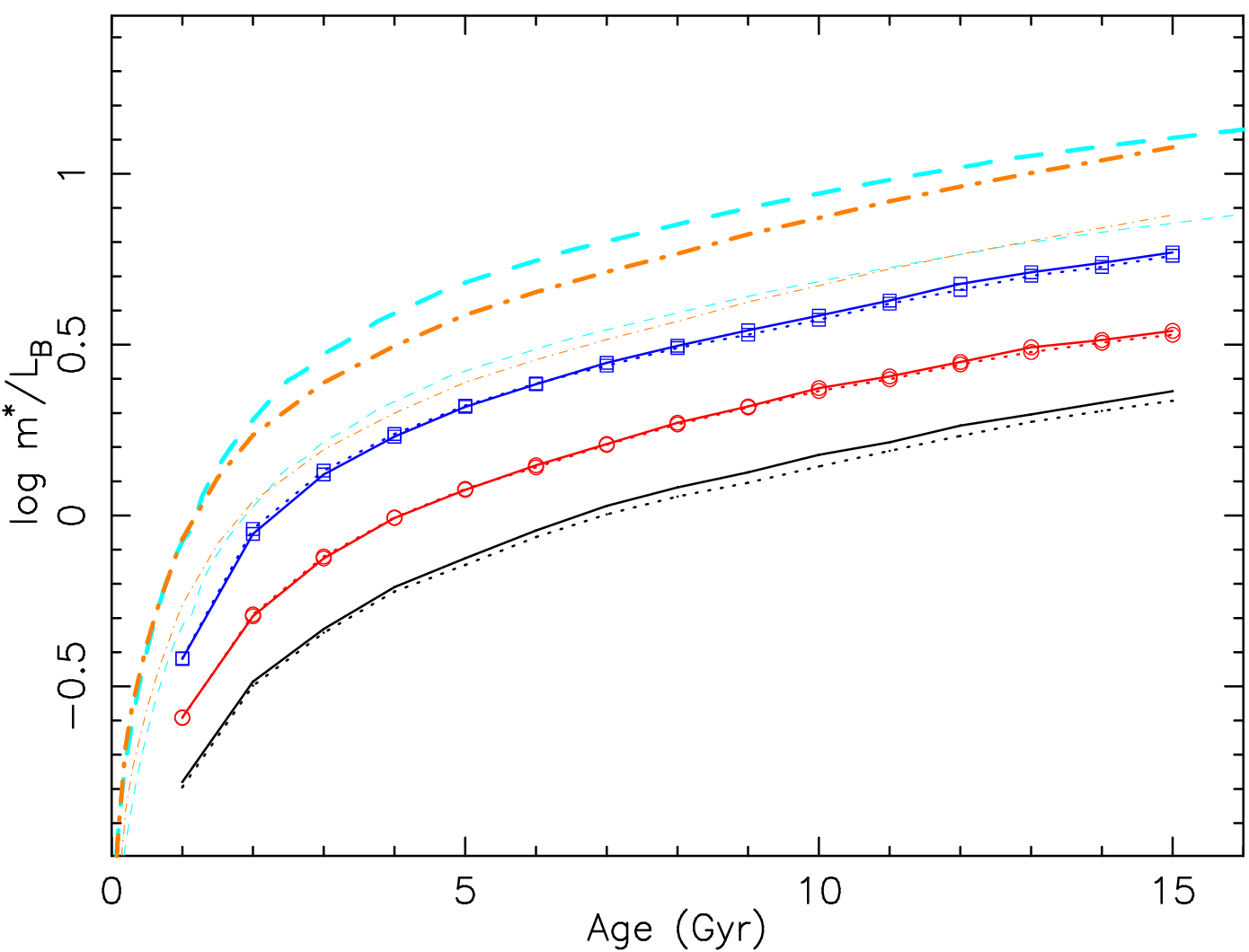}
\includegraphics[bb=81 37 530 700,height=4.00cm,width=4.00cm,clip,angle=-90]{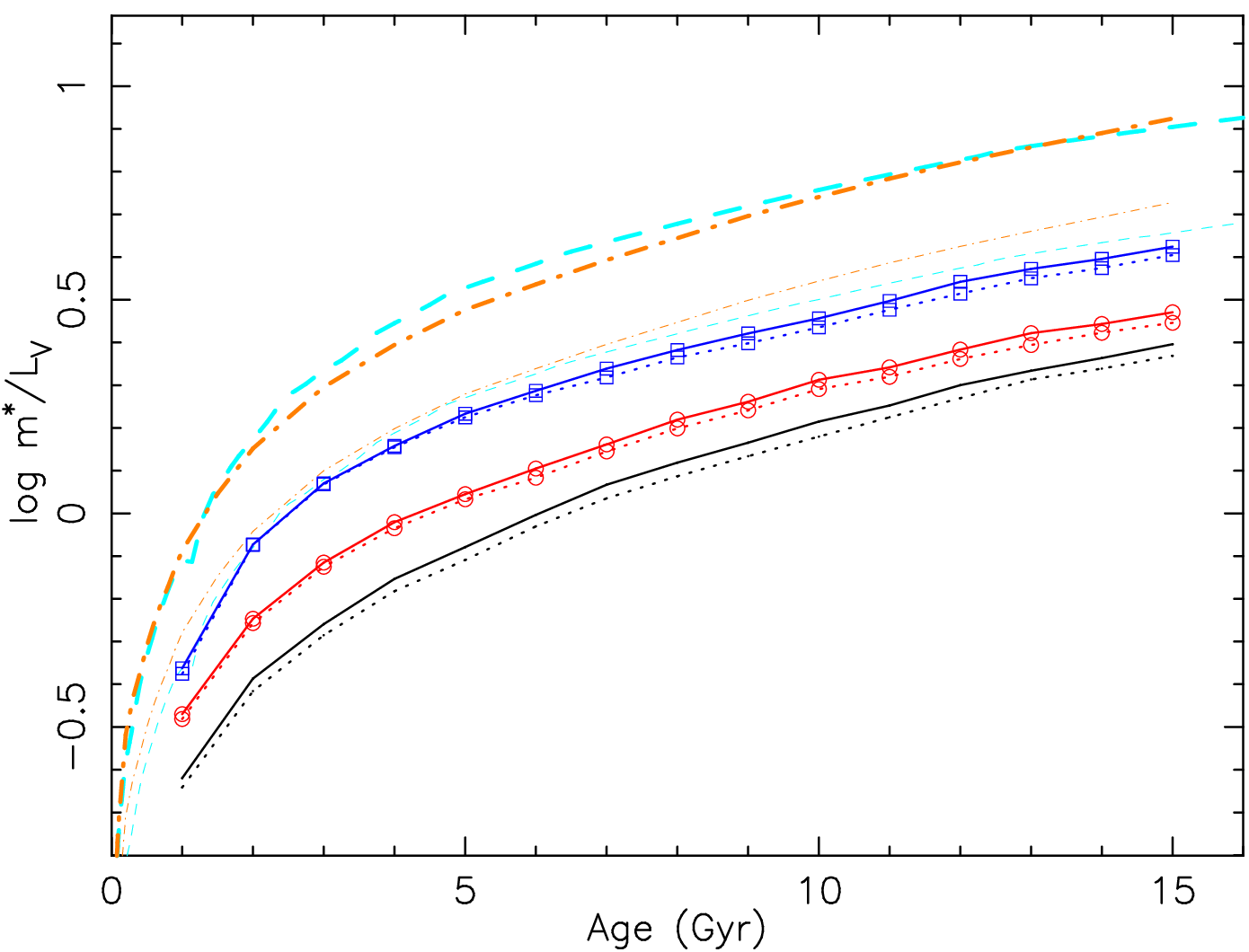}
}
\leftline{
\includegraphics[bb=81 37 580 700,height=4.00cm,width=4.40cm,clip,angle=-90]{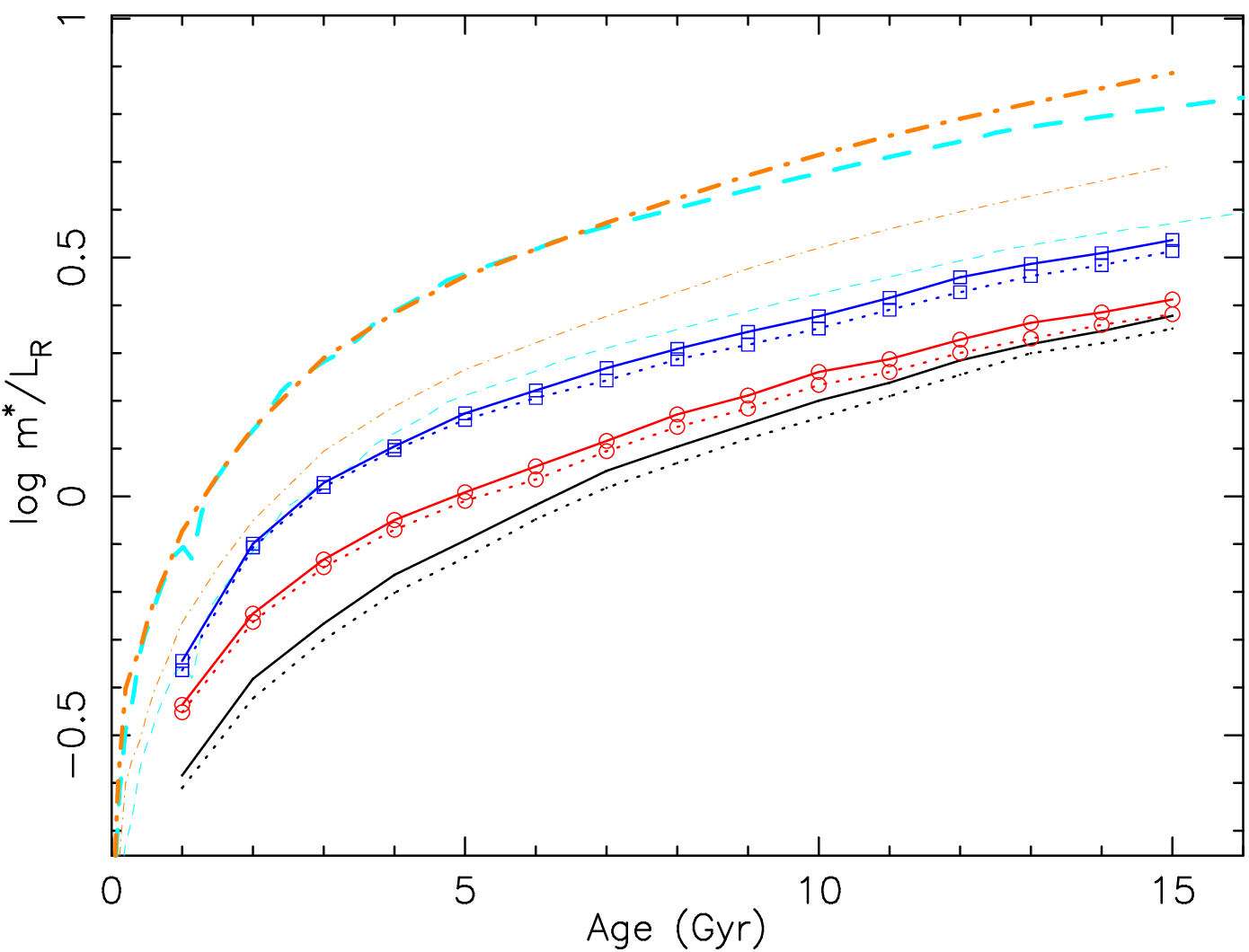}
\includegraphics[bb=81 37 580 700,height=4.00cm,width=4.40cm,clip,angle=-90]{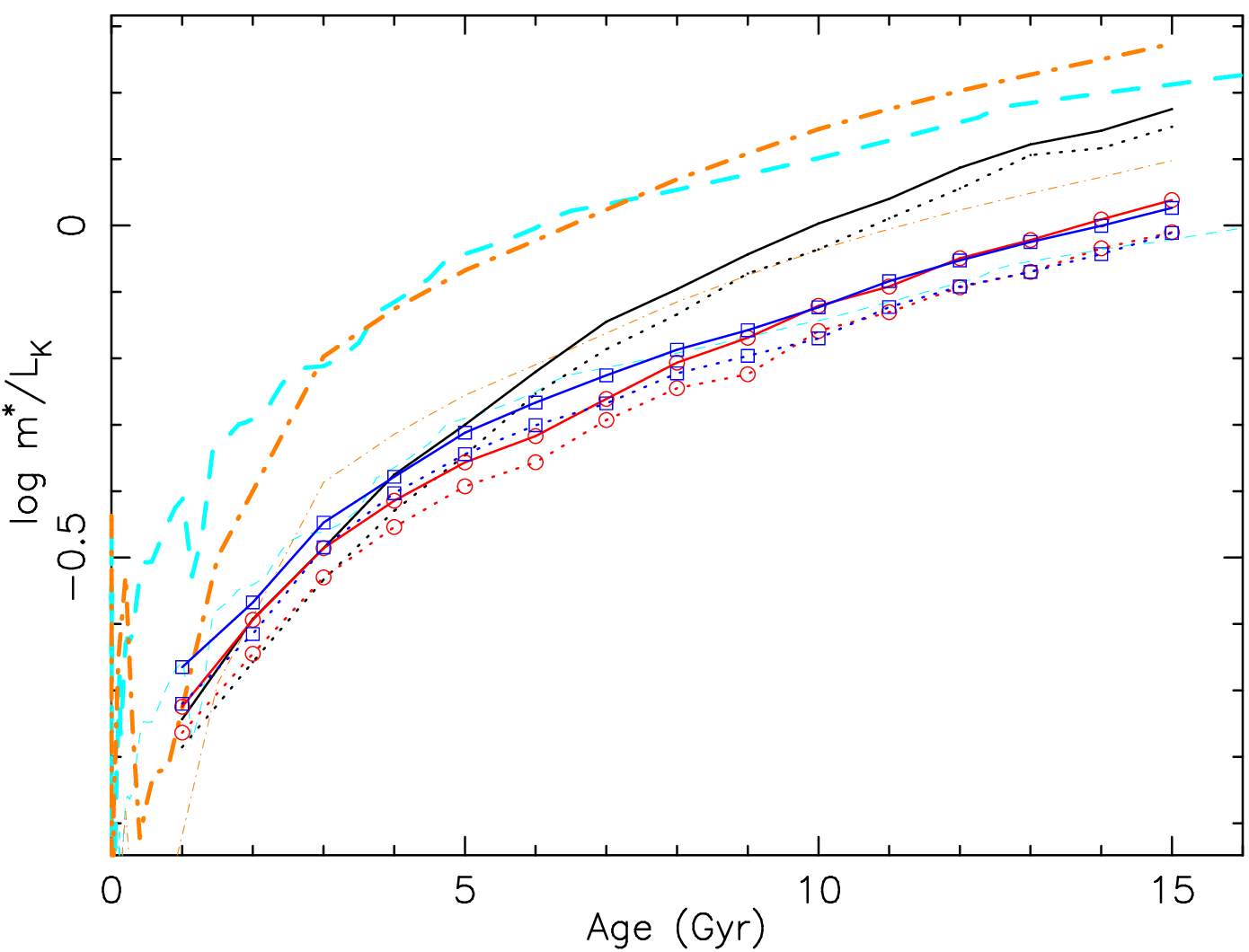}
}
\caption{The ratios of stellar mass to bolometric, $U, B, V, R$ and $K$ luminosities for Models A and B only at $Z=0.02, 0.004$ and 0.0001. Results of BC03-S and BC03-C, M05-S and M05-K are shown. The symbols have the same meanings as in Fig.~\ref{Fig:mags-all}.
$\star$ Note that ${\rm log} (m^{\star}/L_{\rm BOL}), {\rm log} (m^{\star}/L_{\rm U})$, ${\rm log} (m^{\star}/L_{\rm R})$ and ${\rm log} (m^{\star}/L_{\rm K})$ of BC03 are obtained by using their $m^{\star}, BOL, U, R$ and $K (=J-(J-K))$ magnitudes.}
\label{Fig:mvl-all}
\end{figure}

\begin{figure}
\leftline {
\includegraphics[bb=115 97 474 624,height=4.00cm,width=4.00cm,clip,angle=-90]{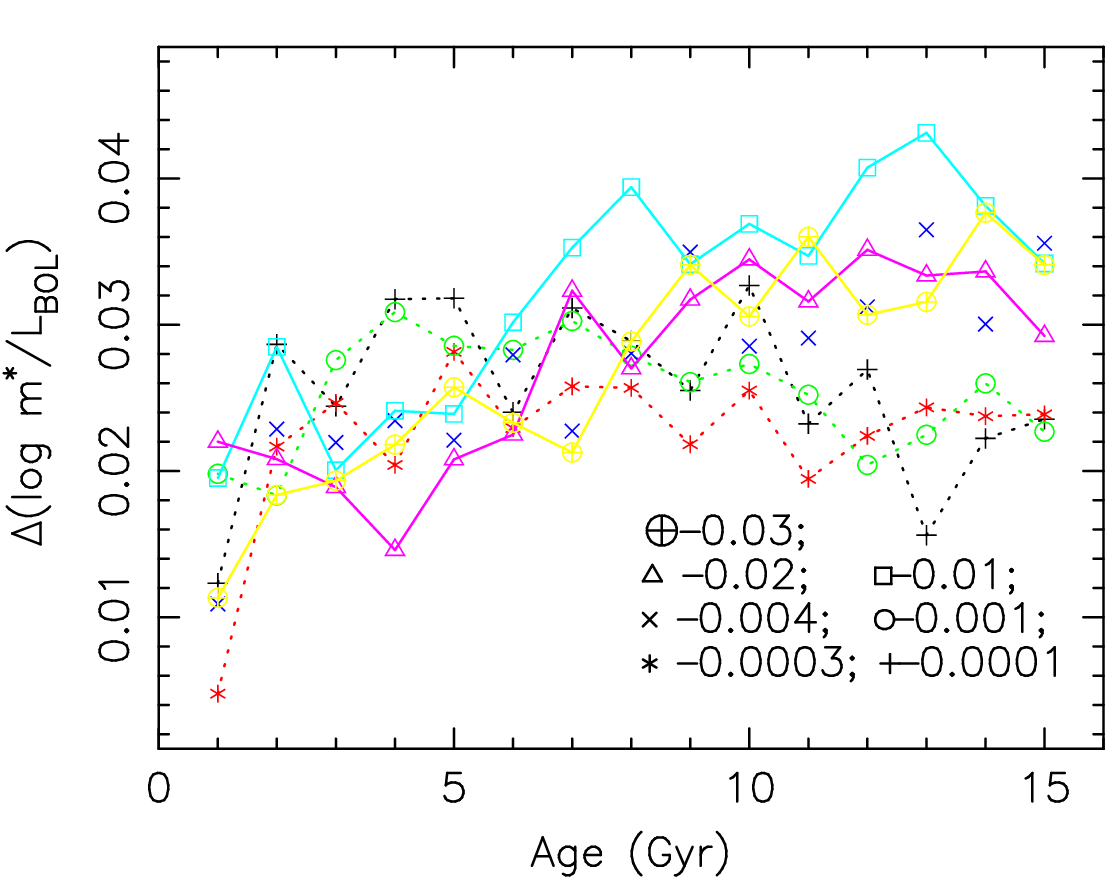}
\includegraphics[bb=115 97 474 624,height=4.00cm,width=4.00cm,clip,angle=-90]{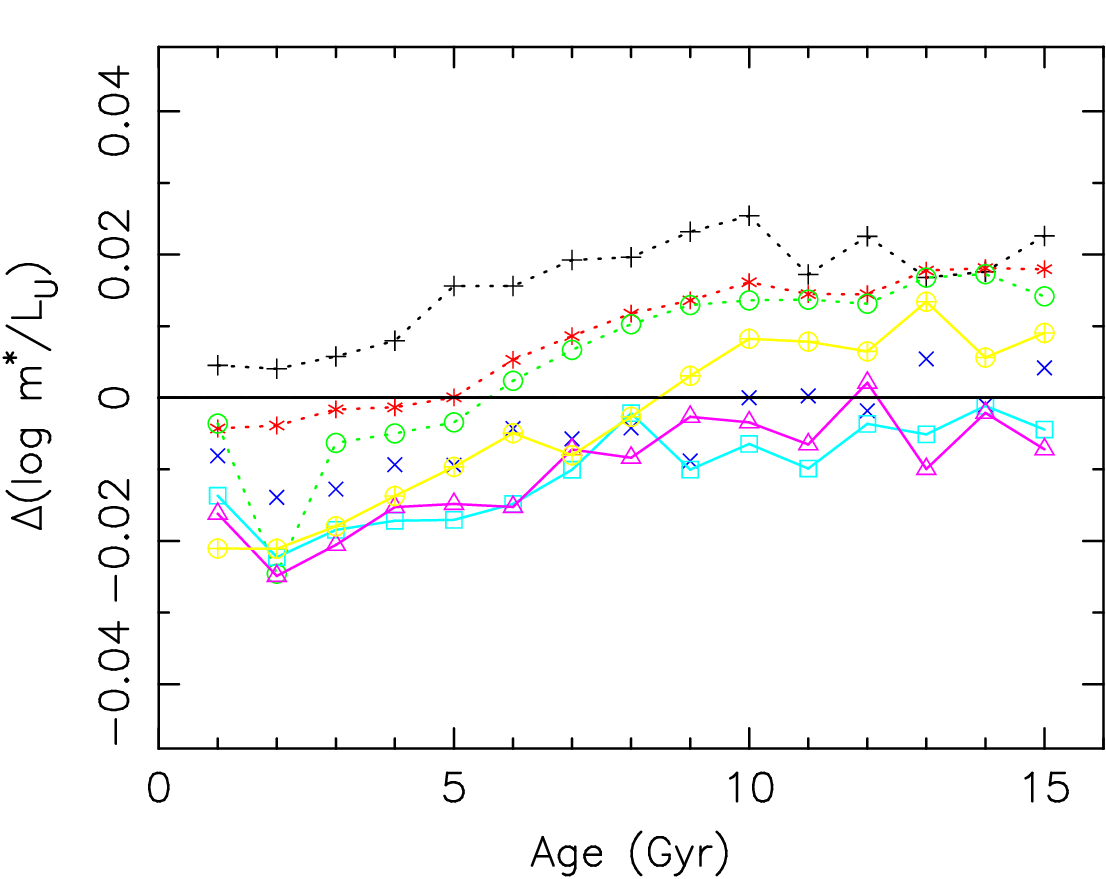}
}
\leftline{
\includegraphics[bb=136 97 474 624,height=4.00cm,width=3.70cm,clip,angle=-90]{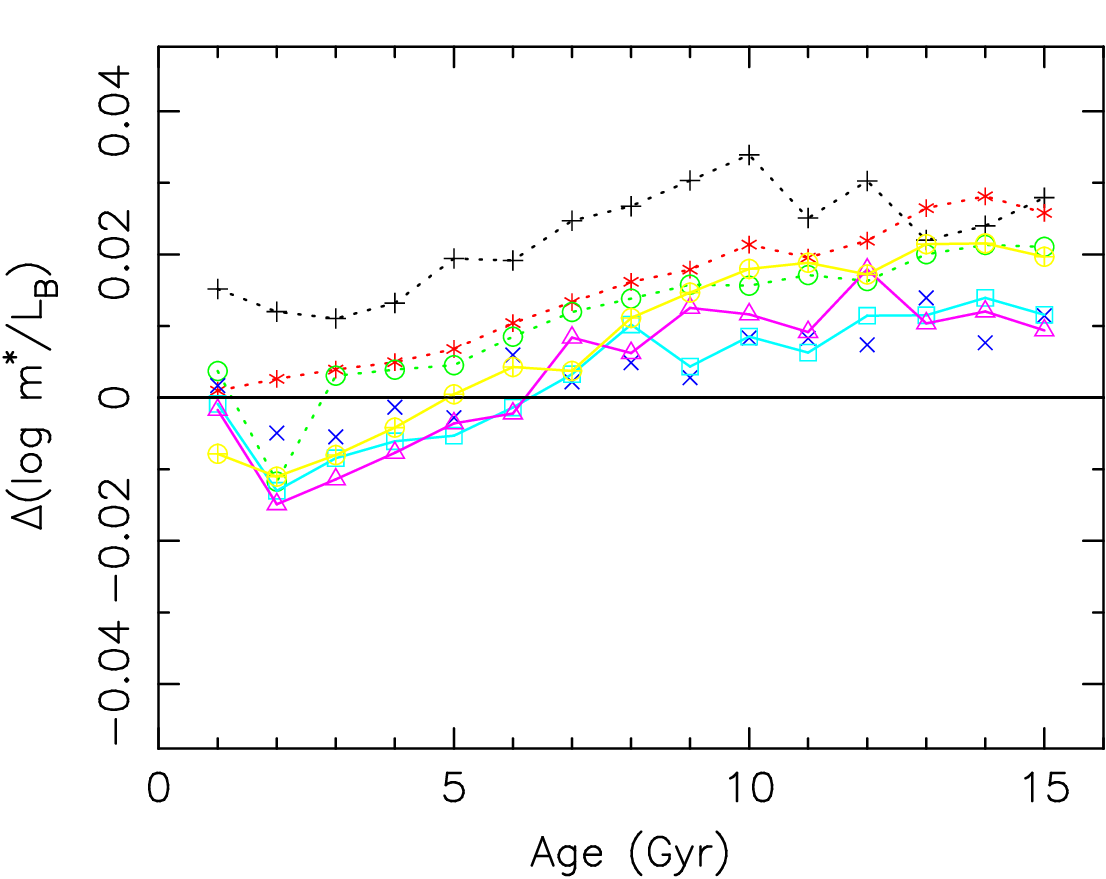}
\includegraphics[bb=136 97 474 624,height=4.00cm,width=3.70cm,clip,angle=-90]{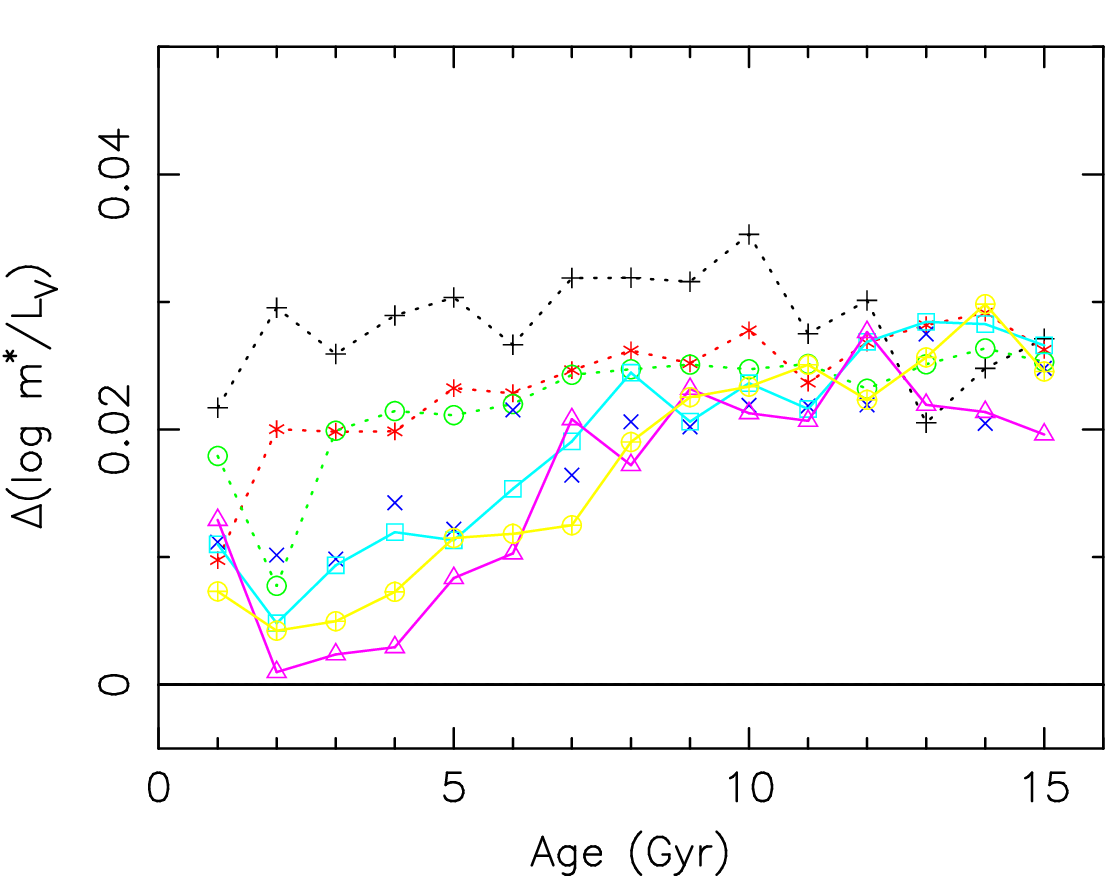}
}
\leftline{
\includegraphics[bb=136 97 537 624,height=4.00cm,width=4.40cm,clip,angle=-90]{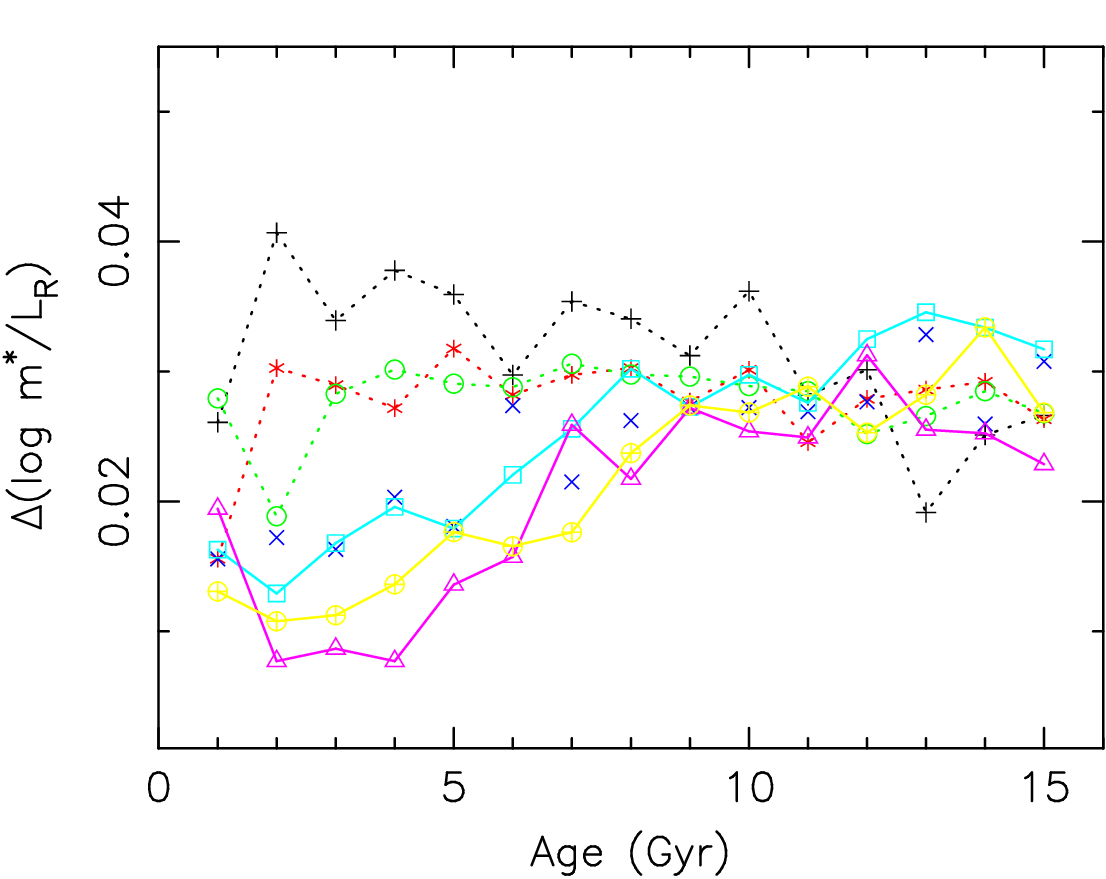}
\includegraphics[bb=136 97 537 624,height=4.00cm,width=4.40cm,clip,angle=-90]{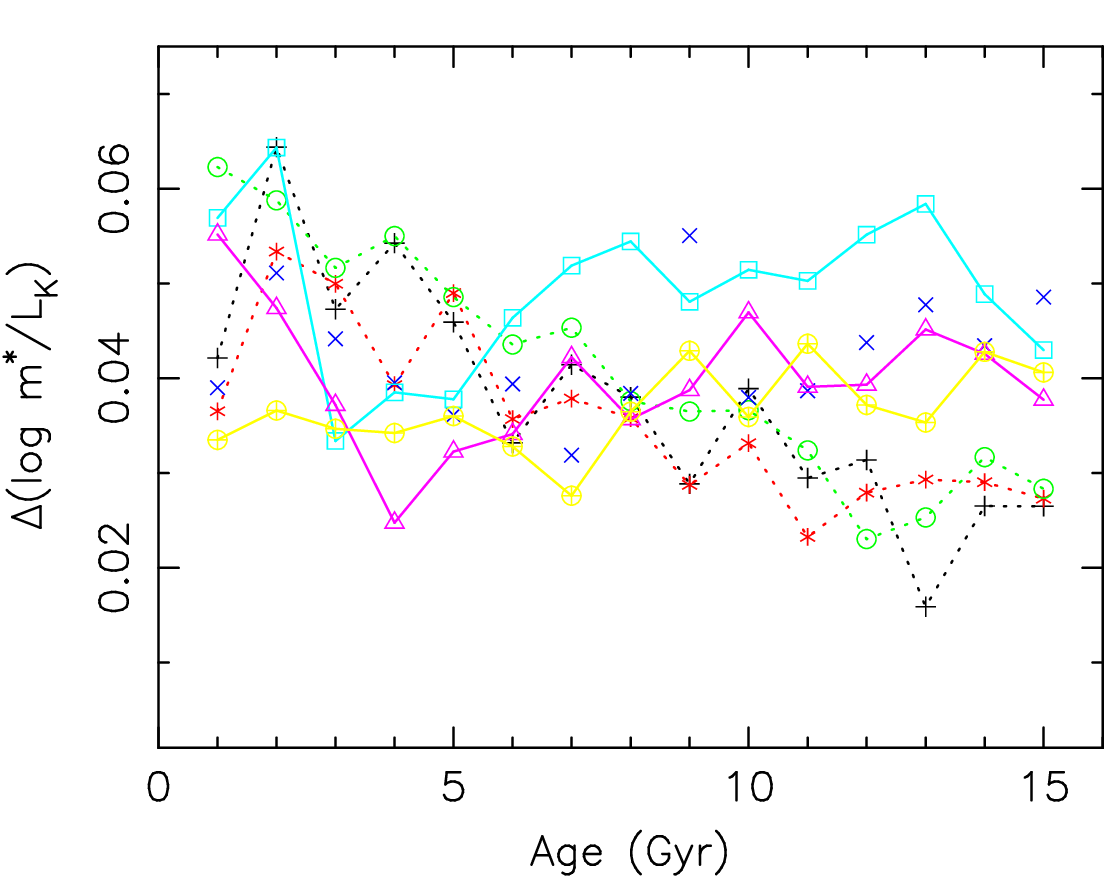}
}
\caption{Differences in the ratios of stellar mass to bolometric, $U$, $B$, $V$, $R$, $K$ luminosities between Models A and B for 7 metallicities. The symbols have the same meanings as in Fig.~\ref{Fig:dmag-t-z}.}
\label{Fig:dmvl-t}
\end{figure}

In this part we present the stellar mass-to-light ratios for various light. Firstly, the ratio of stellar mass to bolometric luminosity ${\rm log} (m^{\star}/L_{\rm BOL})$ increases with age and metallicity. At age $\tau \ge 5$\,Gyr the difference among $Z=0.0001, 0.0003$ and 0.001 is smaller, at age $\tau < 5$\,Gyr the discrepancy between $Z=0.0001$ and 0.0003 becomes to be insignificant. For the sake of clarity, in the top left panel of Fig.~\ref{Fig:mvl-all} we give them only at $Z=0.02, 0.004$ and 0.0001. The ${\rm log} (m^{\star}/L_{\rm BOL})$ increases with age and metallicity means that less light can be emitted by a given mass with increasing age and metallicity.
The ratio of stellar mass to the luminosity in $X$ passband (${\rm log} (m^{\star}/L_X), X=U,B,..,M$) also increases with age, and the variation with metallicity is similar to that of $X$ magnitude, i.e., satisfies U-, R-, J-type to M-type with increasing wavelength. For the sake of its size and clarity, in Fig.~\ref{Fig:mvl-all} we only give the ratios of stellar mass to $U$-, $B$-, $V$-, $R$- and $K$-light except for ${\rm log} (m^{\star}/L_{\rm BOL})$ for Models A and B only at $Z=0.02, 0.004$ and 0.0001.

The ${\rm log} (m^{\star}/L_{\rm BOL})$ of Model A is larger than that of Model B at all ages and metallicities, i.e., Model A can produce less light than Model B for a given mass. The discrepancy in the ratio of mass to bolometric luminosity between Models A and B, $\Delta {\rm log} (m^{\star}/L_{\rm BOL})$ ($\equiv {\rm log} (m^{\star}/L_{\rm BOL})_{\rm A} - {\rm log} (m^{\star}/L_{\rm BOL})_{\rm B}$), increases with age and independent of metallicity (see top left panel of Fig.~\ref{Fig:dmvl-t}). The ratio of $(F_{\rm BOL}/m^{\star})$ of Model A to B ($=10^{- \Delta {\rm log} (m^{\star}/L_{\rm BOL})}$) ranges from $\sim$\,0.94 to 0.97. Comparing with the bolometric flux ratio ($F_{\rm BOL,A}/F_{\rm BOL,B}$, $\sim$ 0.93-0.87) it is greater because less mass of Model A offsets this effect.

The ${\rm log} (m^{\star}/L_{\rm U})$ and ${\rm log} (m^{\star}/L_{\rm B})$ of Model A are less than the corresponding ones for Model B at all ages and $\tau \le 6$\,Gyr at high metallicities, respectively (see top right and middle left panels of Fig.~\ref{Fig:dmvl-t}). That is to say, Model A can produce more $U$- and $B$-light than Model B at these age and metallicity ranges.
Also, the differences in the ratios of stellar mass to $U$- and $B$-light between Models A and B, $\Delta {\rm log} (m^{\star}/L_{\rm U})$ and $\Delta {\rm log} (m^{\star}/L_{\rm B})$, increase with age, decrease with $Z$ at $Z \le 0.004$ while increase with $Z$ at $Z> 0.004$ ({\bf similar to the variation of $m^{\star}$, $\Delta U$,  $\Delta B$ and $\Delta V$ with $Z$}).
For other mass-to-light ratios Model A has larger values than those of Model B. The differences in the ratios of stellar mass to $V$-, $R$- and $I$-light between Models A and B, $\Delta {\rm log} (m^{\star}/L_{\rm V})$, $\Delta {\rm log} (m^{\star}/L_{\rm R})$ and $\Delta {\rm log} (m^{\star}/L_{\rm I})$, increase with age at high metallicity and decrease with $Z$ at early ages. The differences in the ratios of stellar mass to $J$-, $H$- and $K$-light between Models A and B, $\Delta {\rm log} (m^{\star}/L_{\rm J})$, $\Delta {\rm log} (m^{\star}/L_{\rm H})$ and $\Delta {\rm log} (m^{\star}/L_{\rm K})$, are independent of metallicity, but decrease with age at $Z \le 0.004$, i.e., the ratios of mass-to-light-ratio decrease with age.
For the sake of its size, in Fig.~\ref{Fig:dmvl-t} we only give the differences in the ratios of stellar mass to $U$-, $B$-, $V$-, $R$- and $K$-light except for $\Delta {\rm log} (m^{\star}/L_{\rm BOL})$ between Models A and B at $Z=0.0001, 0.003, 0.001, 0.004, 0.01, 0.02$ and 0.03.

From top left panel of Fig.~\ref{Fig:mvl-all} we see that at solar metallicity the ${\rm log} (m^{\star}/L_{\rm BOL})$ of BC03-S is greater by $\sim$\,0.25 dex than those of Models A and B at all ages; BC03-C agrees with Models A and B; the values of M05-S and M05-K are greater than those of Models A and B at all ages. $U, B, V, R$ and $K$ filters used by BC03 have been described in Sect. 3.2.2. The stellar mass-to-light ratios are easily affected by the shape of IMF (BC03; M05).
In the ${\rm log} (m^{\star}/L_{\rm U})$, ${\rm log} (m^{\star}/L_{\rm B})$, ${\rm log} (m^{\star}/L_{\rm V})$, ${\rm log} (m^{\star}/L_{\rm R})$ and ${\rm log} (m^{\star}/L_{\rm K})$ diagrams of Fig.~\ref{Fig:mvl-all} we see that the values presented by BC03 and M05 are greater than those of Models A and B except M05-K at ${\rm log} (m^{\star}/L_{\rm K})$ diagram (lie between that of Models A and B at age $\tau > 8$\,Gyr).

\subsection{Broad-band colours}

\begin{figure*}
\leftline{
\includegraphics[bb=68 37 530 700,height=4.40cm,width=4.40cm,clip,angle=-90]{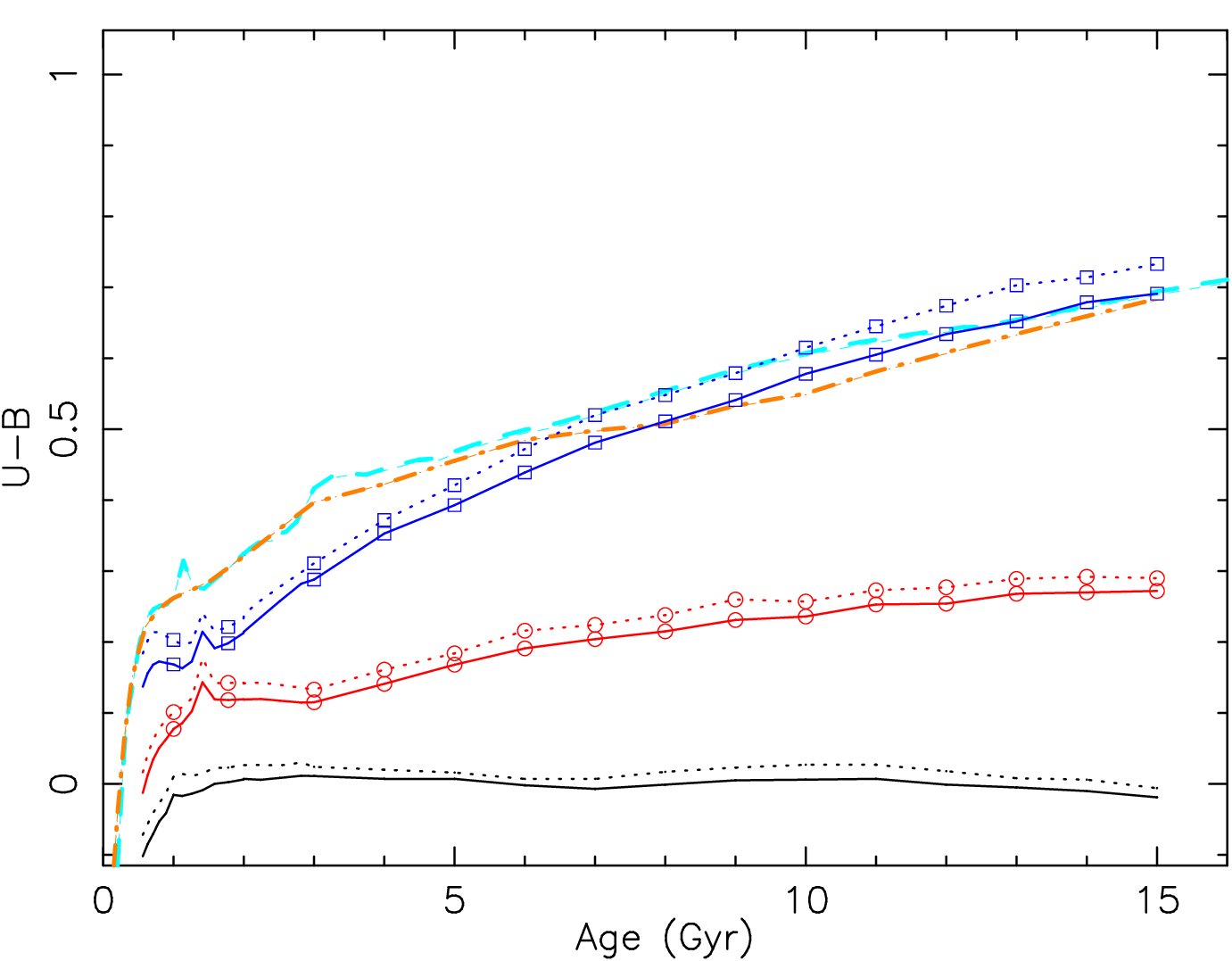}
\includegraphics[bb=68 37 530 700,height=4.40cm,width=4.40cm,clip,angle=-90]{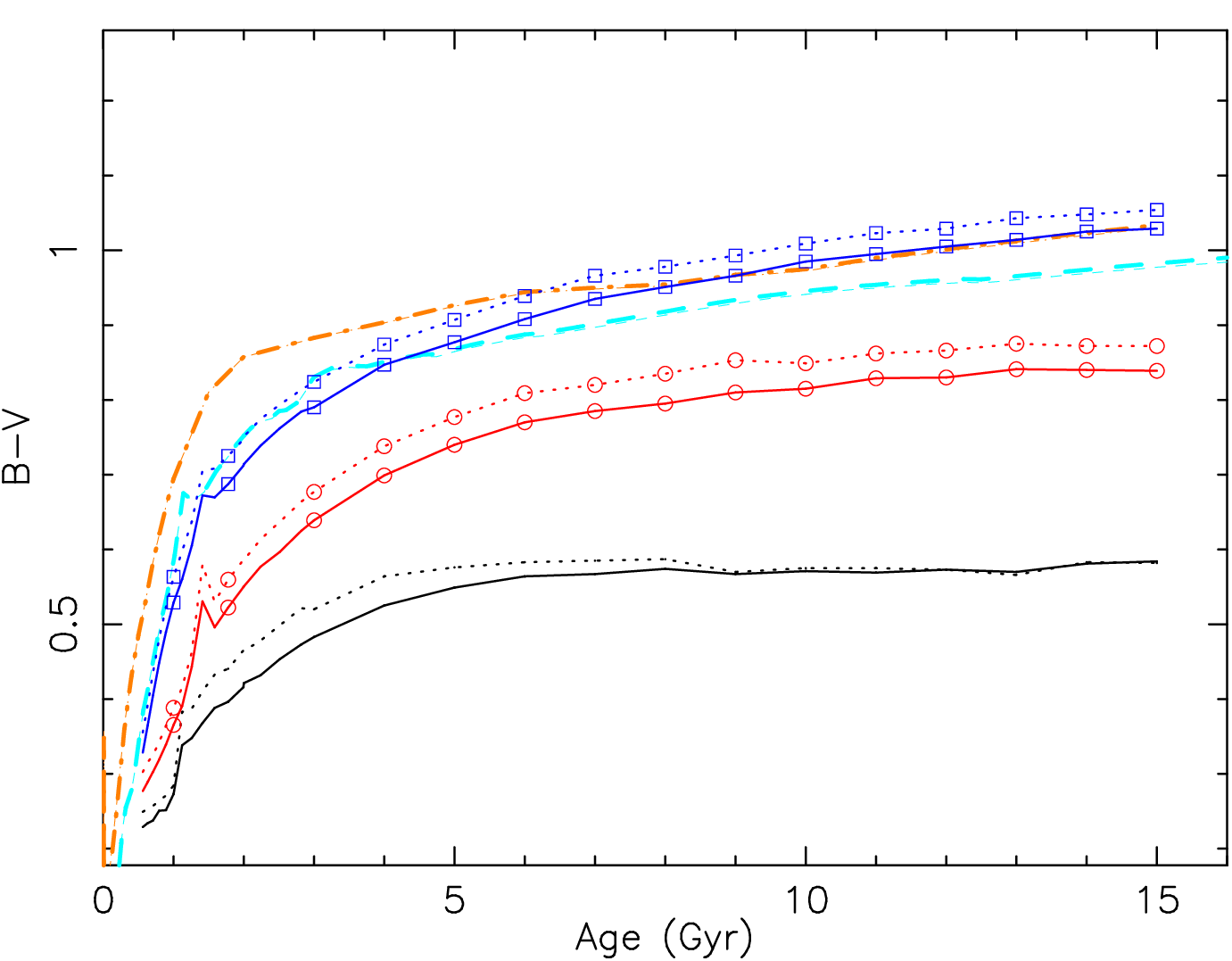}
\includegraphics[bb=68 37 530 700,height=4.40cm,width=4.40cm,clip,angle=-90]{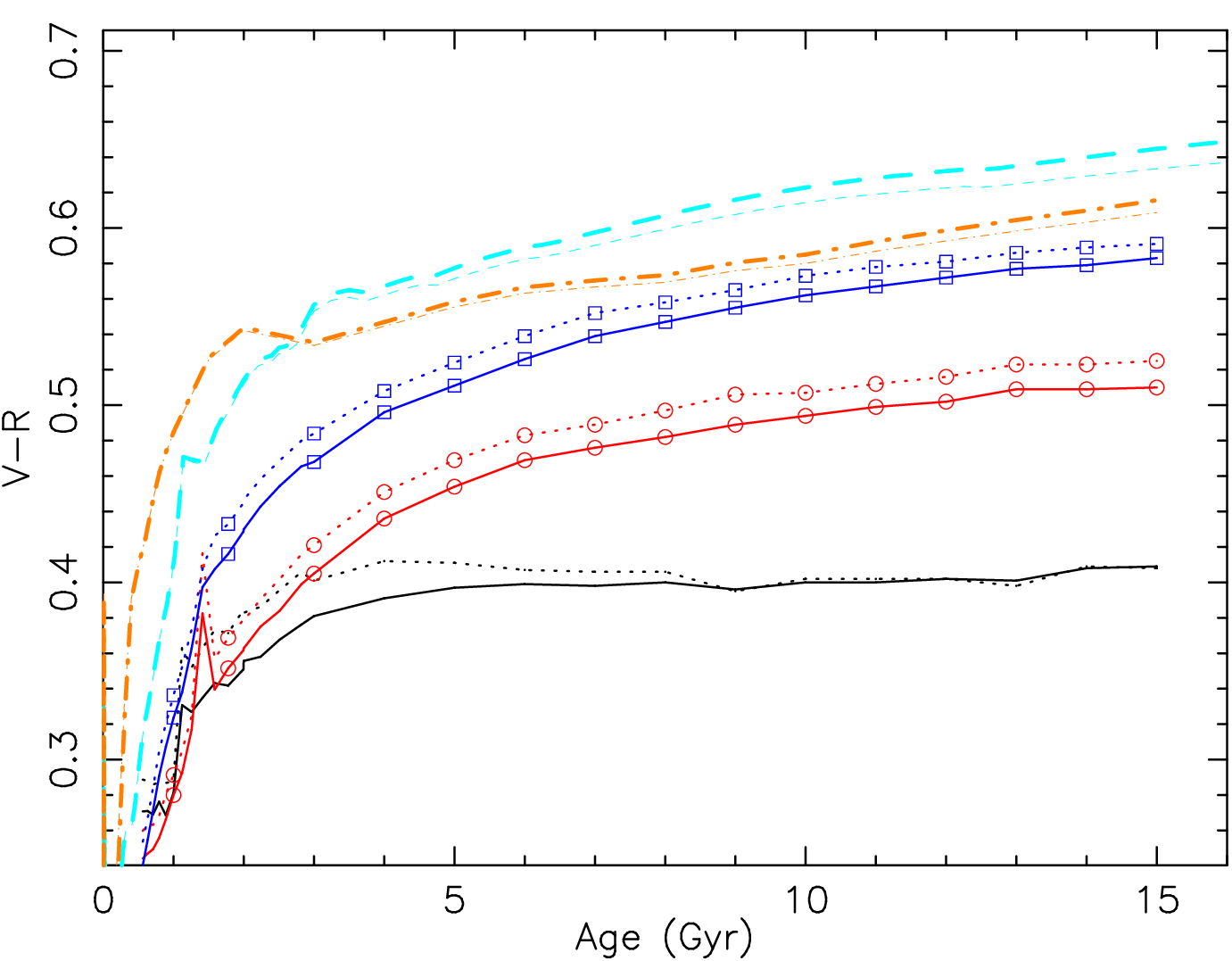}
\includegraphics[bb=68 37 530 700,height=4.40cm,width=4.40cm,clip,angle=-90]{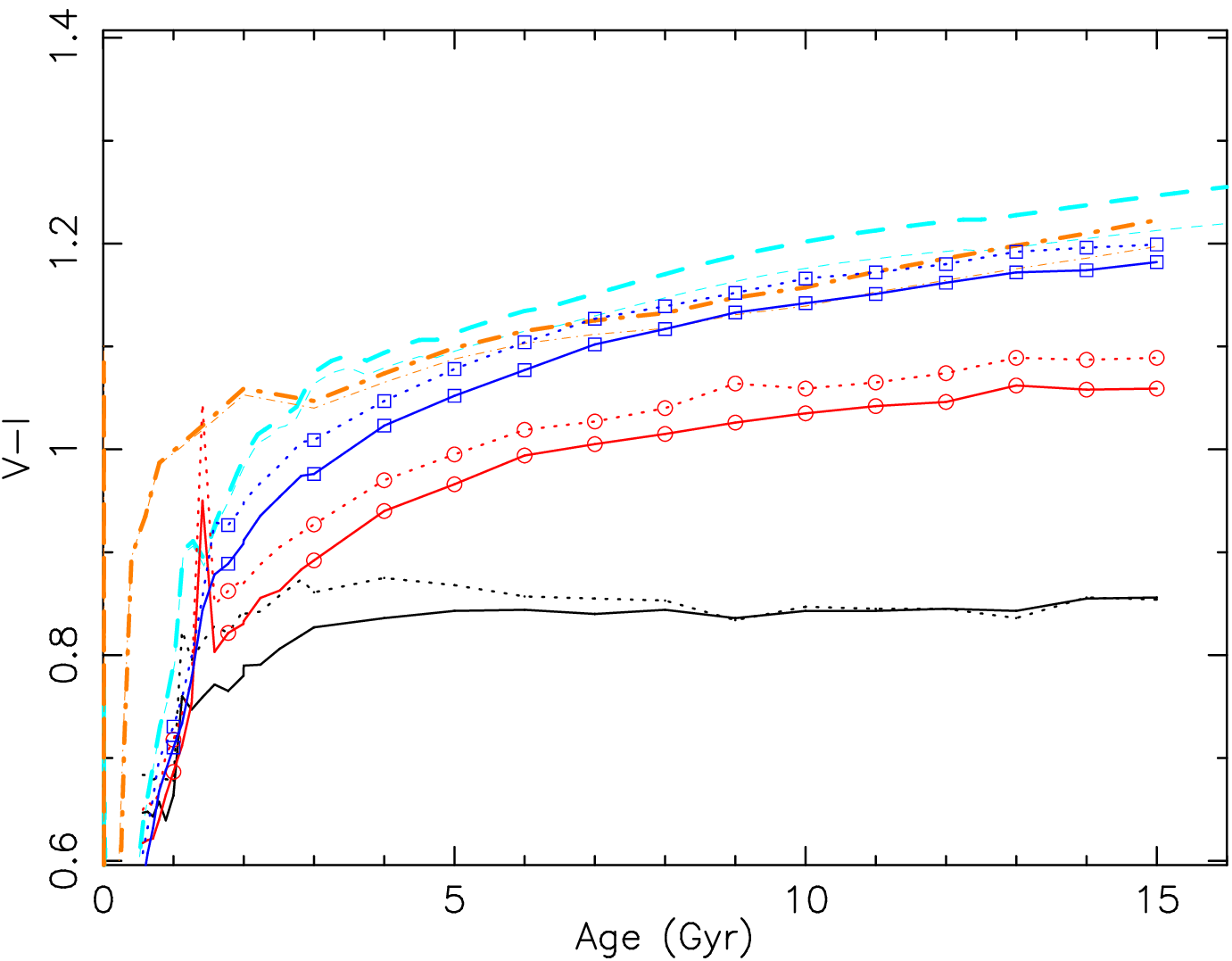}
}
\leftline{
\includegraphics[bb=81 37 531 700,height=4.40cm,width=4.20cm,clip,angle=-90]{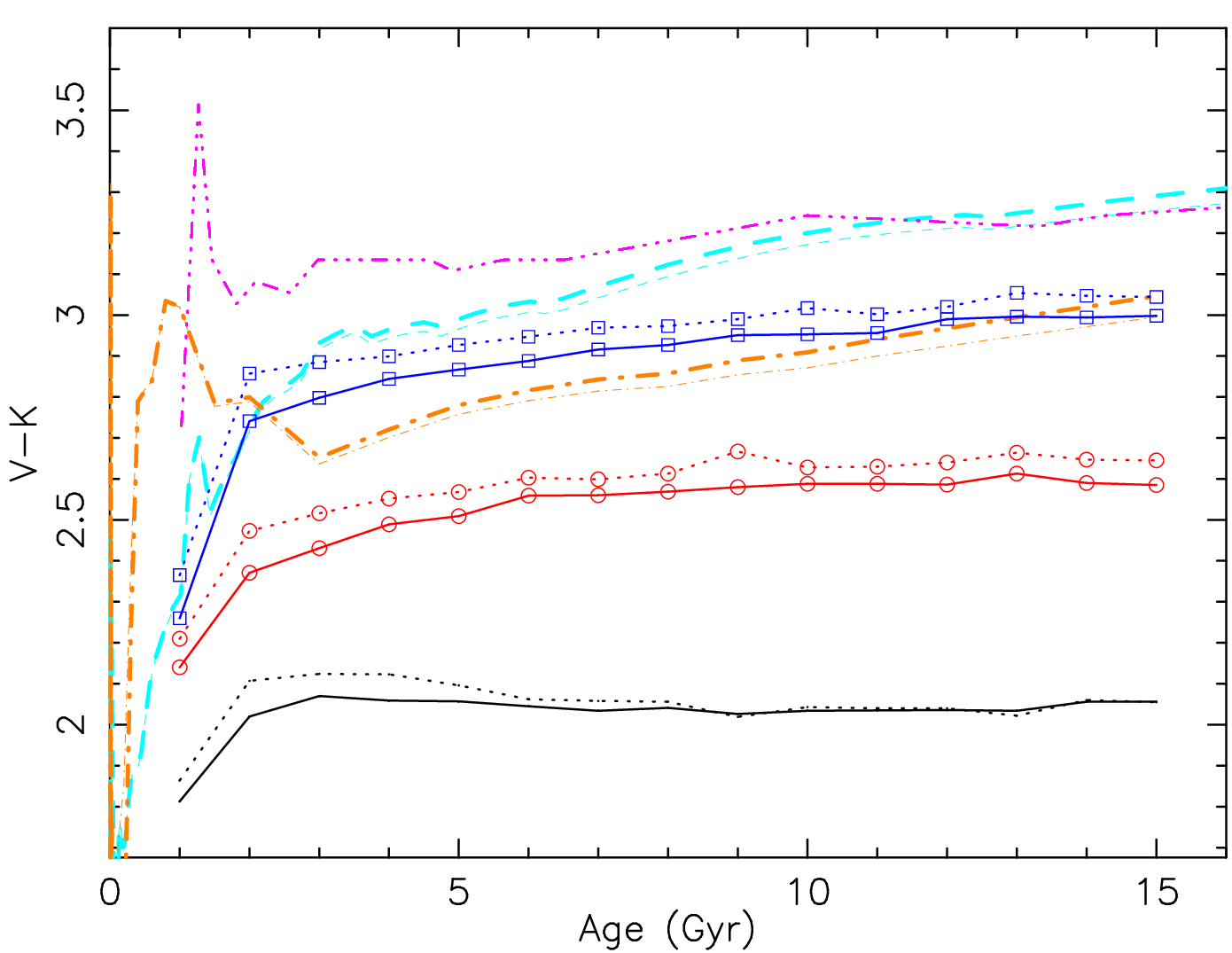}
\includegraphics[bb=81 37 530 700,height=4.40cm,width=4.20cm,clip,angle=-90]{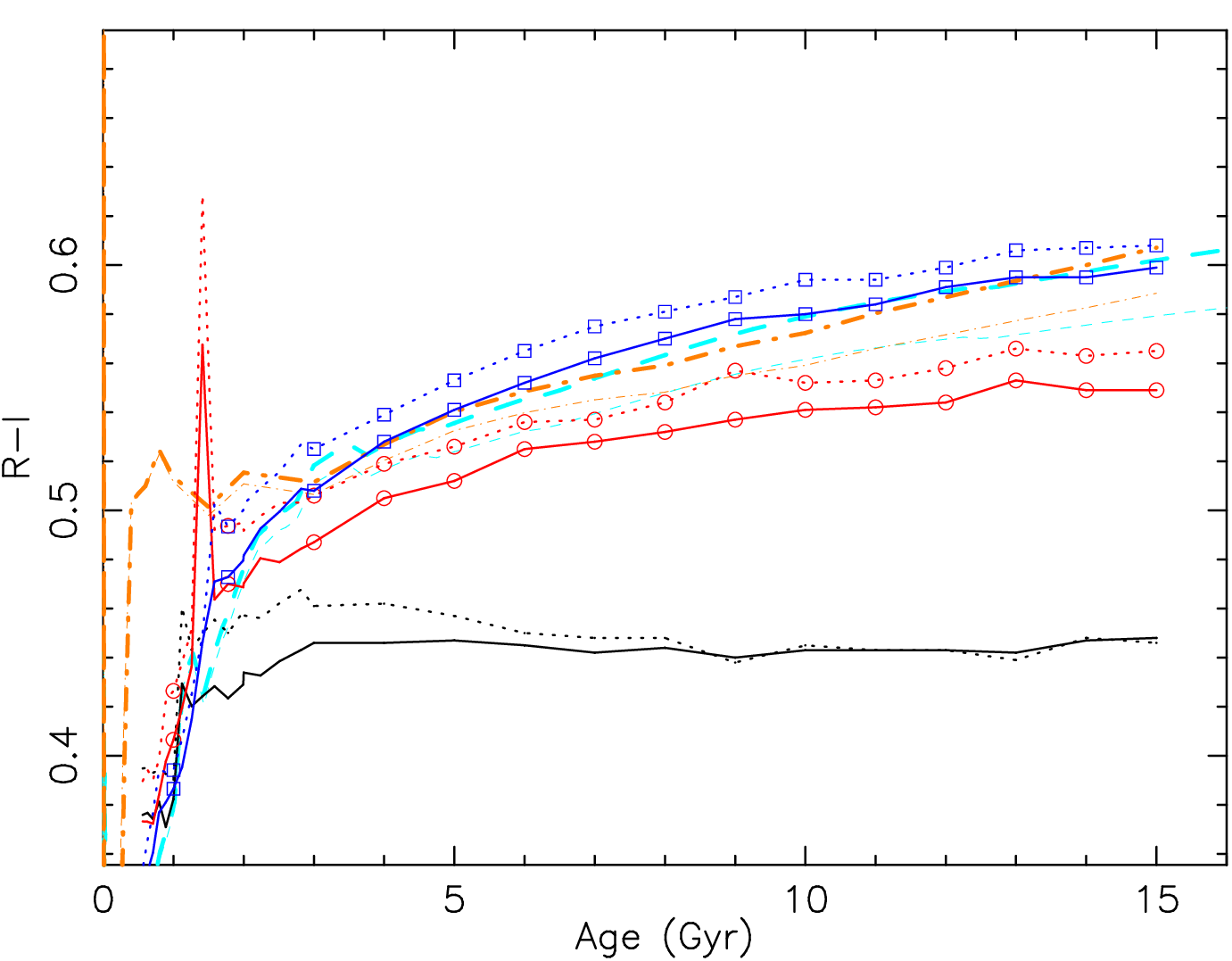}
\includegraphics[bb=81 37 530 700,height=4.40cm,width=4.20cm,clip,angle=-90]{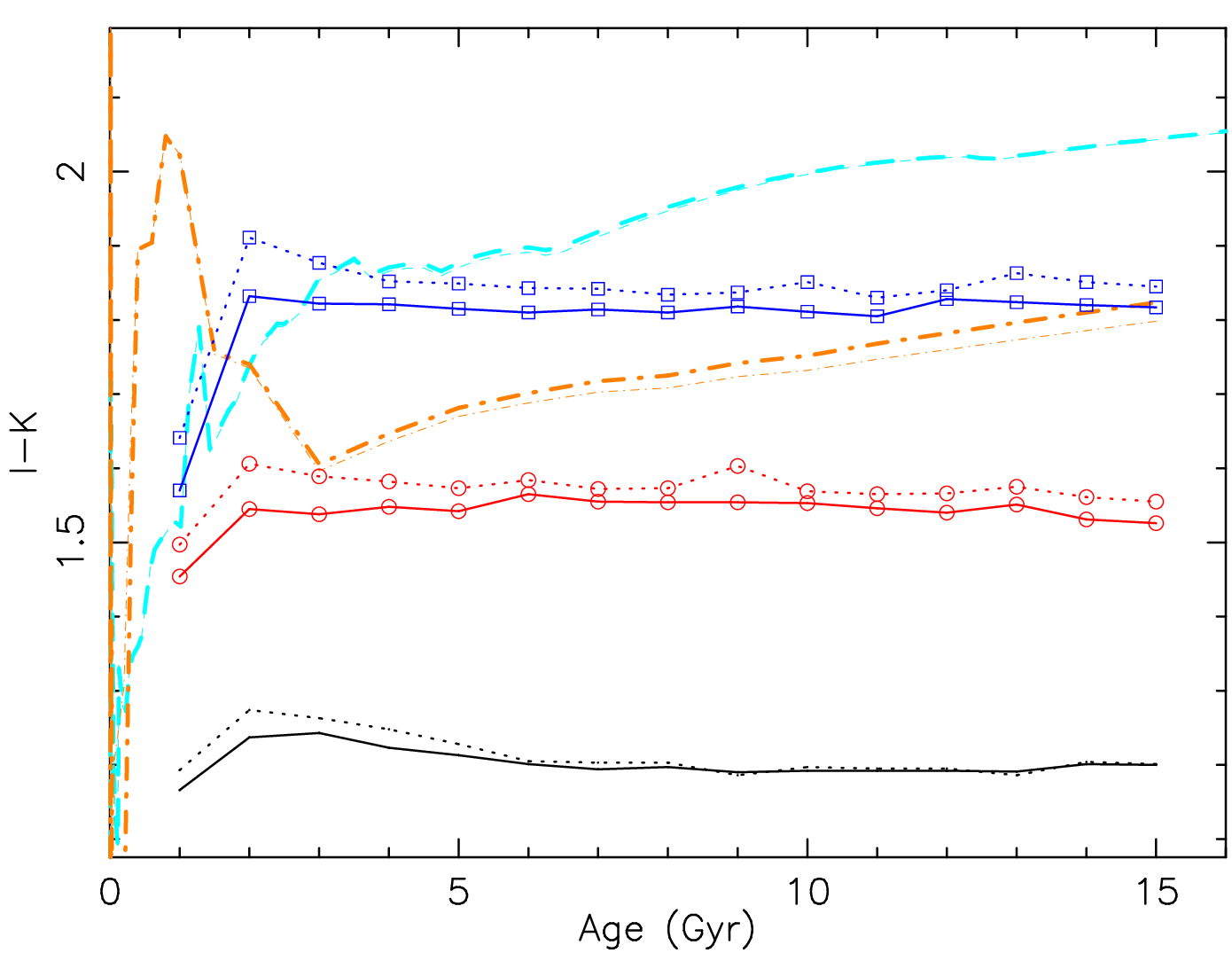}
\includegraphics[bb=81 37 530 700,height=4.40cm,width=4.20cm,clip,angle=-90]{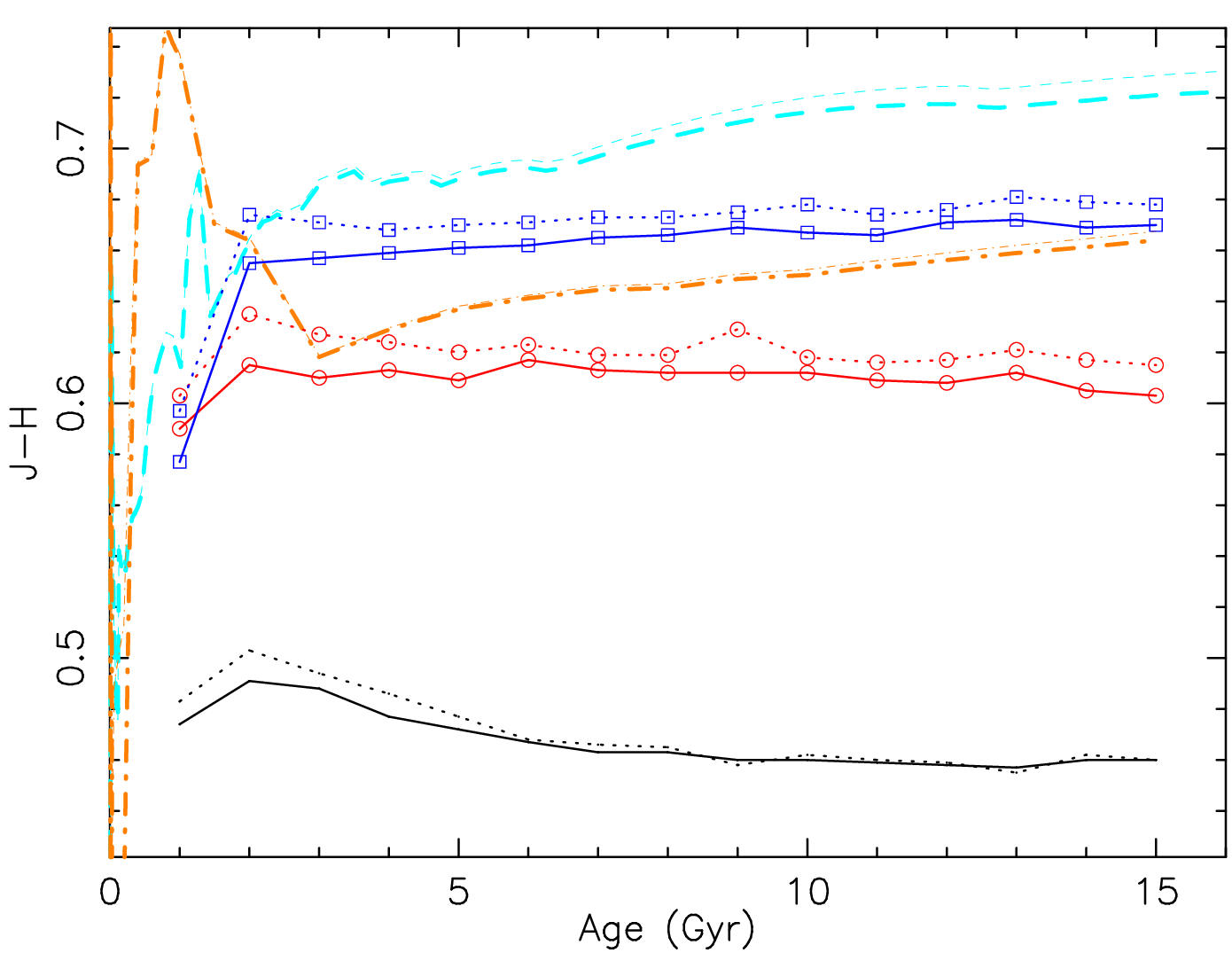}
}
\leftline{
\includegraphics[bb=82 37 580 700,height=4.40cm,width=4.70cm,clip,angle=-90]{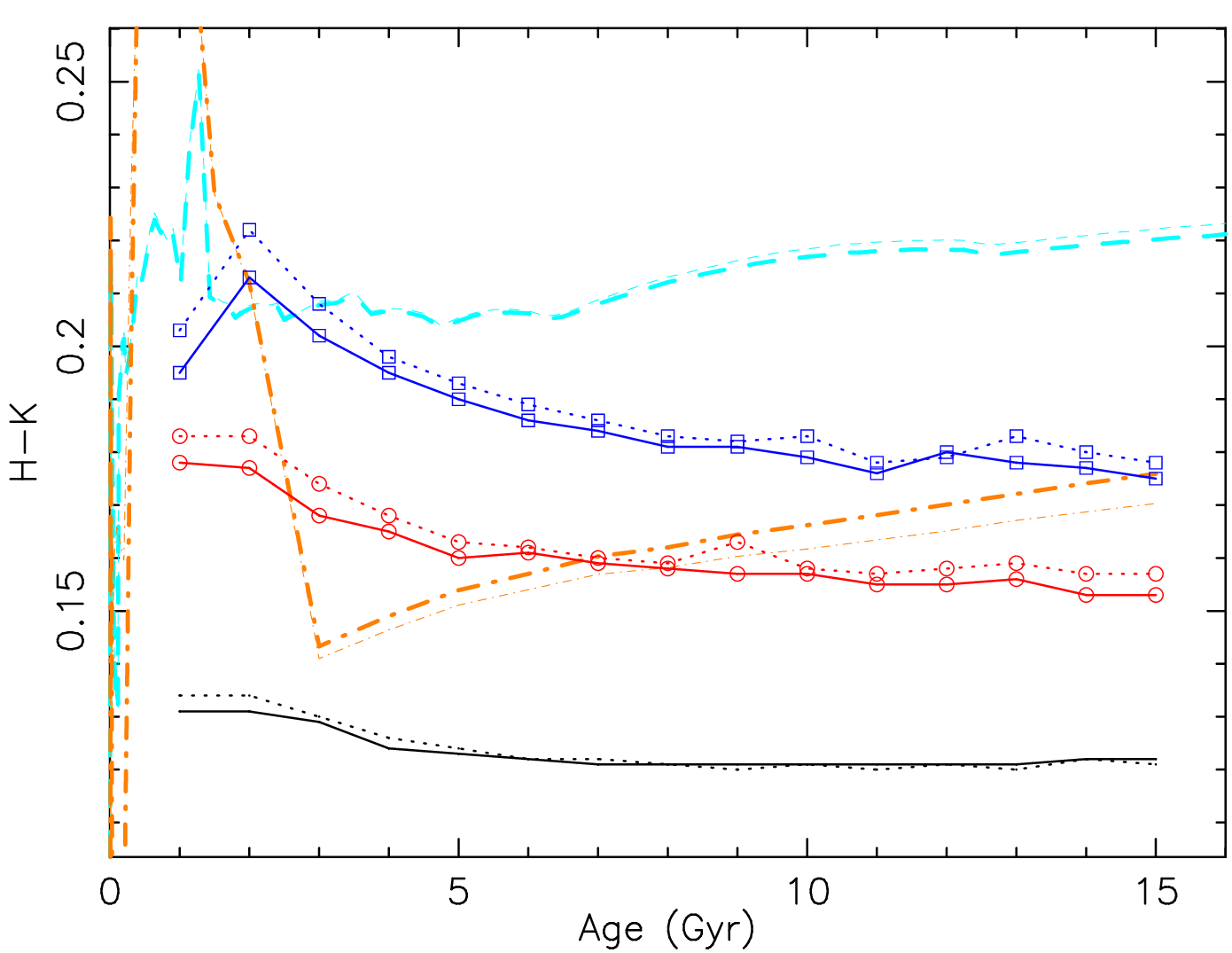}
\includegraphics[bb=82 37 580 700,height=4.40cm,width=4.70cm,clip,angle=-90]{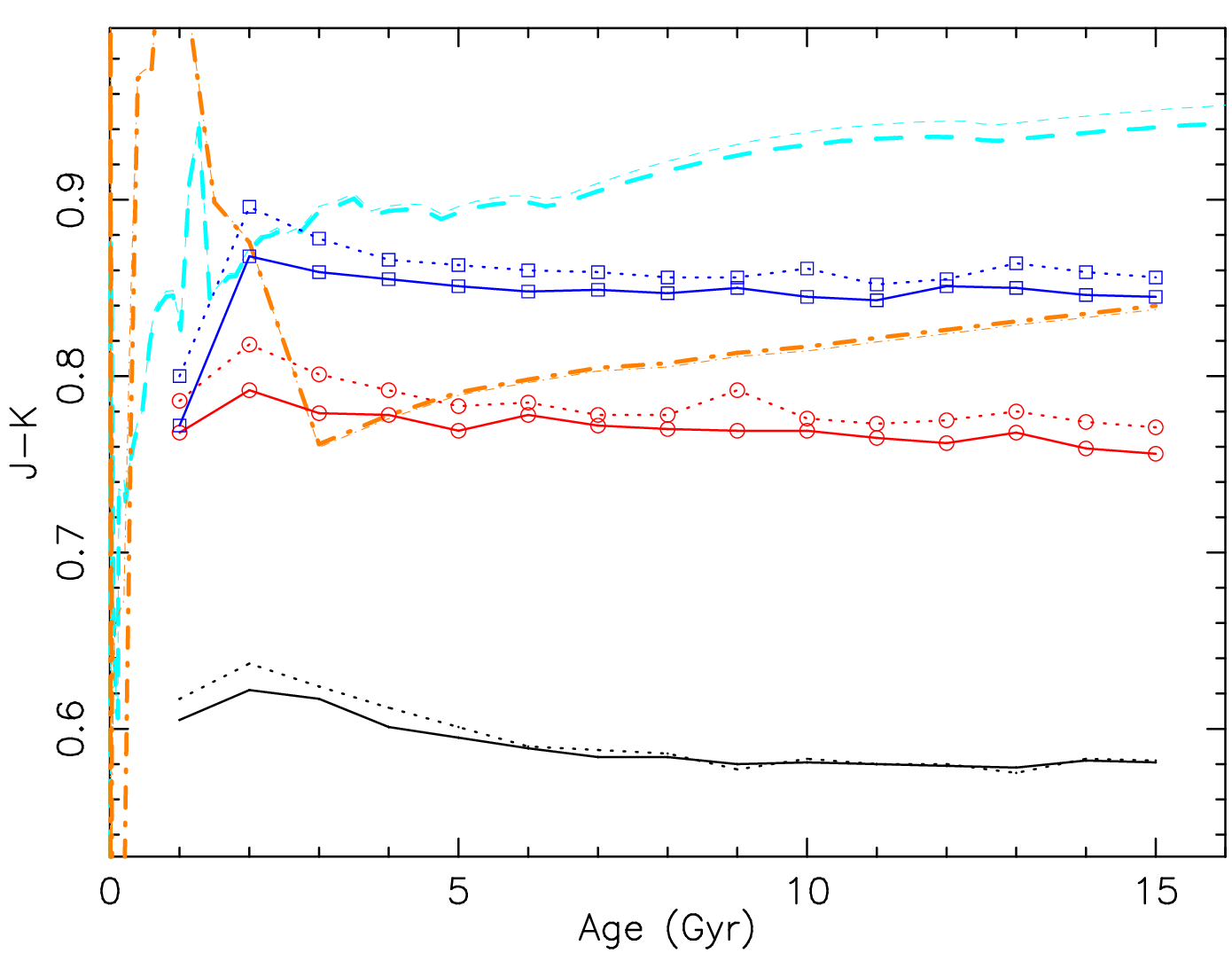}
}
\caption{The integrated colours for Models A and B only at $Z=0.02, 0.004$ and 0.0001. Values obtained by BC03-S, BC03-C, M05-S and M05-K at solar metallicity are also shown. Especially, in $V-K$ diagram the results of B07 at solar metallicity are shown (dash-dot-dot-dot line). The symbols have the same meanings as in Fig.~\ref{Fig:mags-all}.}
\leftline{$\star$ Note that $V-K$, $I-K$ and $H-K$ colours of BC03 are obtained by $V-[J-(J-K)]$, $V-(V-I)-[J-(J-K)]$ and $(J-K)-$}\leftline{$(J-H)$, respectively; $R-I, I-K, J-K$ colours of M05 are obtained by using the corresponding magnitudes.}
\label{Fig:colors-all}
\end{figure*}

\begin{figure*}
\leftline{
\includegraphics[bb=118 95 475 663,height=4.40cm,width=4.40cm,clip,angle=-90]{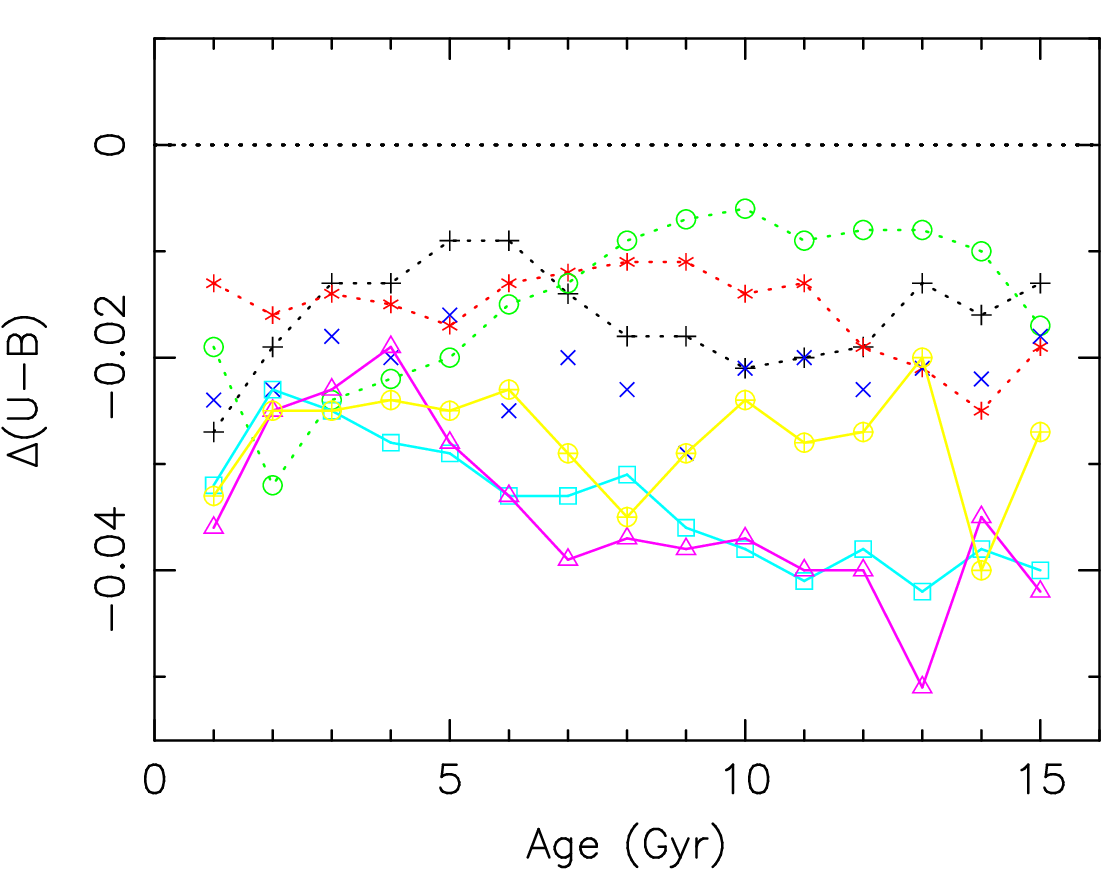}
\includegraphics[bb=118 95 475 663,height=4.40cm,width=4.40cm,clip,angle=-90]{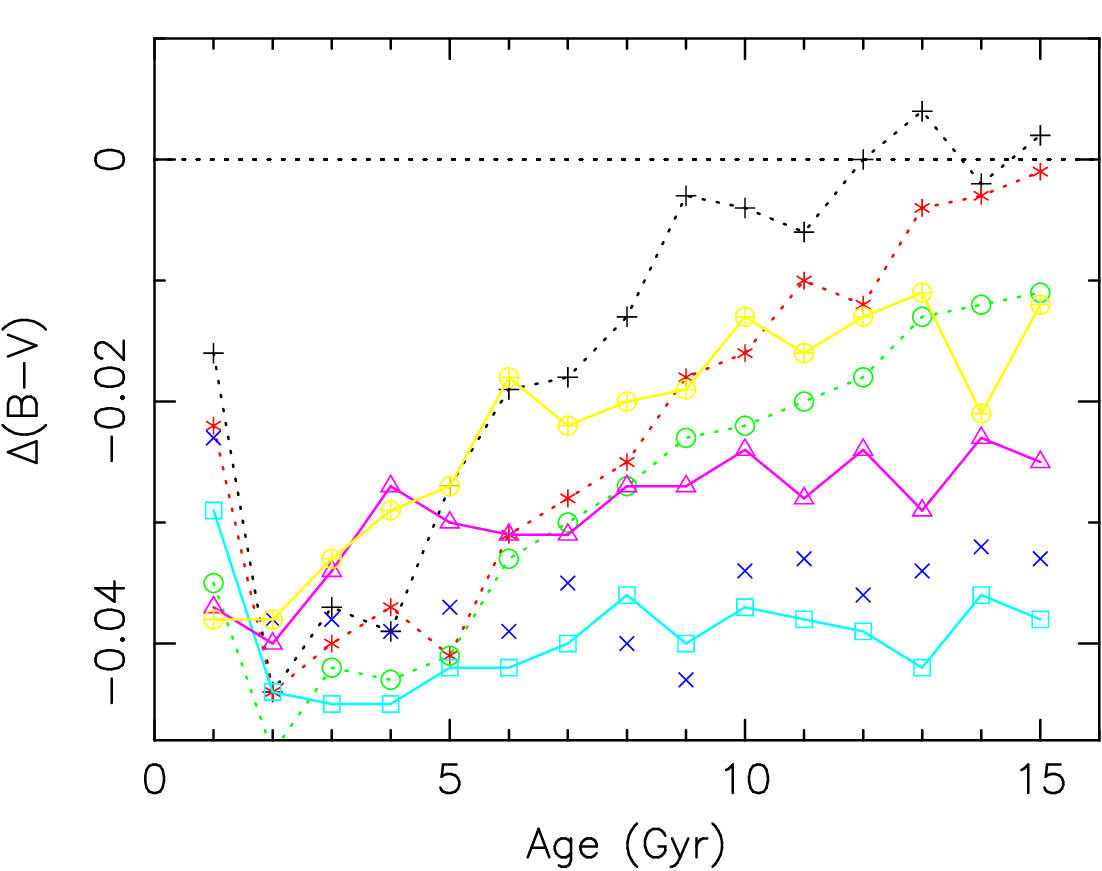}
\includegraphics[bb=118 95 475 663,height=4.40cm,width=4.40cm,clip,angle=-90]{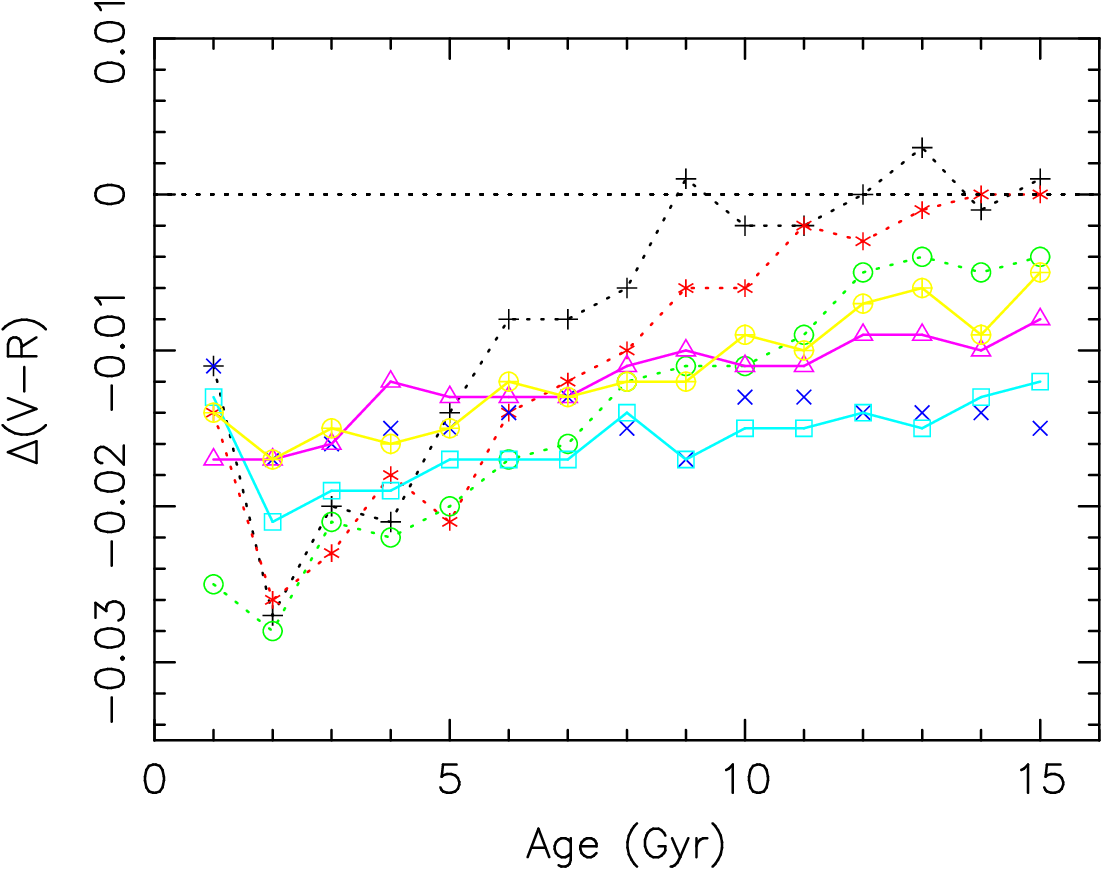}
\includegraphics[bb=118 95 475 663,height=4.40cm,width=4.40cm,clip,angle=-90]{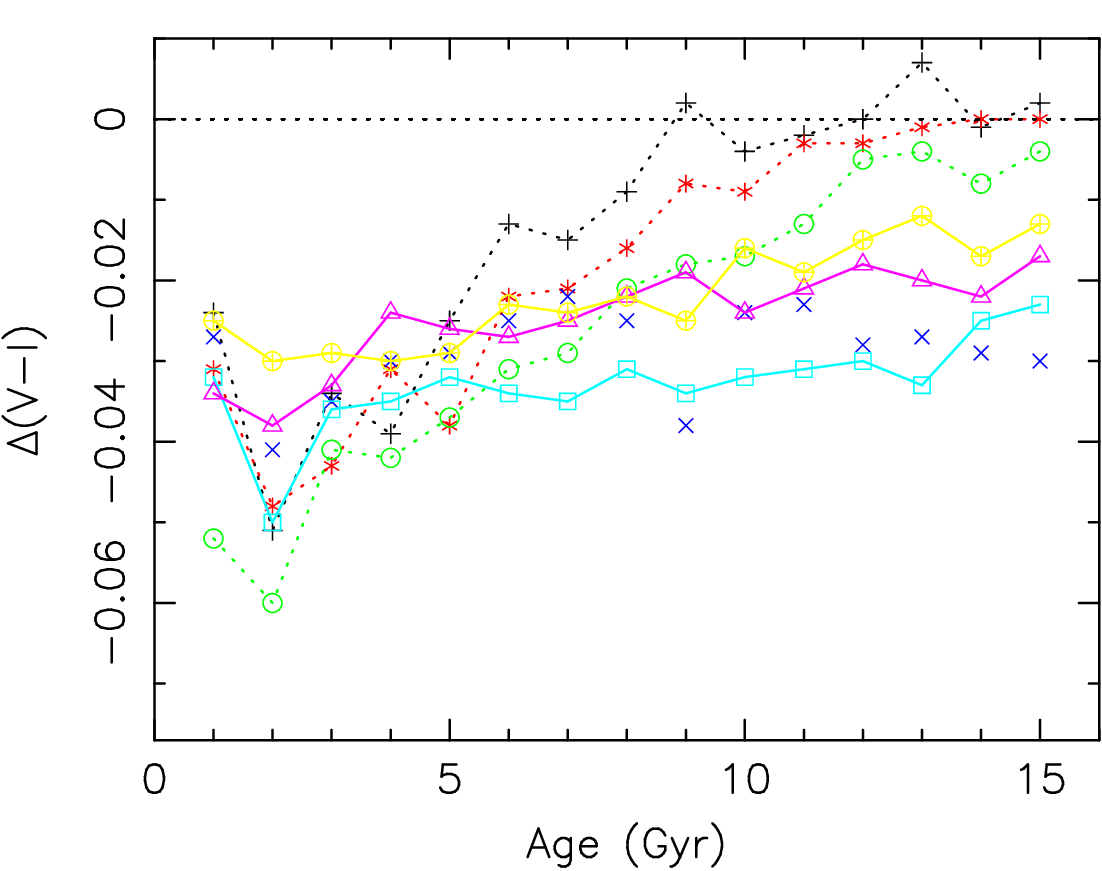}
}
\leftline{
\includegraphics[bb=137 95 474 663,height=4.40cm,width=4.00cm,clip,angle=-90]{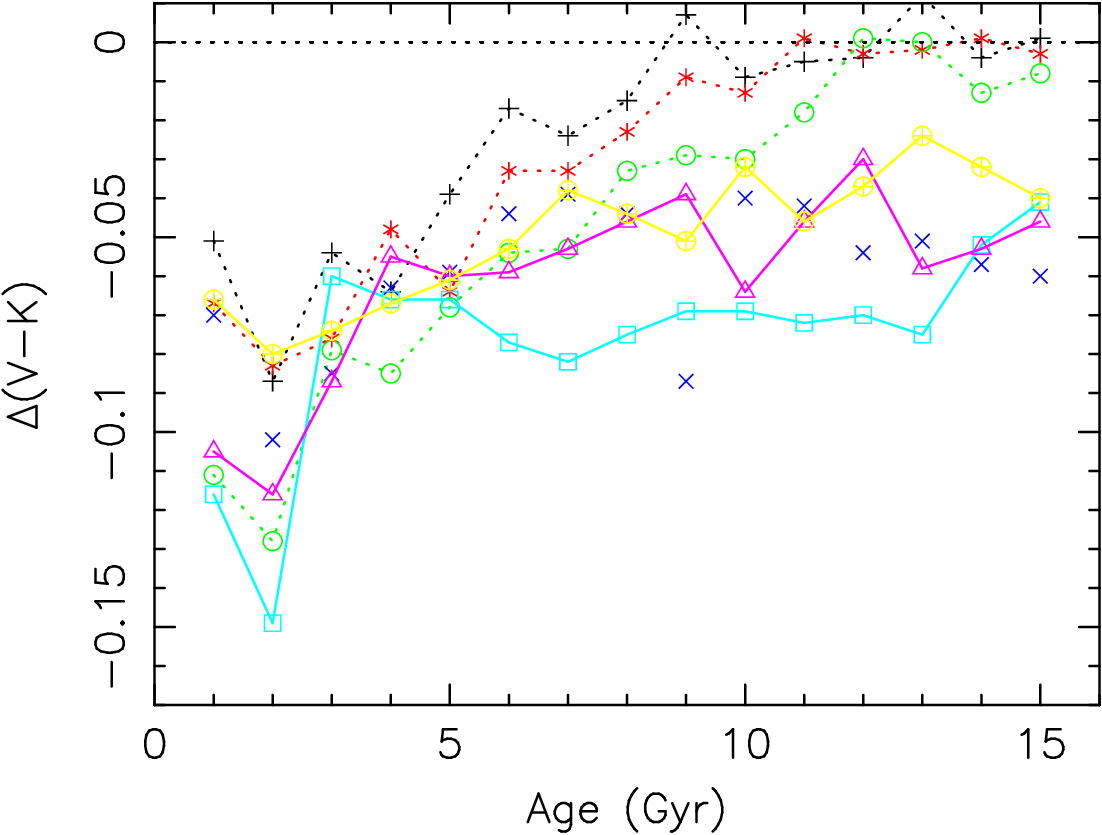}
\includegraphics[bb=137 95 474 663,height=4.40cm,width=4.00cm,clip,angle=-90]{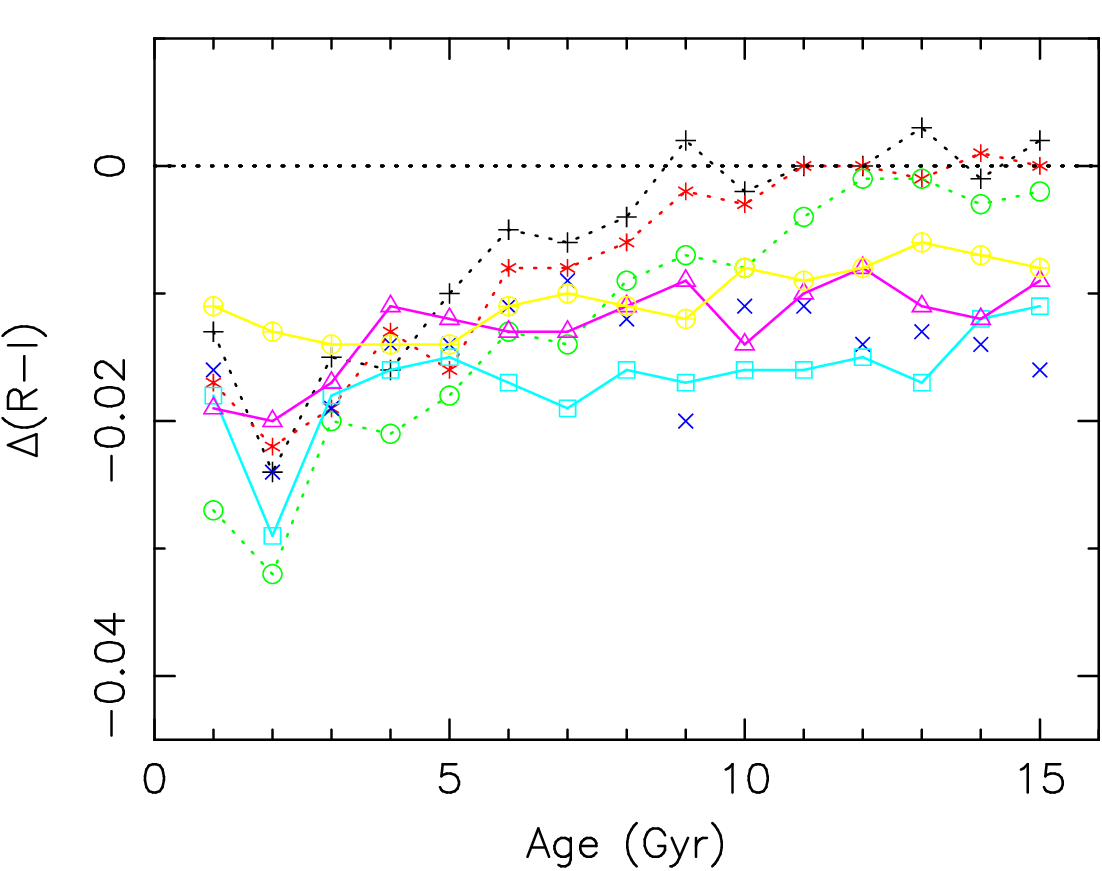}
\includegraphics[bb=137 95 474 663,height=4.40cm,width=4.00cm,clip,angle=-90]{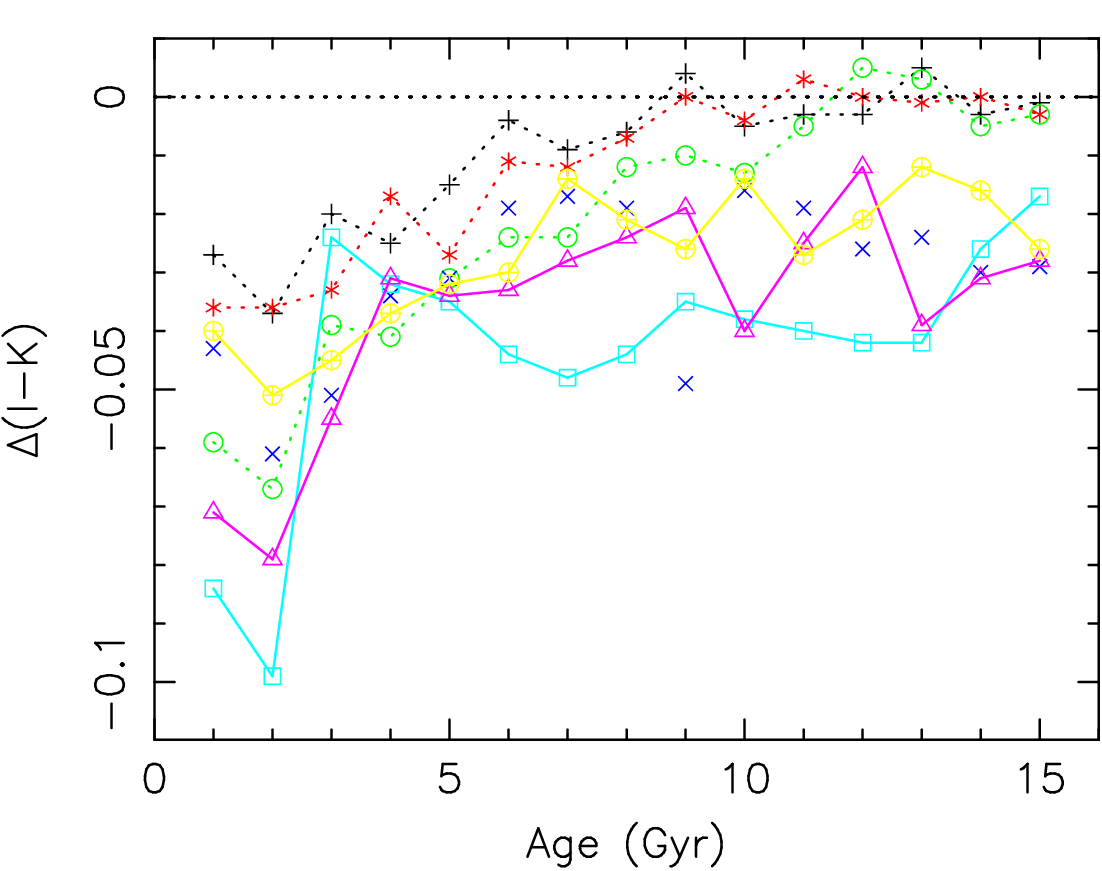}
\includegraphics[bb=137 95 474 663,height=4.40cm,width=4.00cm,clip,angle=-90]{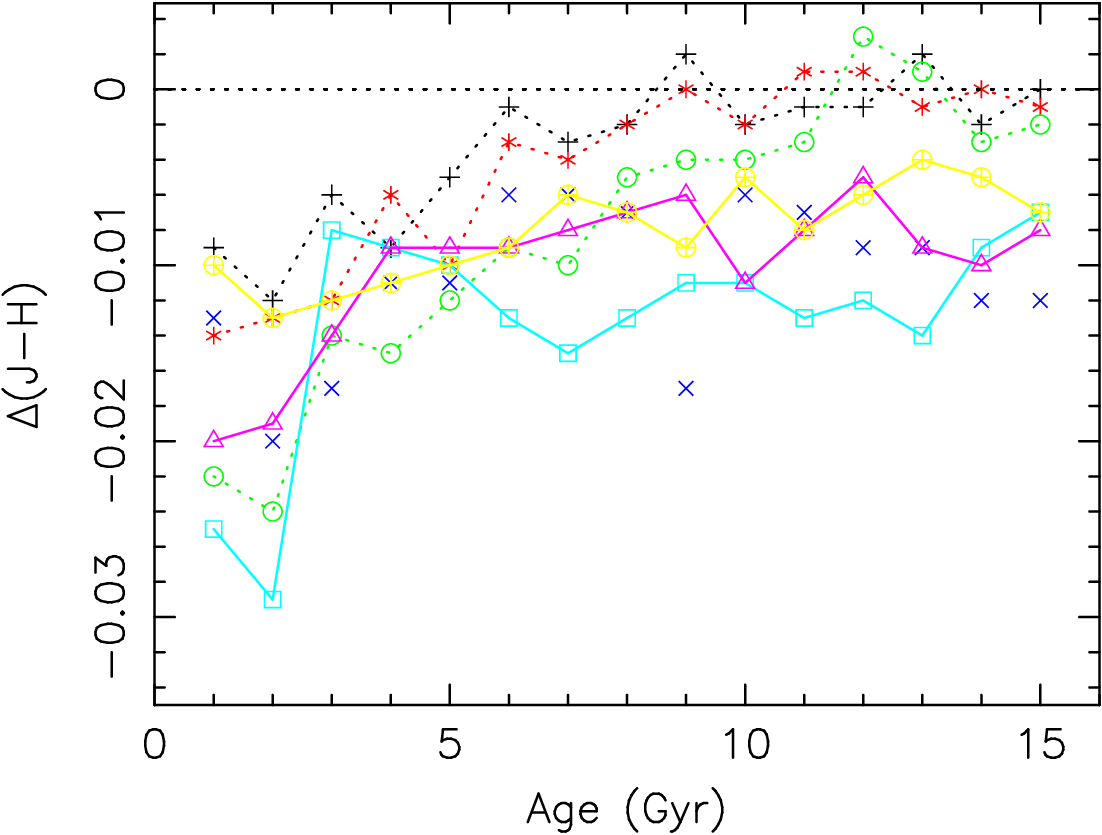}
}
\leftline{
\includegraphics[bb=136 95 538 663,height=4.40cm,width=4.80cm,clip,angle=-90]{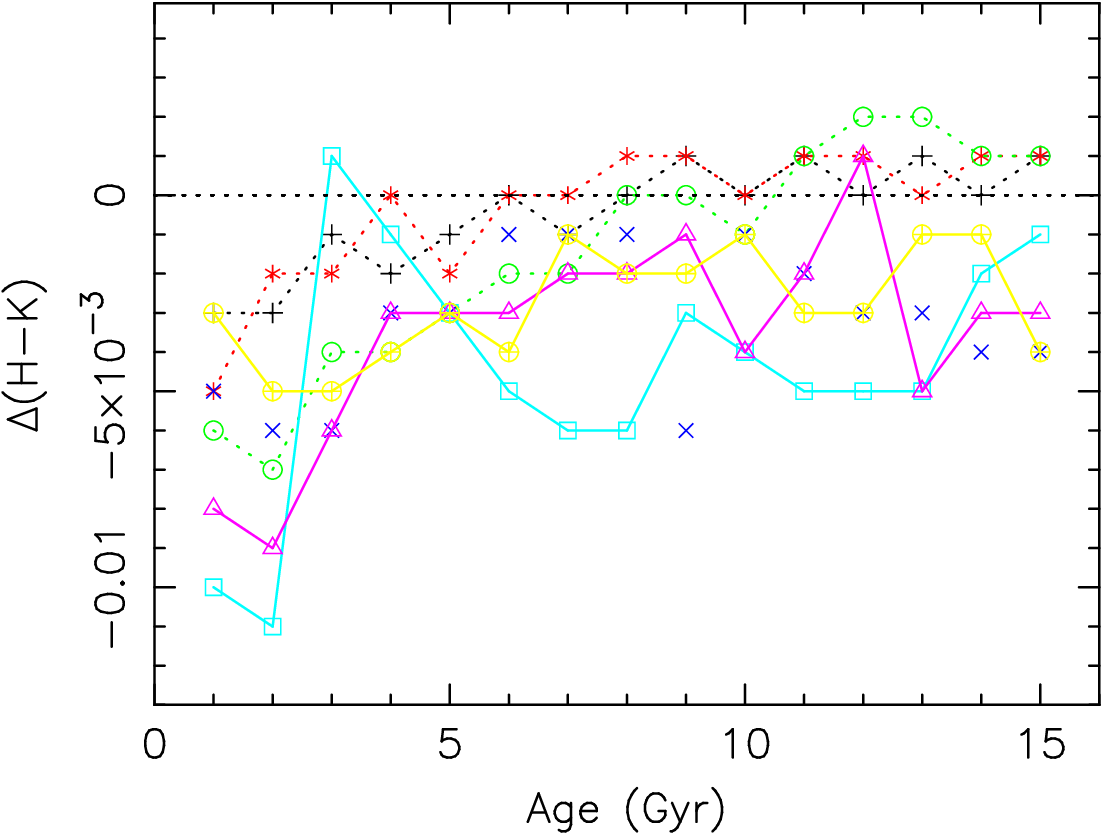}
\includegraphics[bb=136 95 538 663,height=4.40cm,width=4.80cm,clip,angle=-90]{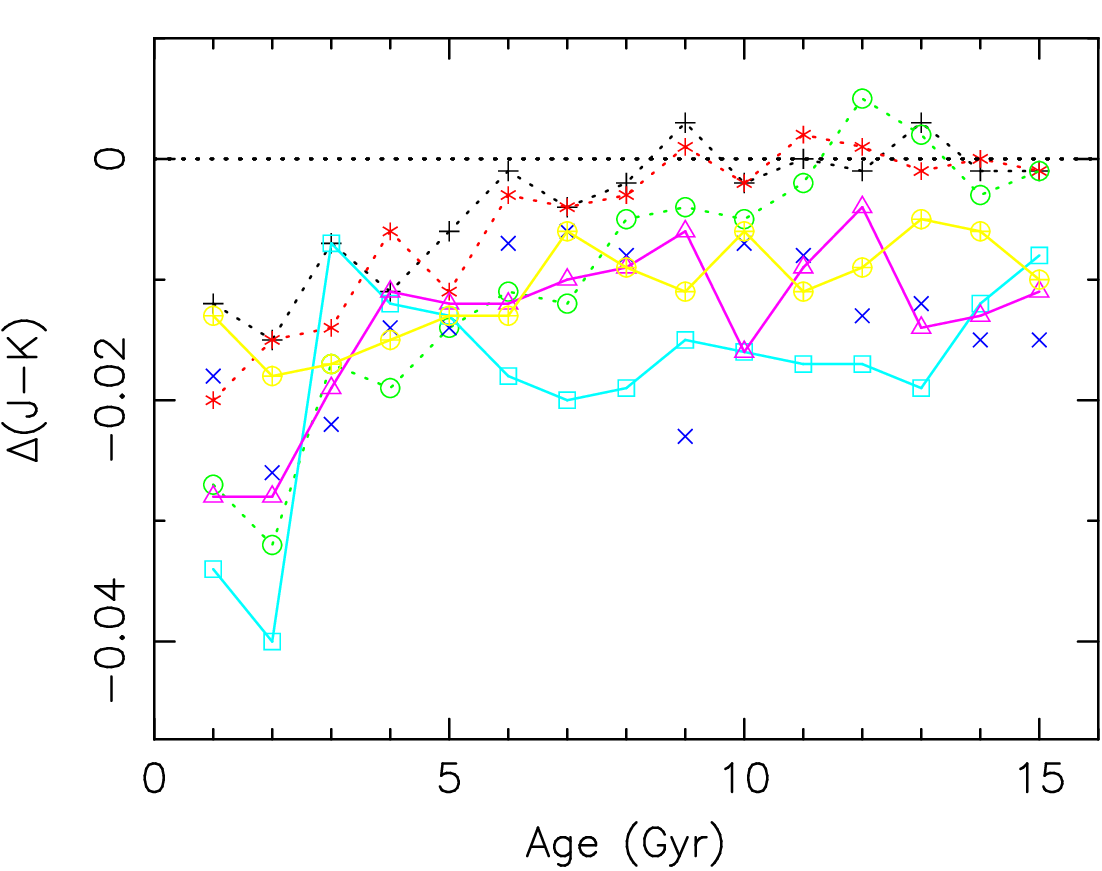}
}
\caption{Differences in the integrated colours between Models A and B for 7 metallicities. The symbols have the same meanings as in Fig.~\ref{Fig:dmag-t-z}.}
\label{Fig:dcol-t-z}
\end{figure*}

\begin{figure}
\centering{
\includegraphics[bb=135 104 533 626,height=6.7cm,width=6.70cm,clip,angle=-90]{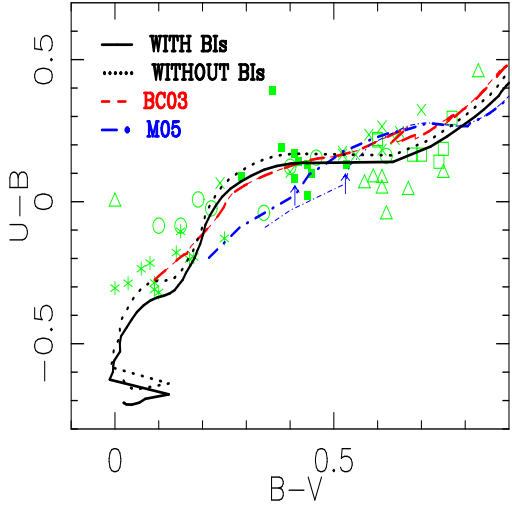}
}
\caption{~($B-V$) versus ($U-B$) colours of star
clusters. The different symbols represent  Magellanic Clouds GCs
with the SWB type in the range 3-7 (3-star, 4-circle, 5-cross, 6-open rectangle,7-triangle). Solid rectangles show the
young star clusters in the merger remnant galaxy NGC 7252. The
black lines show the evolution of Models A (full line) and B (dotted line) at $Z=0.01$, the ages of BSPs are
greater than a few Myr. The red dashed lines are the BC03-S
(thick) and BC03-C (thin) models at $Z=0.008$, note that two lines overlap. The blue dot-dash
lines are the M05-S (thick) and M05-K (thin) models at $Z=0.01$
with age $\rm{log} \tau > 8$\,yr, respectively.}
\label{Fig:bv-ub-MC}
\end{figure}

\begin{figure}
\centering{
\includegraphics[bb=135 104 533 626,height=6.7cm,width=6.7cm,clip,angle=-90]{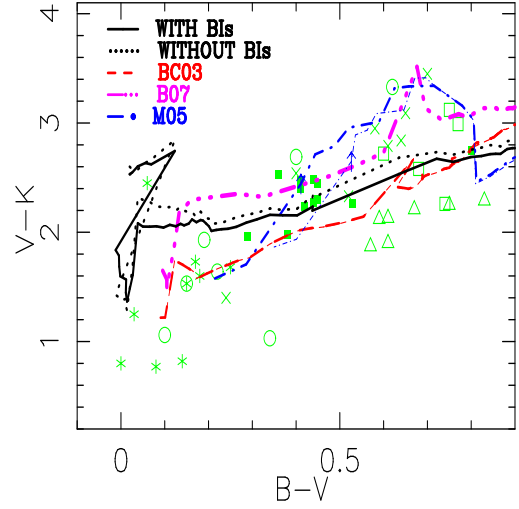}
}
\caption{Similar to Fig.~\ref{Fig:bv-ub-MC}, but for ($B-V$) .vs. ($V-K$) colours. Also, the values of B07 at solar metallicity are also shown (dash-dot-dot-dot line).}
\label{Fig:bv-vk-MC}
\end{figure}

\begin{figure}
\leftline{
\includegraphics[bb=119 95 474 627,height=4.0cm,width=4.0cm,clip,angle=-90]{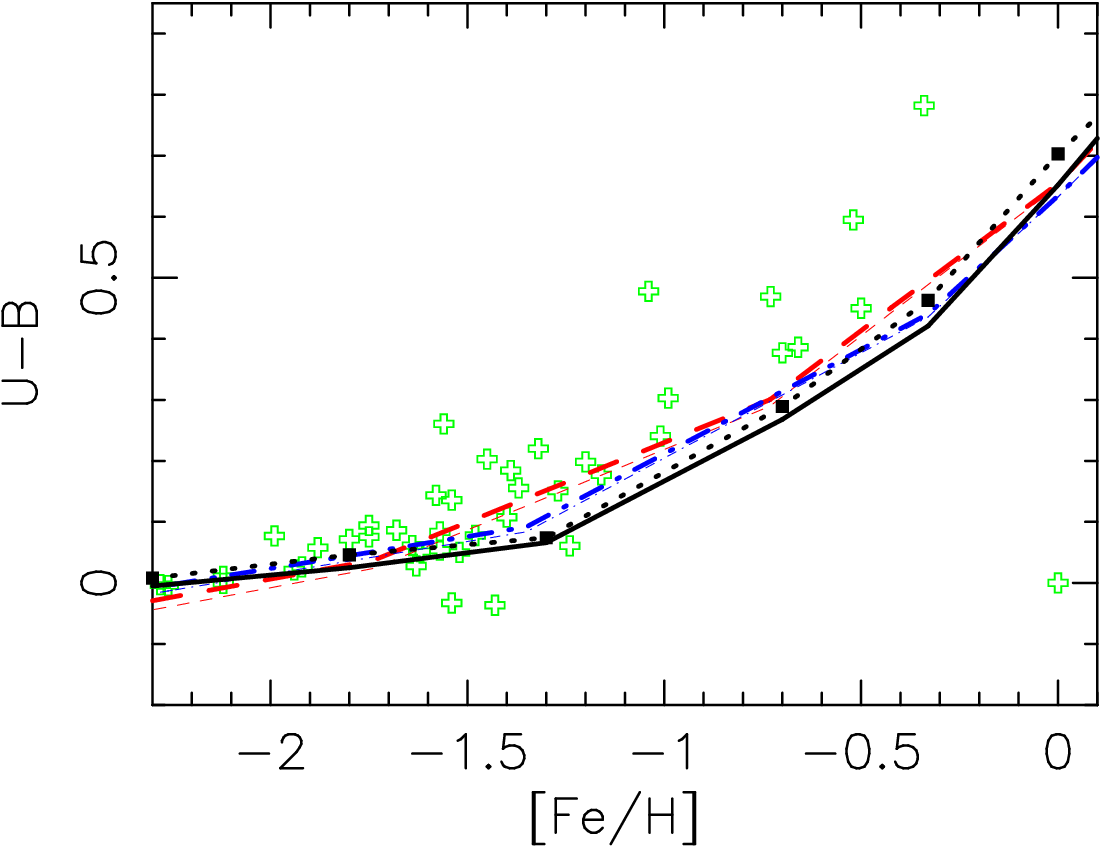}
\includegraphics[bb=119 95 474 627,height=4.0cm,width=4.0cm,clip,angle=-90]{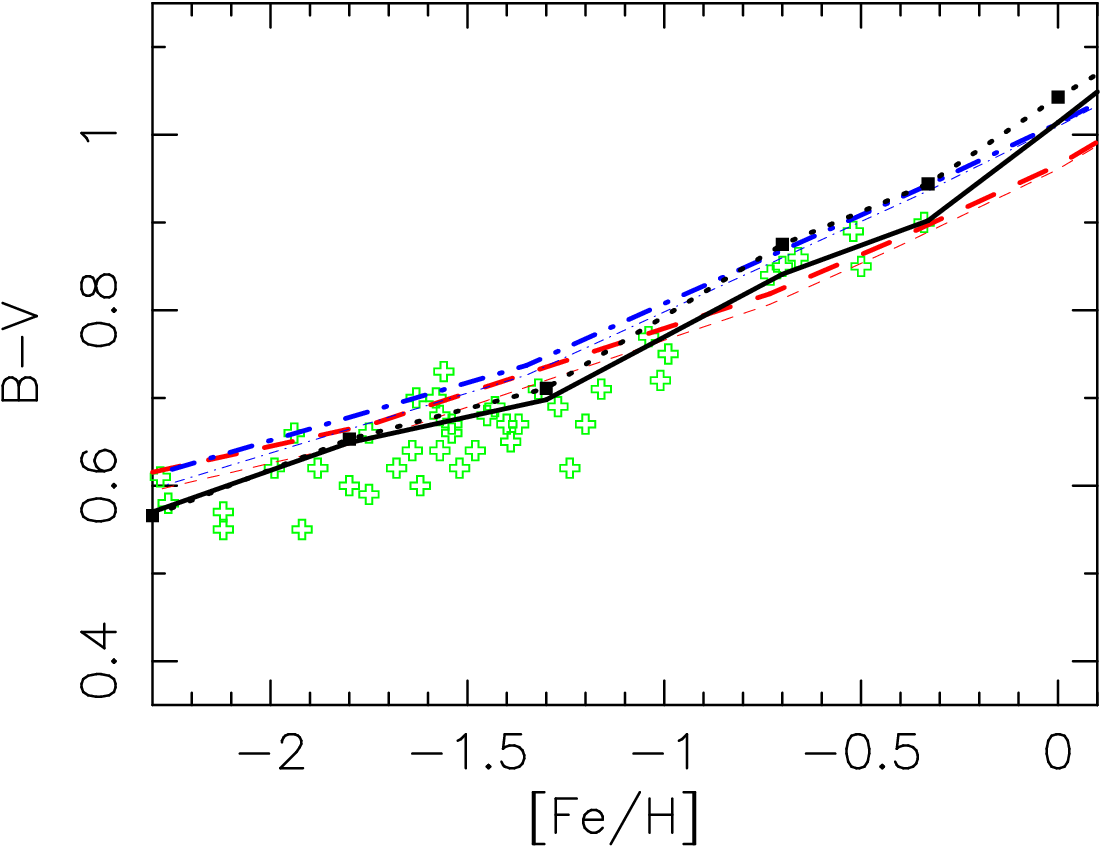}
}
\leftline{
\includegraphics[bb=137 95 474 627,height=4.0cm,width=3.75cm,clip,angle=-90]{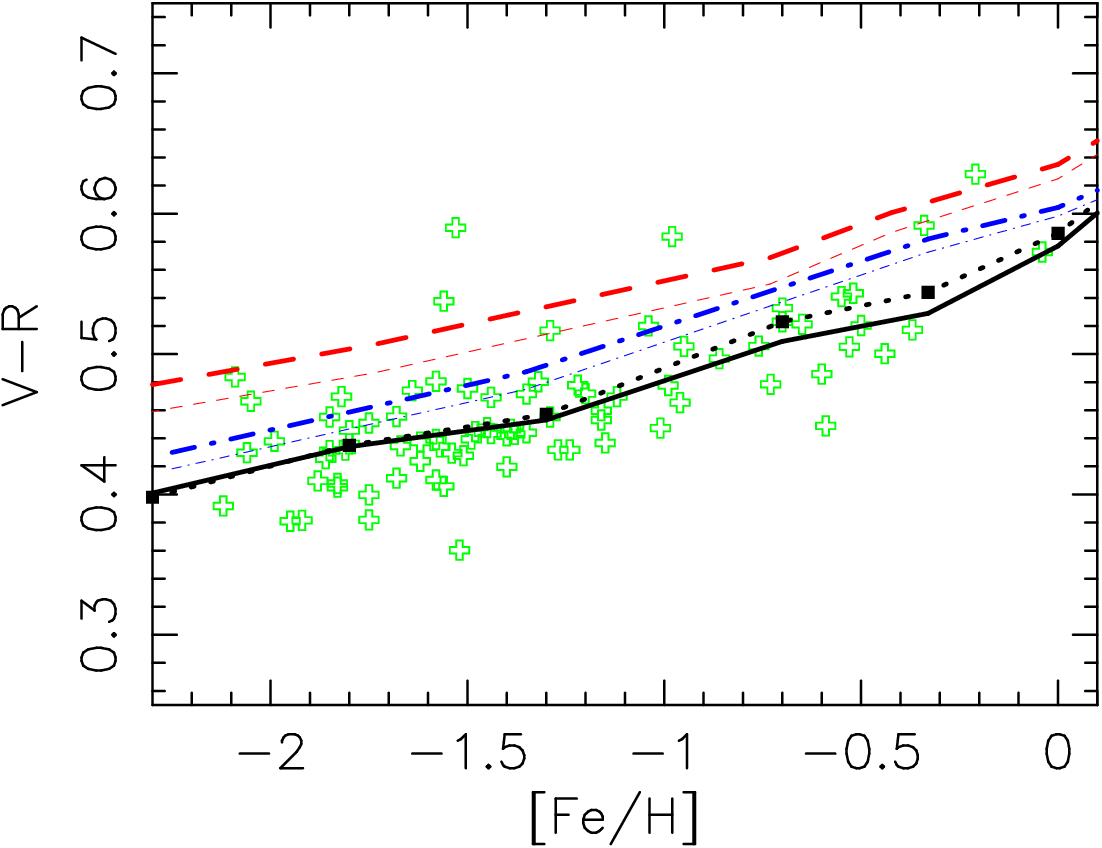}
\includegraphics[bb=137 95 474 627,height=4.00cm,width=3.75cm,clip,angle=-90]{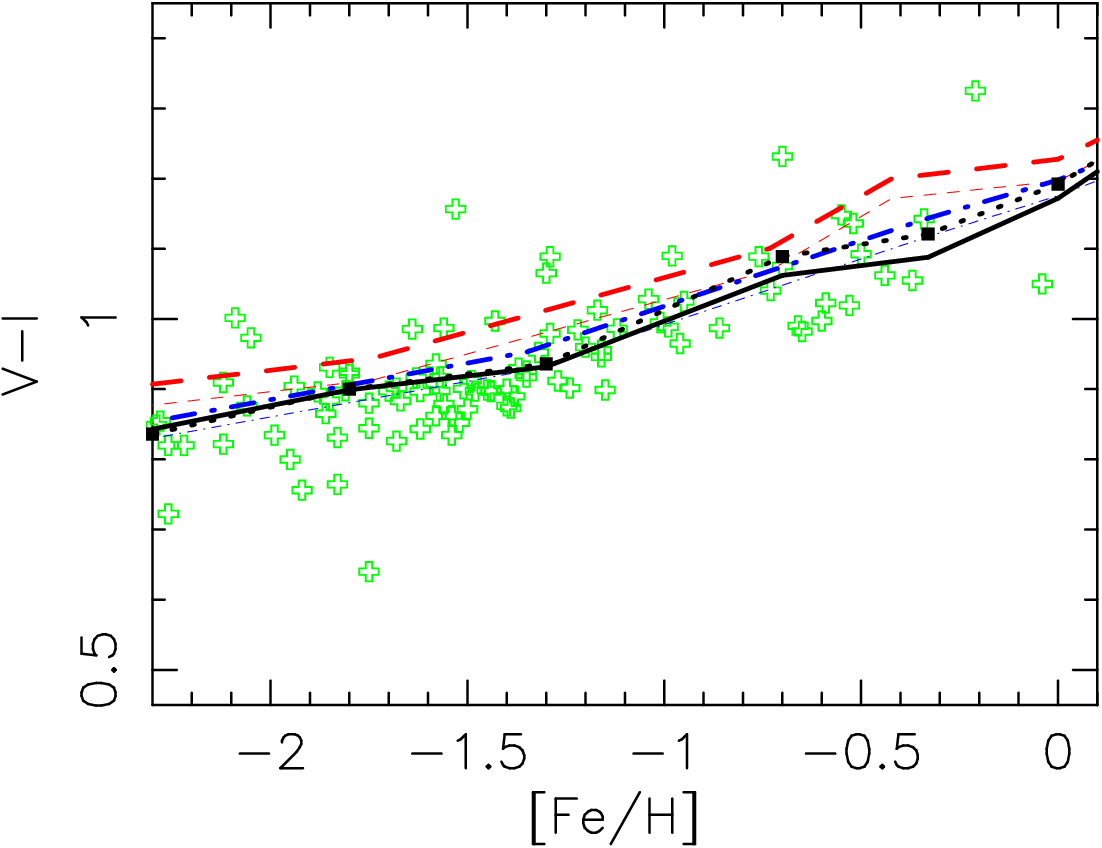}
}
\leftline{
\includegraphics[bb=137 95 563 627,height=4.00cm,width=4.80cm,clip,angle=-90]{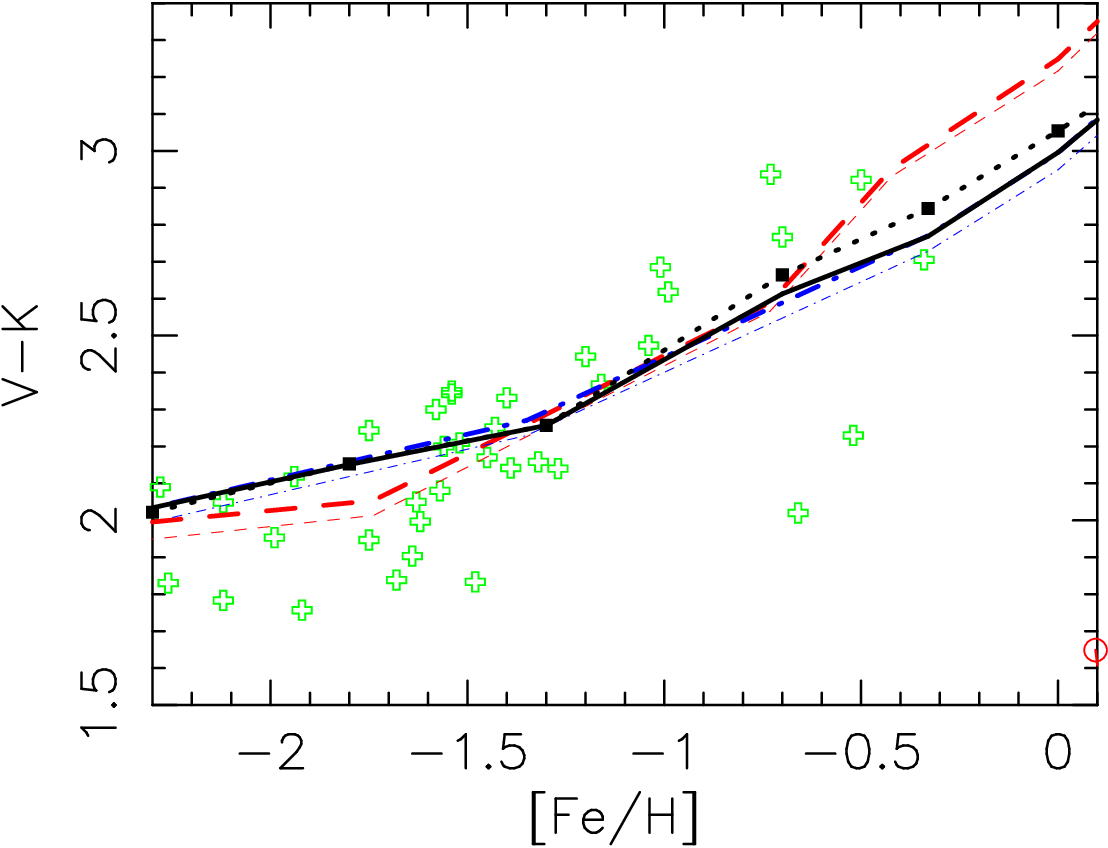}
\includegraphics[bb=137 95 563 627,height=4.00cm,width=4.80cm,clip,angle=-90]{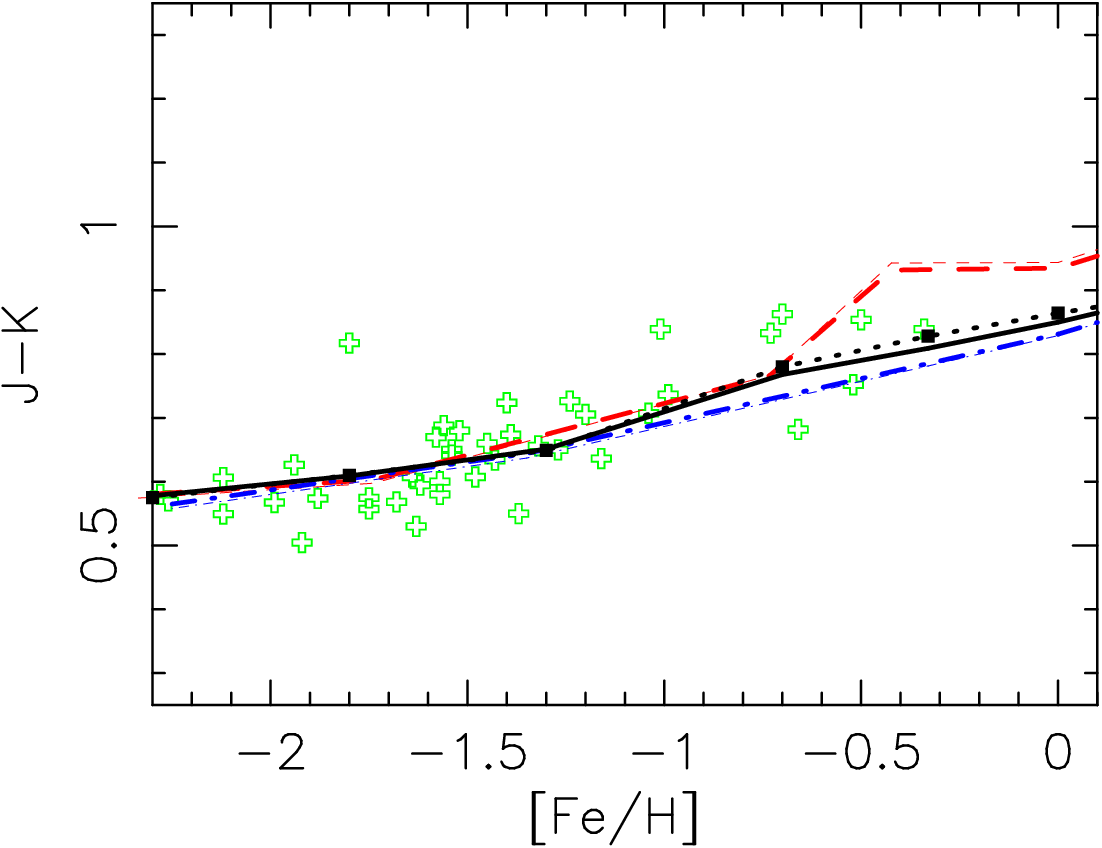}
}
\caption{Comparison of broad-band colors with data of MW GCs. The lines represent the same models as in Fig.~\ref{Fig:bv-ub-MC}. All models have the same age $\tau = 13$\,Gyr}
\label{Fig:col-feh-mw}
\end{figure}

Using the magnitudes we obtain the integrated $U-B$, $B-V$, $V-R$, $V-R$, $V-K$, $R-I$, $I-K$, $J-H$, $H-K$ and $J-K$ colours for Models A and B. For the sake of clarity, in Fig.~\ref{Fig:colors-all} we give them only at $Z=0.02, 0.004$ and 0.0001 . From it we see that all colours increase with age and metallicity for Models A and B.

In Fig.~\ref{Fig:dcol-t-z} we give the differences in these colours between Models A and B for BSPs at $Z=0.0001, 0.0003, 0.001, 0.004, 0.01, 0.02$ and 0.03. From it we see that BIs make the BSPs bluer (smaller) for all metallicities and ages. The absolute value of the difference in the $U-B$ colour between Models A and B, $\Delta (U-B)$ ($\equiv (U-B)_{\rm A} - (U-B)_{\rm B}$), is independent of age and metallicity;
The absolute values of $\Delta (B-V)$, $\Delta (V-R)$, $\Delta (V-I)$, $\Delta (R-I)$, $\Delta (V-K)$ and $\Delta (I-K)$ decrease with age at low metallicities, their variation with metallicity also has an turnoff, increase until $Z \sim$\,0.01 then decrease with metallicity;
The absolute values of $\Delta (H-K)$ and $\Delta (J-K)$ decrease with age at low metallicities and are independent of metallicity.

In Fig.~\ref{Fig:colors-all} we also give the results of BC03-S, BC03-C, M05-S and M05-K at solar metallicity, especially in the $V-K$ diagram the results of B07 are given at solar metallicity.
$U,B,V,R,J$ and $K$ filters used by BC03 have been described in Sect. 3.2.2, $I$ and $H$ filters used by BC03 are in Cousins' \citep{bes83} and 2MASS system, respectively. Different from the above description, Busers' $B3$ filter is used for $B-V$ colour of BC03.
We see that the differences in those colours involving IR passbands are larger, this differences are partly caused by the difference of the filter definition used by each models.
Using the $J,H$ and $K$ definition in different systems (2MASS; Johnson-Cousins; IR filter+Palomar 200 IR detectors+atmosphere) we have transformed the ISEDs of BC03 to $J,H$ and $K$ magnitudes at solar metallicity. By comparison we find that the differences in the filter definition can cause to the discrepancies of $\sim 0.03, 0.04$ and 0.05 mag in $J, H$ and $K$ magnitudes, this will lead to significant differences in $J-H$, $H-K$ and $J-K$ colours because from Fig. \ref{Fig:colors-all} we see that at solar metallicity the variation of $J-H$, $H-K$ and $J-K$ colours from 3 to 15\,Gyr only is $\sim 0.04|0.04, 0.02|0.03$ and $0.04|0.08$ mag for BC03 and M05 models. Moreover, both the difference of the distribution of cooler stars in HR diagram, which caused by the different stellar evolutionary models and the different treatment of TPAGB stars, and the difference in the spectral libraries cause to the differences in those colours involving IR passbands. so great progresses in stellar evolutionary models and spectral libraries are needed. And, it must be cautious when only IR results are used.

\subsubsection{Colour-Age relations by comparing with Magellanic Cloud and NGC 7252}
Combining the results for BSPs of age $\tau < 1$\,Gyr, which are presented in our website or request from the first author, we compare our model results with Magellanic Cloud (MC) GCs with the type of \citet[hereafter SWB]{sea80} in the range of 3-7 and young star clusters in the merger remnant galaxy NGC 7252 in ($B-V$).vs.($U-B$) (Fig.~\ref{Fig:bv-ub-MC}) and ($B-V$).vs.($V-K$) (Fig.~\ref{Fig:bv-vk-MC}) diagrams. These GCs have the known age and can be used to check the colour-age relations. Also in Figs.~\ref{Fig:bv-ub-MC} and \ref{Fig:bv-vk-MC} we plot the values of BC03-S, BC03-C, M05-S and M05-K.
The data of MC GCs are from \citet{van81} for ($U-B$) and ($B-V$); \citet{per83} for ($V-K$); \citet{fro90} for the SWB-type. The reddening $E$(B-V) is taken either from \citet{per83} or from \citet{sch98}. The data of young star clusters in the merger remnant galaxy NGC 7252 are from \citet{mil97} and \citet{mar01}.

From Fig.~\ref{Fig:bv-ub-MC}, we see that Models A and B agree with those of BC03-S and BC03-C (two lines overlap), the curves of M05-S and M05-C underlie those of Models A and B in the $ 0.2< B-V<0.45$ and $0.3< B-V<0.5$ ranges, respectively (correspond to the age of $1-3 \times 10^8$\,yr), while during the following epoch lie above those of Models A and B, BC03-S and BC03-C. Arrows mark the values of M05-S and M05-K at an age of 0.3\,Gyr. It seems that the BC03-S, BC03-C, Models A and B agree with the observations in $(B-V)$.vs.$(U-B)$ diagram.
From Fig.~\ref{Fig:bv-vk-MC}, we see that larger discrepancies exist among models. The curve of B07 models lies above that of BC03 by the amount of $\sim$\,0.5\,mag (different $Z$); the trend of BC03-S, BC03-C, B07, Models A and B is similar and less steeper than those of M05-S and M05-K in the range of $0.28 < B-V < 0.6$ and $0.42 < B-V < 0.6$ ranges, respectively. In the range of $0.6 < B-V < 0.8$ B07 seems to be able to explain the redder MW GCs (correspond to the SWB type 5 and 6) but the age range is narrow.

\subsubsection{Colour-Metallicity relations by comparing with Milky Way}
Milky Way GCs span a wide range of metallicities, but have the nearly same age, so are ideal to calibrate the synthetic colours/metallicity relations (M05).
We compare our model results with those of BC03-S, BC03-C, M05-S, M05-K and the Galactic GCs in colour-metallicity diagrams (Fig.~\ref{Fig:col-feh-mw}). All models have the same age $\tau = 13$\,Gyr. For the Galactic GCs, as in \citet{bar00}, we obtain optical colours, metallicity and reddening $E$(B-V) from the 2003 version of the Harris catalogue (1996) and IR colours from \citet{fro80}, as reported in \citet{bro90}. The cluster's colours are dereddened by using the values of $E$(B-V) given in the catalogue instead of the reddening correction applied by \citet{bro90} and the extinction curve of \citet{car89} for $R_v = 3.1$.

From Fig.~\ref{Fig:col-feh-mw} we see that in $(V-R)$-[Fe/H] diagram larger discrepancy exists among these models, the values of Models A and B are lower than those of BC03-S, BC03-C, M05-S and M05-K, it seems that Model B matches the observations better than Model A, BC03-S, BC03-C, M05-S and M05-K. The $U-B$ colour of all models is lower than that of observations at high metallicities; while greater for $B-V$ colour at low metallicities. All models match the observational $V-K$ and $J-K$ colours well.

\section{Summary and Conclusions}
Using the '{\sl patched}' Monte Carlo simulation we reproduce the samples of binaries in BSPs, present the stellar masses, the integrated magnitudes, mass-to-light ratios and the colours involving infrared passbands for BSPs with (Model A) and without (Model B) binary interactions (BIs). The results presented in this paper are very similar to those calculated from the BSPs composed of $2 \times 10^7$ binaries.

By comparison, we find (i) BIs make the stellar masses of BSPs smaller by the amount of $\sim$\,3.6-4.5 per cent during the past 15\,Gyr. The absolute values of the differences in the stellar masses between Models A and B, increase with age and decreasing metallicity.
(ii) BIs make the magnitudes of BSPs greater ($\sim$\,0.18\,mag at the most) except $U$ magnitude at age $\tau \le 3$\,Gyr and metallicity $Z \ge 0.01$. The differences in the magnitudes between Models A and B ($\Delta X, X=BOL, U, B, ..., M$), increase with age except $R,I,J,H$ and $K$ magnitudes at low metallicities ($Z \la 0.001$). The relations between the differences in the magnitudes and $Z$ have three types:
$\Delta BOL, \Delta J, \Delta H$ and $\Delta K$ are independent of $Z$;
$\Delta U$ and $\Delta B$ correlate with $Z$ and decrease with $Z$ at low metallicities and increase with $Z$ at high metallicities;
$\Delta V$, $\Delta R$ and $\Delta I$ correlate with $Z$ and decrease with $Z$ only at ages $\tau \le 10$\,Gyr.
According to the definition of sensitive parameter \citep{wor94}, we draw a conclusion that BIs make $V$ magnitude less sensitive to age and $R$ and $I$ magnitudes more sensitive to metallicity.
(iii) BIs make the mass-to-light ratios of BSPs greater except those in the $U$ and $B$ passbands at high metallicities. The dependencies of the $\Delta ({\rm log} m^{\star}/L_X)$ ($X=BOL, U, B, ..., M$, the difference in the ratio of stellar mass to the luminosity in $X$ passband between Models A and B) on age and metallicity are similar to those of $X$ magnitude on age and metallicity.
(iv) At last, BIs make the integrated colours smaller except those at late ages and $Z<0.003$.

Comparing the model results with MC GCs in the diagrams of $(B-V)$.vs.$(U-B)$ and $(B-V)$.vs.$(V-K)$, we see that the BC03 and our models agree with the observations in the $(B-V)$.vs.$(U-B)$ diagram; while in the $(B-V)$.vs.$(V-K)$ diagram the models differ from each other.
By comparing the model broad colours with MW GCs in colour-metallicity diagrams, we see that the results of BSPs with BIs (Model A) match the observations better than those of Model B, BC03 and M05 in the $(V-R)-$[Fe/H] diagram.

\section*{acknowledgements}
This work was funded by the Chinese Natural Science Foundation
(Grant Nos 10773026, 10673029, 10433030 \& 10521001) and by Yunnan Natural
Science Foundation (Grant Nos 2005A0035Q \& 2007A113M).

{}

\appendix
\onecolumn

\section[]{The stellar mass, magnitudes, mass-to-light ratios for Model A}
\scriptsize
\begin{longtable}{lrcccccccccccc}
 \hline
 \hline
 Age(yr) &
 $\scriptstyle m^{\star}$&
 $\scriptstyle BOL$&
 $\scriptstyle U $&
 $\scriptstyle B $&
 $\scriptstyle V $&
 $\scriptstyle R $&
 $\scriptstyle I $&
 $\scriptstyle J $&
 $\scriptstyle H $&
 $\scriptstyle K $&
 $\scriptstyle L $&
 $\scriptstyle L2$&
 $\scriptstyle M $ \\
 $\scriptstyle   $&
 $\scriptstyle   $&
 $\scriptstyle m^{\star}/L_{BOL}$&
 $\scriptstyle m^{\star}/L_U $&
 $\scriptstyle m^{\star}/L_B $&
 $\scriptstyle m^{\star}/L_V $&
 $\scriptstyle m^{\star}/L_R $&
 $\scriptstyle m^{\star}/L_I $&
 $\scriptstyle m^{\star}/L_J $&
 $\scriptstyle m^{\star}/L_H $&
 $\scriptstyle m^{\star}/L_K $&
 $\scriptstyle m^{\star}/L_L $&
 $\scriptstyle m^{\star}/L_{L2}$&
 $\scriptstyle m^{\star}/L_M $\\
 \hline
 \endfirsthead
 \caption[]{continued}\\
 \hline
 \hline
 Age(yr) &
 $\scriptstyle m^{\star}$&
 $\scriptstyle BOL$&
 $\scriptstyle U $&
 $\scriptstyle B $&
 $\scriptstyle V $&
 $\scriptstyle R $&
 $\scriptstyle I $&
 $\scriptstyle J $&
 $\scriptstyle H $&
 $\scriptstyle K $&
 $\scriptstyle L $&
 $\scriptstyle L2$&
 $\scriptstyle M $ \\
 $\scriptstyle   $&
 $\scriptstyle   $&
 $\scriptstyle m^{\star}/L_{BOL}$&
 $\scriptstyle m^{\star}/L_U $&
 $\scriptstyle m^{\star}/L_B $&
 $\scriptstyle m^{\star}/L_V $&
 $\scriptstyle m^{\star}/L_R $&
 $\scriptstyle m^{\star}/L_I $&
 $\scriptstyle m^{\star}/L_J $&
 $\scriptstyle m^{\star}/L_H $&
 $\scriptstyle m^{\star}/L_K $&
 $\scriptstyle m^{\star}/L_L $&
 $\scriptstyle m^{\star}/L_{L2}$&
 $\scriptstyle m^{\star}/L_M $\\
 \hline
 \endhead
 \multicolumn{14}{r}{continued on the next page}
 \endfoot
 \hline
 \endlastfoot

 \vspace*{1pt} \\
 \multicolumn{14}{c}{$Z = 0.0001$} \\
   1.0 $\times 10^9$  &     0.662  &     3.274  &     3.995  &     4.012  &     3.747  &     3.474  &     3.100  &     2.539  &     2.065  &     1.934  &     1.823  &     1.811  &     1.215  \\
                   &            &     0.172  &     0.145  &     0.166  &     0.240  &     0.261  &     0.252  &     0.224  &     0.197  &     0.181  &     0.167  &     0.165  &     0.128  \\
   2.0 $\times 10^9$  &     0.620  &     3.976  &     4.822  &     4.816  &     4.400  &     4.048  &     3.618  &     3.003  &     2.512  &     2.381  &     2.267  &     2.255  &     1.655  \\
                   &            &     0.307  &     0.291  &     0.326  &     0.411  &     0.415  &     0.381  &     0.322  &     0.279  &     0.255  &     0.235  &     0.233  &     0.180  \\
   3.0 $\times 10^9$  &     0.600  &     4.345  &     5.248  &     5.237  &     4.754  &     4.373  &     3.927  &     3.301  &     2.812  &     2.684  &     2.571  &     2.559  &     1.963  \\
                   &            &     0.417  &     0.417  &     0.466  &     0.551  &     0.542  &     0.490  &     0.410  &     0.356  &     0.327  &     0.301  &     0.299  &     0.231  \\
   4.0 $\times 10^9$  &     0.587  &     4.658  &     5.574  &     5.568  &     5.043  &     4.653  &     4.207  &     3.585  &     3.108  &     2.984  &     2.873  &     2.861  &     2.276  \\
                   &            &     0.545  &     0.551  &     0.618  &     0.703  &     0.685  &     0.620  &     0.521  &     0.457  &     0.421  &     0.389  &     0.386  &     0.302  \\
   5.0 $\times 10^9$  &     0.578  &     4.877  &     5.802  &     5.795  &     5.246  &     4.850  &     4.403  &     3.784  &     3.313  &     3.190  &     3.079  &     3.067  &     2.486  \\
                   &            &     0.656  &     0.668  &     0.750  &     0.834  &     0.809  &     0.731  &     0.617  &     0.543  &     0.501  &     0.463  &     0.459  &     0.361  \\
   6.0 $\times 10^9$  &     0.571  &     5.088  &     6.010  &     6.012  &     5.447  &     5.048  &     4.603  &     3.991  &     3.524  &     3.402  &     3.292  &     3.280  &     2.703  \\
                   &            &     0.787  &     0.799  &     0.904  &     0.992  &     0.959  &     0.868  &     0.737  &     0.651  &     0.602  &     0.556  &     0.552  &     0.435  \\
   7.0 $\times 10^9$  &     0.565  &     5.281  &     6.195  &     6.202  &     5.635  &     5.238  &     4.795  &     4.185  &     3.722  &     3.601  &     3.491  &     3.479  &     2.904  \\
                   &            &     0.931  &     0.940  &     1.067  &     1.167  &     1.131  &     1.026  &     0.873  &     0.774  &     0.716  &     0.662  &     0.656  &     0.518  \\
   8.0 $\times 10^9$  &     0.561  &     5.418  &     6.346  &     6.347  &     5.773  &     5.373  &     4.929  &     4.316  &     3.853  &     3.732  &     3.622  &     3.609  &     3.034  \\
                   &            &     1.047  &     1.071  &     1.209  &     1.315  &     1.270  &     1.151  &     0.977  &     0.867  &     0.802  &     0.741  &     0.734  &     0.580  \\
   9.0 $\times 10^9$  &     0.557  &     5.547  &     6.469  &     6.464  &     5.897  &     5.501  &     5.061  &     4.451  &     3.992  &     3.871  &     3.761  &     3.748  &     3.175  \\
                   &            &     1.171  &     1.191  &     1.338  &     1.464  &     1.420  &     1.291  &     1.098  &     0.977  &     0.905  &     0.836  &     0.829  &     0.655  \\
  10.0 $\times 10^9$  &     0.554  &     5.675  &     6.605  &     6.598  &     6.027  &     5.628  &     5.185  &     4.574  &     4.114  &     3.993  &     3.882  &     3.870  &     3.295  \\
                   &            &     1.311  &     1.342  &     1.505  &     1.641  &     1.586  &     1.439  &     1.223  &     1.088  &     1.007  &     0.930  &     0.921  &     0.728  \\
  11.0 $\times 10^9$  &     0.551  &     5.772  &     6.703  &     6.696  &     6.127  &     5.727  &     5.284  &     4.671  &     4.213  &     4.092  &     3.980  &     3.968  &     3.394  \\
                   &            &     1.425  &     1.461  &     1.638  &     1.789  &     1.729  &     1.568  &     1.331  &     1.185  &     1.097  &     1.012  &     1.003  &     0.793  \\
  12.0 $\times 10^9$  &     0.548  &     5.893  &     6.822  &     6.823  &     6.250  &     5.849  &     5.405  &     4.793  &     4.335  &     4.214  &     4.102  &     4.089  &     3.515  \\
                   &            &     1.587  &     1.623  &     1.833  &     1.996  &     1.926  &     1.747  &     1.481  &     1.320  &     1.222  &     1.127  &     1.117  &     0.883  \\
  13.0 $\times 10^9$  &     0.546  &     5.981  &     6.904  &     6.910  &     6.340  &     5.939  &     5.496  &     4.884  &     4.426  &     4.306  &     4.193  &     4.180  &     3.606  \\
                   &            &     1.714  &     1.744  &     1.977  &     2.158  &     2.084  &     1.891  &     1.604  &     1.431  &     1.325  &     1.221  &     1.210  &     0.957  \\
  14.0 $\times 10^9$  &     0.544  &     6.047  &     6.989  &     6.999  &     6.417  &     6.010  &     5.562  &     4.944  &     4.483  &     4.362  &     4.249  &     4.236  &     3.659  \\
                   &            &     1.814  &     1.878  &     2.139  &     2.310  &     2.217  &     2.003  &     1.690  &     1.503  &     1.390  &     1.280  &     1.269  &     1.000  \\
  15.0 $\times 10^9$  &     0.543  &     6.126  &     7.068  &     7.087  &     6.502  &     6.094  &     5.646  &     5.027  &     4.568  &     4.446  &     4.332  &     4.319  &     3.742  \\
                   &            &     1.945  &     2.013  &     2.311  &     2.490  &     2.387  &     2.156  &     1.819  &     1.619  &     1.497  &     1.379  &     1.366  &     1.077  \\

\vspace*{1pt} \\
\multicolumn{14}{c}{$Z$ = 0.0003} \\

   1.0 $\times 10^9$  &     0.662  &     3.344  &     4.040  &     4.100  &     3.798  &     3.520  &     3.153  &     2.600  &     2.114  &     1.987  &     1.876  &     1.865  &     1.289  \\
                   &            &     0.183  &     0.151  &     0.180  &     0.252  &     0.272  &     0.265  &     0.237  &     0.206  &     0.190  &     0.175  &     0.174  &     0.137  \\
   2.0 $\times 10^9$  &     0.619  &     4.039  &     4.899  &     4.888  &     4.454  &     4.099  &     3.664  &     3.032  &     2.511  &     2.378  &     2.262  &     2.250  &     1.651  \\
                   &            &     0.325  &     0.312  &     0.348  &     0.431  &     0.434  &     0.396  &     0.330  &     0.278  &     0.254  &     0.234  &     0.232  &     0.179  \\
   3.0 $\times 10^9$  &     0.599  &     4.379  &     5.335  &     5.316  &     4.797  &     4.403  &     3.943  &     3.291  &     2.768  &     2.636  &     2.520  &     2.509  &     1.914  \\
                   &            &     0.429  &     0.451  &     0.500  &     0.571  &     0.555  &     0.496  &     0.406  &     0.340  &     0.312  &     0.287  &     0.284  &     0.221  \\
   4.0 $\times 10^9$  &     0.586  &     4.661  &     5.645  &     5.632  &     5.070  &     4.662  &     4.199  &     3.548  &     3.033  &     2.905  &     2.791  &     2.780  &     2.194  \\
                   &            &     0.545  &     0.586  &     0.653  &     0.718  &     0.689  &     0.614  &     0.503  &     0.425  &     0.391  &     0.360  &     0.357  &     0.279  \\
   5.0 $\times 10^9$  &     0.576  &     4.898  &     5.915  &     5.901  &     5.295  &     4.873  &     4.408  &     3.759  &     3.251  &     3.127  &     3.014  &     3.003  &     2.424  \\
                   &            &     0.666  &     0.739  &     0.823  &     0.869  &     0.824  &     0.732  &     0.601  &     0.511  &     0.472  &     0.435  &     0.431  &     0.340  \\
   6.0 $\times 10^9$  &     0.569  &     5.104  &     6.120  &     6.126  &     5.498  &     5.068  &     4.602  &     3.957  &     3.455  &     3.333  &     3.221  &     3.210  &     2.638  \\
                   &            &     0.795  &     0.881  &     1.000  &     1.035  &     0.973  &     0.864  &     0.711  &     0.608  &     0.563  &     0.519  &     0.515  &     0.408  \\
   7.0 $\times 10^9$  &     0.563  &     5.289  &     6.298  &     6.313  &     5.676  &     5.245  &     4.780  &     4.139  &     3.643  &     3.522  &     3.412  &     3.400  &     2.833  \\
                   &            &     0.933  &     1.028  &     1.177  &     1.207  &     1.133  &     1.007  &     0.832  &     0.716  &     0.663  &     0.612  &     0.608  &     0.484  \\
   8.0 $\times 10^9$  &     0.558  &     5.444  &     6.450  &     6.468  &     5.828  &     5.396  &     4.933  &     4.295  &     3.803  &     3.684  &     3.573  &     3.562  &     2.998  \\
                   &            &     1.068  &     1.173  &     1.346  &     1.377  &     1.292  &     1.151  &     0.953  &     0.824  &     0.763  &     0.705  &     0.699  &     0.558  \\
   9.0 $\times 10^9$  &     0.554  &     5.571  &     6.592  &     6.606  &     5.958  &     5.522  &     5.058  &     4.417  &     3.925  &     3.806  &     3.695  &     3.683  &     3.118  \\
                   &            &     1.192  &     1.328  &     1.517  &     1.540  &     1.441  &     1.281  &     1.059  &     0.915  &     0.848  &     0.783  &     0.777  &     0.619  \\
  10.0 $\times 10^9$  &     0.551  &     5.711  &     6.743  &     6.752  &     6.097  &     5.660  &     5.195  &     4.555  &     4.064  &     3.945  &     3.834  &     3.822  &     3.257  \\
                   &            &     1.347  &     1.516  &     1.724  &     1.741  &     1.626  &     1.445  &     1.195  &     1.034  &     0.958  &     0.884  &     0.877  &     0.699  \\
  11.0 $\times 10^9$  &     0.548  &     5.811  &     6.835  &     6.837  &     6.190  &     5.757  &     5.295  &     4.658  &     4.171  &     4.051  &     3.940  &     3.928  &     3.364  \\
                   &            &     1.470  &     1.642  &     1.855  &     1.885  &     1.768  &     1.576  &     1.307  &     1.134  &     1.051  &     0.970  &     0.962  &     0.768  \\
  12.0 $\times 10^9$  &     0.545  &     5.919  &     6.949  &     6.942  &     6.298  &     5.866  &     5.404  &     4.765  &     4.277  &     4.157  &     4.045  &     4.033  &     3.468  \\
                   &            &     1.615  &     1.814  &     2.033  &     2.073  &     1.946  &     1.734  &     1.436  &     1.245  &     1.153  &     1.064  &     1.055  &     0.841  \\
  13.0 $\times 10^9$  &     0.543  &     6.003  &     7.057  &     7.032  &     6.384  &     5.949  &     5.485  &     4.842  &     4.353  &     4.233  &     4.121  &     4.109  &     3.542  \\
                   &            &     1.739  &     1.996  &     2.201  &     2.235  &     2.092  &     1.861  &     1.535  &     1.330  &     1.232  &     1.136  &     1.126  &     0.897  \\
  14.0 $\times 10^9$  &     0.541  &     6.072  &     7.116  &     7.088  &     6.455  &     6.025  &     5.562  &     4.921  &     4.433  &     4.313  &     4.200  &     4.188  &     3.621  \\
                   &            &     1.846  &     2.100  &     2.308  &     2.378  &     2.235  &     1.991  &     1.645  &     1.427  &     1.321  &     1.217  &     1.207  &     0.961  \\
  15.0 $\times 10^9$  &     0.539  &     6.171  &     7.221  &     7.195  &     6.556  &     6.123  &     5.659  &     5.015  &     4.528  &     4.408  &     4.295  &     4.283  &     3.715  \\
                   &            &     2.016  &     2.306  &     2.538  &     2.600  &     2.438  &     2.169  &     1.788  &     1.551  &     1.437  &     1.324  &     1.313  &     1.044  \\

\vspace*{1pt} \\
\multicolumn{14}{c}{$Z$ = 0.001} \\

   1.0 $\times 10^9$  &     0.663  &     3.530  &     4.297  &     4.332  &     3.985  &     3.681  &     3.287  &     2.682  &     2.144  &     2.005  &     1.889  &     1.879  &     1.262  \\
                   &            &     0.218  &     0.192  &     0.223  &     0.299  &     0.316  &     0.300  &     0.256  &     0.212  &     0.193  &     0.177  &     0.176  &     0.134  \\
   2.0 $\times 10^9$  &     0.617  &     4.135  &     5.093  &     5.040  &     4.579  &     4.224  &     3.775  &     3.096  &     2.524  &     2.372  &     2.248  &     2.237  &     1.591  \\
                   &            &     0.353  &     0.371  &     0.399  &     0.481  &     0.485  &     0.437  &     0.349  &     0.280  &     0.252  &     0.230  &     0.228  &     0.169  \\
   3.0 $\times 10^9$  &     0.595  &     4.514  &     5.593  &     5.518  &     4.940  &     4.539  &     4.064  &     3.373  &     2.804  &     2.657  &     2.534  &     2.523  &     1.889  \\
                   &            &     0.483  &     0.567  &     0.597  &     0.647  &     0.625  &     0.550  &     0.434  &     0.349  &     0.316  &     0.288  &     0.286  &     0.214  \\
   4.0 $\times 10^9$  &     0.580  &     4.740  &     5.876  &     5.803  &     5.172  &     4.748  &     4.263  &     3.566  &     2.999  &     2.855  &     2.734  &     2.723  &     2.096  \\
                   &            &     0.580  &     0.719  &     0.758  &     0.782  &     0.739  &     0.645  &     0.506  &     0.408  &     0.370  &     0.338  &     0.336  &     0.253  \\
   5.0 $\times 10^9$  &     0.570  &     4.930  &     6.101  &     6.036  &     5.371  &     4.930  &     4.438  &     3.738  &     3.173  &     3.031  &     2.911  &     2.900  &     2.281  \\
                   &            &     0.679  &     0.868  &     0.922  &     0.923  &     0.859  &     0.744  &     0.583  &     0.470  &     0.427  &     0.391  &     0.388  &     0.294  \\
   6.0 $\times 10^9$  &     0.562  &     5.123  &     6.304  &     6.248  &     5.563  &     5.111  &     4.617  &     3.920  &     3.360  &     3.221  &     3.103  &     3.092  &     2.480  \\
                   &            &     0.799  &     1.032  &     1.106  &     1.085  &     1.001  &     0.865  &     0.680  &     0.551  &     0.502  &     0.460  &     0.457  &     0.349  \\
   7.0 $\times 10^9$  &     0.556  &     5.308  &     6.507  &     6.453  &     5.752  &     5.292  &     4.793  &     4.097  &     3.539  &     3.402  &     3.284  &     3.273  &     2.668  \\
                   &            &     0.938  &     1.230  &     1.321  &     1.278  &     1.168  &     1.007  &     0.791  &     0.643  &     0.586  &     0.538  &     0.534  &     0.410  \\
   8.0 $\times 10^9$  &     0.551  &     5.464  &     6.657  &     6.605  &     5.900  &     5.438  &     4.943  &     4.251  &     3.700  &     3.566  &     3.449  &     3.439  &     2.842  \\
                   &            &     1.073  &     1.400  &     1.506  &     1.451  &     1.325  &     1.145  &     0.903  &     0.739  &     0.675  &     0.620  &     0.616  &     0.477  \\
   9.0 $\times 10^9$  &     0.546  &     5.597  &     6.778  &     6.731  &     6.027  &     5.566  &     5.076  &     4.390  &     3.846  &     3.715  &     3.600  &     3.590  &     3.001  \\
                   &            &     1.203  &     1.553  &     1.678  &     1.618  &     1.479  &     1.284  &     1.018  &     0.839  &     0.769  &     0.707  &     0.702  &     0.548  \\
  10.0 $\times 10^9$  &     0.543  &     5.715  &     6.894  &     6.840  &     6.135  &     5.676  &     5.190  &     4.507  &     3.969  &     3.841  &     3.727  &     3.717  &     3.137  \\
                   &            &     1.332  &     1.715  &     1.842  &     1.776  &     1.626  &     1.416  &     1.126  &     0.933  &     0.858  &     0.790  &     0.784  &     0.617  \\
  11.0 $\times 10^9$  &     0.540  &     5.810  &     6.994  &     6.937  &     6.231  &     5.771  &     5.285  &     4.601  &     4.064  &     3.937  &     3.822  &     3.812  &     3.232  \\
                   &            &     1.445  &     1.871  &     2.002  &     1.928  &     1.764  &     1.537  &     1.221  &     1.013  &     0.931  &     0.857  &     0.851  &     0.670  \\
  12.0 $\times 10^9$  &     0.537  &     5.916  &     7.098  &     7.038  &     6.335  &     5.877  &     5.391  &     4.709  &     4.173  &     4.045  &     3.931  &     3.920  &     3.341  \\
                   &            &     1.585  &     2.049  &     2.186  &     2.112  &     1.934  &     1.687  &     1.342  &     1.113  &     1.024  &     0.942  &     0.935  &     0.736  \\
  13.0 $\times 10^9$  &     0.534  &     6.025  &     7.195  &     7.129  &     6.431  &     5.977  &     5.499  &     4.825  &     4.299  &     4.174  &     4.061  &     4.050  &     3.480  \\
                   &            &     1.745  &     2.229  &     2.367  &     2.296  &     2.112  &     1.854  &     1.487  &     1.244  &     1.148  &     1.058  &     1.050  &     0.833  \\
  14.0 $\times 10^9$  &     0.532  &     6.111  &     7.280  &     7.204  &     6.513  &     6.063  &     5.586  &     4.913  &     4.387  &     4.262  &     4.149  &     4.138  &     3.568  \\
                   &            &     1.882  &     2.401  &     2.526  &     2.467  &     2.276  &     2.001  &     1.605  &     1.344  &     1.240  &     1.142  &     1.134  &     0.899  \\
  15.0 $\times 10^9$  &     0.530  &     6.156  &     7.339  &     7.244  &     6.558  &     6.110  &     5.634  &     4.958  &     4.432  &     4.307  &     4.194  &     4.183  &     3.612  \\
                   &            &     1.953  &     2.526  &     2.610  &     2.562  &     2.369  &     2.083  &     1.667  &     1.395  &     1.287  &     1.186  &     1.177  &     0.933  \\

\vspace*{1pt} \\
\multicolumn{14}{c}{$Z$ = 0.004} \\

   1.0 $\times 10^9$  &     0.669  &     3.705  &     4.548  &     4.472  &     4.108  &     3.830  &     3.423  &     2.737  &     2.147  &     1.969  &     1.829  &     1.814  &     1.028  \\
                   &            &     0.258  &     0.244  &     0.257  &     0.339  &     0.366  &     0.343  &     0.272  &     0.215  &     0.189  &     0.170  &     0.168  &     0.109  \\
   2.0 $\times 10^9$  &     0.621  &     4.319  &     5.412  &     5.295  &     4.749  &     4.390  &     3.923  &     3.170  &     2.555  &     2.378  &     2.237  &     2.221  &     1.403  \\
                   &            &     0.421  &     0.501  &     0.507  &     0.567  &     0.568  &     0.505  &     0.376  &     0.290  &     0.255  &     0.229  &     0.226  &     0.143  \\
   3.0 $\times 10^9$  &     0.597  &     4.671  &     5.875  &     5.759  &     5.121  &     4.715  &     4.228  &     3.469  &     2.858  &     2.690  &     2.553  &     2.539  &     1.755  \\
                   &            &     0.560  &     0.738  &     0.749  &     0.767  &     0.738  &     0.643  &     0.476  &     0.369  &     0.327  &     0.295  &     0.291  &     0.190  \\
   4.0 $\times 10^9$  &     0.582  &     4.915  &     6.225  &     6.084  &     5.385  &     4.950  &     4.445  &     3.675  &     3.062  &     2.897  &     2.762  &     2.748  &     1.978  \\
                   &            &     0.683  &     0.993  &     0.984  &     0.954  &     0.892  &     0.764  &     0.561  &     0.433  &     0.385  &     0.348  &     0.344  &     0.227  \\
   5.0 $\times 10^9$  &     0.570  &     5.091  &     6.478  &     6.310  &     5.571  &     5.116  &     4.604  &     3.832  &     3.222  &     3.062  &     2.931  &     2.917  &     2.172  \\
                   &            &     0.788  &     1.230  &     1.189  &     1.110  &     1.020  &     0.868  &     0.635  &     0.493  &     0.440  &     0.398  &     0.395  &     0.267  \\
   6.0 $\times 10^9$  &     0.562  &     5.235  &     6.698  &     6.507  &     5.737  &     5.268  &     4.743  &     3.956  &     3.340  &     3.178  &     3.045  &     3.032  &     2.282  \\
                   &            &     0.886  &     1.482  &     1.403  &     1.273  &     1.155  &     0.972  &     0.702  &     0.541  &     0.482  &     0.436  &     0.432  &     0.290  \\
   7.0 $\times 10^9$  &     0.554  &     5.389  &     6.880  &     6.676  &     5.892  &     5.416  &     4.887  &     4.104  &     3.491  &     3.332  &     3.201  &     3.188  &     2.461  \\
                   &            &     1.008  &     1.731  &     1.619  &     1.450  &     1.306  &     1.095  &     0.794  &     0.613  &     0.548  &     0.497  &     0.492  &     0.338  \\
   8.0 $\times 10^9$  &     0.549  &     5.541  &     7.059  &     6.844  &     6.049  &     5.566  &     5.034  &     4.250  &     3.638  &     3.480  &     3.350  &     3.337  &     2.621  \\
                   &            &     1.147  &     2.020  &     1.869  &     1.657  &     1.485  &     1.240  &     0.899  &     0.695  &     0.622  &     0.564  &     0.559  &     0.388  \\
   9.0 $\times 10^9$  &     0.544  &     5.649  &     7.204  &     6.973  &     6.163  &     5.674  &     5.138  &     4.353  &     3.740  &     3.583  &     3.454  &     3.441  &     2.735  \\
                   &            &     1.257  &     2.288  &     2.086  &     1.826  &     1.626  &     1.353  &     0.979  &     0.757  &     0.678  &     0.615  &     0.610  &     0.427  \\
  10.0 $\times 10^9$  &     0.540  &     5.781  &     7.351  &     7.115  &     6.300  &     5.806  &     5.265  &     4.481  &     3.869  &     3.712  &     3.583  &     3.570  &     2.874  \\
                   &            &     1.408  &     2.599  &     2.360  &     2.055  &     1.822  &     1.510  &     1.094  &     0.846  &     0.757  &     0.687  &     0.681  &     0.482  \\
  11.0 $\times 10^9$  &     0.536  &     5.862  &     7.460  &     7.208  &     6.379  &     5.880  &     5.338  &     4.556  &     3.947  &     3.791  &     3.664  &     3.651  &     2.969  \\
                   &            &     1.506  &     2.857  &     2.553  &     2.196  &     1.938  &     1.604  &     1.165  &     0.903  &     0.809  &     0.736  &     0.729  &     0.522  \\
  12.0 $\times 10^9$  &     0.533  &     5.971  &     7.574  &     7.320  &     6.490  &     5.988  &     5.444  &     4.667  &     4.059  &     3.904  &     3.777  &     3.765  &     3.094  \\
                   &            &     1.656  &     3.153  &     2.815  &     2.418  &     2.128  &     1.759  &     1.281  &     0.995  &     0.893  &     0.812  &     0.805  &     0.582  \\
  13.0 $\times 10^9$  &     0.530  &     6.060  &     7.700  &     7.433  &     6.591  &     6.082  &     5.529  &     4.746  &     4.134  &     3.978  &     3.851  &     3.838  &     3.174  \\
                   &            &     1.788  &     3.523  &     3.106  &     2.641  &     2.308  &     1.892  &     1.372  &     1.061  &     0.951  &     0.864  &     0.857  &     0.624  \\
  14.0 $\times 10^9$  &     0.528  &     6.128  &     7.761  &     7.491  &     6.651  &     6.141  &     5.592  &     4.820  &     4.214  &     4.061  &     3.936  &     3.924  &     3.275  \\
                   &            &     1.894  &     3.709  &     3.262  &     2.776  &     2.426  &     1.996  &     1.461  &     1.137  &     1.021  &     0.930  &     0.923  &     0.681  \\
  15.0 $\times 10^9$  &     0.525  &     6.202  &     7.835  &     7.563  &     6.724  &     6.214  &     5.665  &     4.895  &     4.292  &     4.139  &     4.014  &     4.002  &     3.357  \\
                   &            &     2.019  &     3.952  &     3.472  &     2.957  &     2.583  &     2.125  &     1.559  &     1.216  &     1.093  &     0.995  &     0.987  &     0.732  \\

\vspace*{1pt} \\
\multicolumn{14}{c}{$Z$ = 0.01} \\

   1.0 $\times 10^9$  &     0.676  &     3.843  &     4.773  &     4.637  &     4.200  &     3.914  &     3.534  &     2.780  &     2.195  &     2.001  &     1.849  &     1.829  &     0.893  \\
                   &            &     0.296  &     0.303  &     0.301  &     0.372  &     0.399  &     0.384  &     0.286  &     0.227  &     0.196  &     0.174  &     0.172  &     0.097  \\
   2.0 $\times 10^9$  &     0.626  &     4.467  &     5.713  &     5.574  &     4.939  &     4.546  &     4.079  &     3.231  &     2.601  &     2.405  &     2.250  &     2.232  &     1.289  \\
                   &            &     0.487  &     0.667  &     0.662  &     0.680  &     0.662  &     0.587  &     0.401  &     0.305  &     0.264  &     0.234  &     0.230  &     0.130  \\
   3.0 $\times 10^9$  &     0.602  &     4.812  &     6.257  &     6.063  &     5.342  &     4.905  &     4.404  &     3.513  &     2.867  &     2.669  &     2.512  &     2.491  &     1.541  \\
                   &            &     0.643  &     1.058  &     0.998  &     0.948  &     0.886  &     0.762  &     0.500  &     0.375  &     0.323  &     0.286  &     0.281  &     0.157  \\
   4.0 $\times 10^9$  &     0.586  &     5.076  &     6.587  &     6.357  &     5.598  &     5.142  &     4.636  &     3.771  &     3.138  &     2.954  &     2.806  &     2.791  &     1.913  \\
                   &            &     0.798  &     1.396  &     1.274  &     1.168  &     1.073  &     0.918  &     0.617  &     0.468  &     0.409  &     0.365  &     0.361  &     0.216  \\
   5.0 $\times 10^9$  &     0.574  &     5.251  &     6.866  &     6.593  &     5.794  &     5.319  &     4.797  &     3.924  &     3.288  &     3.106  &     2.960  &     2.945  &     2.093  \\
                   &            &     0.919  &     1.769  &     1.552  &     1.372  &     1.237  &     1.043  &     0.697  &     0.527  &     0.461  &     0.412  &     0.408  &     0.249  \\
   6.0 $\times 10^9$  &     0.565  &     5.412  &     7.085  &     6.785  &     5.962  &     5.474  &     4.944  &     4.077  &     3.443  &     3.265  &     3.123  &     3.109  &     2.292  \\
                   &            &     1.049  &     2.130  &     1.823  &     1.575  &     1.404  &     1.176  &     0.789  &     0.598  &     0.525  &     0.471  &     0.466  &     0.295  \\
   7.0 $\times 10^9$  &     0.557  &     5.520  &     7.250  &     6.925  &     6.082  &     5.584  &     5.048  &     4.178  &     3.543  &     3.366  &     3.226  &     3.212  &     2.410  \\
                   &            &     1.143  &     2.447  &     2.046  &     1.737  &     1.534  &     1.277  &     0.855  &     0.647  &     0.569  &     0.511  &     0.506  &     0.324  \\
   8.0 $\times 10^9$  &     0.551  &     5.650  &     7.413  &     7.070  &     6.214  &     5.710  &     5.171  &     4.304  &     3.670  &     3.496  &     3.357  &     3.343  &     2.554  \\
                   &            &     1.275  &     2.811  &     2.313  &     1.939  &     1.704  &     1.414  &     0.949  &     0.719  &     0.634  &     0.570  &     0.565  &     0.366  \\
   9.0 $\times 10^9$  &     0.546  &     5.783  &     7.575  &     7.216  &     6.349  &     5.840  &     5.298  &     4.435  &     3.804  &     3.632  &     3.494  &     3.481  &     2.708  \\
                   &            &     1.427  &     3.231  &     2.620  &     2.176  &     1.902  &     1.574  &     1.061  &     0.806  &     0.712  &     0.641  &     0.635  &     0.418  \\
  10.0 $\times 10^9$  &     0.541  &     5.907  &     7.737  &     7.364  &     6.487  &     5.972  &     5.422  &     4.552  &     3.915  &     3.742  &     3.603  &     3.589  &     2.808  \\
                   &            &     1.585  &     3.721  &     2.977  &     2.450  &     2.128  &     1.750  &     1.171  &     0.885  &     0.781  &     0.702  &     0.696  &     0.454  \\
  11.0 $\times 10^9$  &     0.537  &     5.996  &     7.854  &     7.467  &     6.585  &     6.066  &     5.512  &     4.637  &     3.998  &     3.824  &     3.685  &     3.671  &     2.887  \\
                   &            &     1.709  &     4.113  &     3.251  &     2.662  &     2.305  &     1.887  &     1.258  &     0.949  &     0.836  &     0.752  &     0.745  &     0.485  \\
  12.0 $\times 10^9$  &     0.534  &     6.080  &     7.976  &     7.567  &     6.673  &     6.148  &     5.589  &     4.715  &     4.077  &     3.903  &     3.765  &     3.752  &     2.978  \\
                   &            &     1.835  &     4.572  &     3.542  &     2.868  &     2.469  &     2.013  &     1.343  &     1.013  &     0.894  &     0.805  &     0.797  &     0.524  \\
  13.0 $\times 10^9$  &     0.531  &     6.164  &     8.081  &     7.660  &     6.758  &     6.230  &     5.670  &     4.799  &     4.161  &     3.989  &     3.852  &     3.838  &     3.068  \\
                   &            &     1.971  &     5.011  &     3.835  &     3.084  &     2.648  &     2.157  &     1.442  &     1.089  &     0.962  &     0.866  &     0.858  &     0.567  \\
  14.0 $\times 10^9$  &     0.528  &     6.230  &     8.169  &     7.732  &     6.824  &     6.293  &     5.732  &     4.862  &     4.226  &     4.055  &     3.918  &     3.905  &     3.144  \\
                   &            &     2.084  &     5.405  &     4.076  &     3.260  &     2.792  &     2.271  &     1.520  &     1.150  &     1.017  &     0.916  &     0.907  &     0.604  \\
  15.0 $\times 10^9$  &     0.526  &     6.290  &     8.255  &     7.803  &     6.888  &     6.354  &     5.791  &     4.919  &     4.283  &     4.111  &     3.974  &     3.960  &     3.194  \\
                   &            &     2.190  &     5.822  &     4.332  &     3.440  &     2.940  &     2.386  &     1.595  &     1.206  &     1.065  &     0.960  &     0.950  &     0.629  \\

\vspace*{1pt} \\
\multicolumn{14}{c}{$Z$ = 0.02} \\

   1.0 $\times 10^9$  &     0.681  &     4.006  &     5.046  &     4.881  &     4.359  &     4.041  &     3.669  &     2.871  &     2.294  &     2.099  &     1.944  &     1.915  &     0.955  \\
                   &            &     0.346  &     0.392  &     0.380  &     0.434  &     0.452  &     0.438  &     0.313  &     0.250  &     0.216  &     0.192  &     0.187  &     0.104  \\
   2.0 $\times 10^9$  &     0.631  &     4.615  &     6.092  &     5.879  &     5.167  &     4.738  &     4.257  &     3.294  &     2.639  &     2.426  &     2.260  &     2.228  &     1.222  \\
                   &            &     0.562  &     0.953  &     0.883  &     0.846  &     0.796  &     0.697  &     0.428  &     0.318  &     0.271  &     0.238  &     0.231  &     0.123  \\
   3.0 $\times 10^9$  &     0.607  &     4.983  &     6.644  &     6.356  &     5.567  &     5.099  &     4.591  &     3.627  &     2.971  &     2.769  &     2.609  &     2.579  &     1.623  \\
                   &            &     0.759  &     1.523  &     1.319  &     1.176  &     1.067  &     0.912  &     0.560  &     0.416  &     0.357  &     0.315  &     0.307  &     0.171  \\
   4.0 $\times 10^9$  &     0.590  &     5.204  &     7.015  &     6.662  &     5.815  &     5.319  &     4.792  &     3.825  &     3.166  &     2.971  &     2.815  &     2.787  &     1.863  \\
                   &            &     0.905  &     2.087  &     1.701  &     1.438  &     1.273  &     1.068  &     0.654  &     0.484  &     0.419  &     0.371  &     0.362  &     0.207  \\
   5.0 $\times 10^9$  &     0.578  &     5.399  &     7.295  &     6.903  &     6.026  &     5.514  &     4.974  &     4.010  &     3.349  &     3.159  &     3.005  &     2.979  &     2.078  \\
                   &            &     1.061  &     2.647  &     2.080  &     1.711  &     1.493  &     1.238  &     0.759  &     0.562  &     0.488  &     0.433  &     0.424  &     0.248  \\
   6.0 $\times 10^9$  &     0.569  &     5.537  &     7.525  &     7.086  &     6.178  &     5.652  &     5.101  &     4.138  &     3.476  &     3.290  &     3.139  &     3.114  &     2.237  \\
                   &            &     1.185  &     3.216  &     2.422  &     1.936  &     1.667  &     1.368  &     0.840  &     0.621  &     0.541  &     0.482  &     0.472  &     0.282  \\
   7.0 $\times 10^9$  &     0.561  &     5.665  &     7.739  &     7.258  &     6.323  &     5.783  &     5.221  &     4.256  &     3.591  &     3.407  &     3.257  &     3.234  &     2.368  \\
                   &            &     1.316  &     3.865  &     2.799  &     2.183  &     1.856  &     1.508  &     0.925  &     0.681  &     0.595  &     0.530  &     0.520  &     0.314  \\
   8.0 $\times 10^9$  &     0.555  &     5.776  &     7.906  &     7.394  &     6.443  &     5.896  &     5.327  &     4.363  &     3.698  &     3.516  &     3.368  &     3.346  &     2.497  \\
                   &            &     1.441  &     4.455  &     3.139  &     2.412  &     2.036  &     1.643  &     1.009  &     0.743  &     0.650  &     0.580  &     0.570  &     0.350  \\
   9.0 $\times 10^9$  &     0.549  &     5.869  &     8.059  &     7.518  &     6.552  &     5.997  &     5.419  &     4.450  &     3.782  &     3.601  &     3.453  &     3.432  &     2.587  \\
                   &            &     1.554  &     5.077  &     3.481  &     2.638  &     2.210  &     1.771  &     1.082  &     0.795  &     0.696  &     0.621  &     0.611  &     0.376  \\
  10.0 $\times 10^9$  &     0.544  &     5.965  &     8.213  &     7.635  &     6.650  &     6.087  &     5.507  &     4.542  &     3.875  &     3.697  &     3.550  &     3.529  &     2.695  \\
                   &            &     1.682  &     5.800  &     3.844  &     2.862  &     2.382  &     1.904  &     1.167  &     0.858  &     0.753  &     0.673  &     0.662  &     0.412  \\
  11.0 $\times 10^9$  &     0.540  &     6.072  &     8.360  &     7.755  &     6.760  &     6.192  &     5.608  &     4.646  &     3.980  &     3.804  &     3.658  &     3.639  &     2.818  \\
                   &            &     1.842  &     6.590  &     4.257  &     3.143  &     2.603  &     2.074  &     1.275  &     0.938  &     0.825  &     0.738  &     0.727  &     0.458  \\
  12.0 $\times 10^9$  &     0.536  &     6.175  &     8.519  &     7.884  &     6.879  &     6.308  &     5.717  &     4.740  &     4.069  &     3.889  &     3.742  &     3.722  &     2.884  \\
                   &            &     2.012  &     7.573  &     4.764  &     3.484  &     2.875  &     2.276  &     1.381  &     1.011  &     0.886  &     0.791  &     0.779  &     0.483  \\
  13.0 $\times 10^9$  &     0.533  &     6.251  &     8.628  &     7.976  &     6.962  &     6.385  &     5.789  &     4.815  &     4.144  &     3.965  &     3.820  &     3.800  &     2.976  \\
                   &            &     2.143  &     8.328  &     5.152  &     3.735  &     3.067  &     2.418  &     1.470  &     1.076  &     0.945  &     0.845  &     0.832  &     0.522  \\
  14.0 $\times 10^9$  &     0.530  &     6.317  &     8.730  &     8.051  &     7.026  &     6.447  &     5.852  &     4.878  &     4.209  &     4.032  &     3.887  &     3.868  &     3.048  \\
                   &            &     2.266  &     9.093  &     5.490  &     3.942  &     3.230  &     2.547  &     1.549  &     1.137  &     0.999  &     0.893  &     0.880  &     0.555  \\
  15.0 $\times 10^9$  &     0.527  &     6.389  &     8.824  &     8.133  &     7.104  &     6.521  &     5.922  &     4.950  &     4.281  &     4.105  &     3.961  &     3.943  &     3.133  \\
                   &            &     2.409  &     9.863  &     5.888  &     4.211  &     3.439  &     2.703  &     1.647  &     1.208  &     1.063  &     0.952  &     0.938  &     0.597  \\

\vspace*{1pt} \\
\multicolumn{14}{c}{$Z$ = 0.03} \\

   1.0 $\times 10^9$  &     0.684  &     4.025  &     5.173  &     4.991  &     4.421  &     4.075  &     3.689  &     2.824  &     2.241  &     2.021  &     1.859  &     1.819  &     0.822  \\
                   &            &     0.354  &     0.443  &     0.423  &     0.461  &     0.469  &     0.448  &     0.301  &     0.239  &     0.202  &     0.178  &     0.172  &     0.092  \\
   2.0 $\times 10^9$  &     0.634  &     4.696  &     6.294  &     6.016  &     5.262  &     4.811  &     4.336  &     3.340  &     2.700  &     2.469  &     2.303  &     2.252  &     1.225  \\
                   &            &     0.609  &     1.154  &     1.008  &     0.928  &     0.857  &     0.754  &     0.449  &     0.339  &     0.283  &     0.249  &     0.238  &     0.124  \\
   3.0 $\times 10^9$  &     0.610  &     5.068  &     6.885  &     6.520  &     5.685  &     5.189  &     4.676  &     3.675  &     3.027  &     2.806  &     2.645  &     2.598  &     1.617  \\
                   &            &     0.824  &     1.912  &     1.541  &     1.318  &     1.166  &     0.991  &     0.588  &     0.440  &     0.371  &     0.328  &     0.314  &     0.171  \\
   4.0 $\times 10^9$  &     0.593  &     5.303  &     7.250  &     6.836  &     5.962  &     5.441  &     4.905  &     3.891  &     3.234  &     3.015  &     2.855  &     2.811  &     1.846  \\
                   &            &     0.996  &     2.605  &     2.007  &     1.655  &     1.431  &     1.191  &     0.698  &     0.518  &     0.438  &     0.387  &     0.372  &     0.205  \\
   5.0 $\times 10^9$  &     0.581  &     5.512  &     7.557  &     7.093  &     6.183  &     5.643  &     5.090  &     4.092  &     3.438  &     3.227  &     3.072  &     3.033  &     2.112  \\
                   &            &     1.183  &     3.384  &     2.490  &     1.988  &     1.687  &     1.384  &     0.823  &     0.612  &     0.522  &     0.463  &     0.447  &     0.257  \\
   6.0 $\times 10^9$  &     0.571  &     5.640  &     7.809  &     7.287  &     6.337  &     5.776  &     5.208  &     4.212  &     3.555  &     3.350  &     3.199  &     3.166  &     2.278  \\
                   &            &     1.310  &     4.196  &     2.927  &     2.252  &     1.877  &     1.517  &     0.904  &     0.671  &     0.574  &     0.511  &     0.497  &     0.294  \\
   7.0 $\times 10^9$  &     0.564  &     5.780  &     8.019  &     7.463  &     6.497  &     5.927  &     5.348  &     4.343  &     3.682  &     3.476  &     3.326  &     3.293  &     2.410  \\
                   &            &     1.469  &     5.021  &     3.395  &     2.574  &     2.127  &     1.702  &     1.006  &     0.744  &     0.637  &     0.566  &     0.551  &     0.328  \\
   8.0 $\times 10^9$  &     0.557  &     5.871  &     8.189  &     7.593  &     6.605  &     6.023  &     5.434  &     4.427  &     3.764  &     3.561  &     3.412  &     3.382  &     2.517  \\
                   &            &     1.578  &     5.807  &     3.784  &     2.810  &     2.297  &     1.821  &     1.074  &     0.793  &     0.680  &     0.606  &     0.591  &     0.358  \\
   9.0 $\times 10^9$  &     0.551  &     5.943  &     8.357  &     7.719  &     6.708  &     6.115  &     5.514  &     4.486  &     3.815  &     3.609  &     3.458  &     3.428  &     2.552  \\
                   &            &     1.670  &     6.706  &     4.205  &     3.058  &     2.475  &     1.939  &     1.122  &     0.823  &     0.704  &     0.626  &     0.611  &     0.366  \\
  10.0 $\times 10^9$  &     0.546  &     6.031  &     8.514  &     7.830  &     6.801  &     6.200  &     5.594  &     4.570  &     3.901  &     3.698  &     3.548  &     3.520  &     2.662  \\
                   &            &     1.795  &     7.679  &     4.619  &     3.303  &     2.652  &     2.069  &     1.202  &     0.882  &     0.757  &     0.674  &     0.659  &     0.401  \\
  11.0 $\times 10^9$  &     0.542  &     6.133  &     8.678  &     7.956  &     6.908  &     6.298  &     5.688  &     4.669  &     4.002  &     3.802  &     3.654  &     3.627  &     2.786  \\
                   &            &     1.955  &     8.863  &     5.141  &     3.612  &     2.880  &     2.238  &     1.306  &     0.960  &     0.826  &     0.737  &     0.721  &     0.446  \\
  12.0 $\times 10^9$  &     0.538  &     6.211  &     8.828  &     8.069  &     7.003  &     6.387  &     5.770  &     4.740  &     4.070  &     3.869  &     3.720  &     3.693  &     2.846  \\
                   &            &     2.086  &    10.105  &     5.667  &     3.917  &     3.103  &     2.398  &     1.384  &     1.015  &     0.872  &     0.778  &     0.760  &     0.468  \\
  13.0 $\times 10^9$  &     0.535  &     6.252  &     8.954  &     8.149  &     7.065  &     6.441  &     5.817  &     4.771  &     4.096  &     3.893  &     3.742  &     3.713  &     2.855  \\
                   &            &     2.151  &    11.267  &     6.057  &     4.117  &     3.239  &     2.486  &     1.415  &     1.033  &     0.886  &     0.789  &     0.770  &     0.469  \\
  14.0 $\times 10^9$  &     0.531  &     6.381  &     9.084  &     8.278  &     7.196  &     6.573  &     5.946  &     4.899  &     4.225  &     4.022  &     3.873  &     3.845  &     2.991  \\
                   &            &     2.408  &    12.623  &     6.783  &     4.619  &     3.634  &     2.783  &     1.582  &     1.156  &     0.992  &     0.884  &     0.864  &     0.528  \\
  15.0 $\times 10^9$  &     0.528  &     6.428  &     9.187  &     8.347  &     7.251  &     6.624  &     5.993  &     4.941  &     4.266  &     4.064  &     3.914  &     3.886  &     3.033  \\
                   &            &     2.502  &    13.808  &     7.190  &     4.835  &     3.789  &     2.891  &     1.636  &     1.195  &     1.026  &     0.913  &     0.892  &     0.546  \\

\label{licktab}
\end{longtable}

\bsp
\label{lastpage}

\begin{thebibliography}{}

 \bibitem[\protect\citeauthoryear{Barmby \& Huchra}{2000}]{bar00}
 Barmby P. \& Huchra J. P., 2000, ApJ, 521, L29

 \bibitem[\protect\citeauthoryear{Brodie \& Huchra}{1990}]{bro90}
 Brodie J. \& Huchra J. P., 1990, apJ, 362, 503

 \bibitem[\protect\citeauthoryear{Bruzual}{2007}]{bru07}
 Bruzual G. A., 2007, IAUs, 241, 125, Stellar populations as Building Blocks of Galaxies

\bibitem[\protect\citeauthoryear{Bruzual \& Charlot}{2003}]{bru03}
 Bruzual G. A. \& Charlot S., 2003, MNRAS, 344, 1000

 \bibitem[\protect\citeauthoryear{Cardelli, Clayton \& Mathis}{1989}]{car89}
 Cardelli J. a., Clayton G. C. \& Mathis J. S., 1989, ApJ, 345, 245

 \bibitem[\protect\citeauthoryear{Chabrier}{2003}]{cha03}
 Chabrier G., 2003, PASP, 115, 763

 \bibitem[\protect\citeauthoryear{Bessell}{1983}]{bes83}
 Bessell M. S., 1983, PASP, 95, 480

 \bibitem[\protect\citeauthoryear{Eggleton, Fitchett \& Tout}{1989}]{egg89}
 Eggleton P. P., Fitchett M. J. \& Tout C. A., 1989, ApJ, 347, 998

 \bibitem[\protect\citeauthoryear{Frogel, Persson \& Cohen}{1980}]{fro80}
 Frogel J., Persson S. E. \& Cohen J. G., 1980, ApJ, 240, 785

 \bibitem[\protect\citeauthoryear{Frogel, Mould \& Blanco}{1990}]{fro90}
 Frogel J., Mould J. \& Blanco V. M., 1990, ApJ, 352, 96


 \bibitem[\protect\citeauthoryear{Harris}{1996}]{har96}
 Harris W., 1996, AJ, 112, 1487

 \bibitem[\protect\citeauthoryear{Hurley, Pols \& Tout}{2000}]{hur00}
 Hurley J. R., Pols O. R. \& Tout C. A., 2000, MNRAS, 315, 543

 \bibitem[\protect\citeauthoryear{Hurley, Tout \& Pols}{2002}]{hur02}
 Hurley J. R., Tout C. A. \& Pols O. R., 2002, MNRAS, 329, 897

 \bibitem[\protect\citeauthoryear{Kroupa}{2001}]{kro01}
 Kroupa P., 2001, MNRAS, 322, 231

 \bibitem[\protect\citeauthoryear{Maraston et al.}{2001}]{mar01}
 Maraston C., Kissler-Patig M., Brodie J. P., Barmby P., Huchra J. P., 2001, A\&A, 370, 176

 \bibitem[\protect\citeauthoryear{Maraston}{2005}]{mar05}
 Maraston C., 2005, MNRAS, 362, 799

 \bibitem[\protect\citeauthoryear{Maraston et al.}{2006}]{mar06}
 Maraston C., Daddi E., Renzini A., et al. 2006, ApJ, 652, 85

 \bibitem[\protect\citeauthoryear{Marigo}{2007}]{mar07a}
 Marigo P., 2007, (astro-ph/0701536)


 \bibitem[\protect\citeauthoryear{Miller \& Scalo}{1979}]{mil79}
 Miller G. E. \& Scalo J. M., 1979, ApJS, 41, 513

 \bibitem[\protect\citeauthoryear{Miller et al.}{1997}]{mil97}
 Miller B. W., Whitemore B. C., Schweizer F. \& Fall S. M., 1997, AJ, 114, 2381


 \bibitem[\protect\citeauthoryear{Persson et al.}{1983}]{per83}
 Persson S. E., Aaronson M., Cohen J. G., Frogel J. \& Matthews K., 1983, ApJ, 266,105

 \bibitem[\protect\citeauthoryear{Rettura et al.}{2006}]{ret06}
 Rettura A., Posati P., Strazzullo V., 2006, A\&A, 458, 717

 \bibitem[\protect\citeauthoryear{Salpeter}{1955}]{sal55}
 Salpeter E. E., 1955, ApJ, 121, 161

 \bibitem[\protect\citeauthoryear{Searale, Wilkinson \& Bagnuolo}{1980}]{sea80}
 Searale L., Wilkinson A. \& Bagnuolo W. G., 1980, ApJ, 239, 803

 \bibitem[\protect\citeauthoryear{Schlegel, Finkbeiner \& Davis}{1998}]{sch98}
 Schlegel D., Finkbeiner D. P. \& Davis M., 1998, ApJ, 500, 525

 \bibitem[\protect\citeauthoryear{van den Bergh}{1981}]{van81}
 van den Bergh S., 1981, A\&AS, 46, 79

 \bibitem[\protect\citeauthoryear{Worthey}{1994}]{wor94}
 Worthey G., 1994, ApJS, 95, 107

 \bibitem[\protect\citeauthoryear{Zhang et al.}{2005a}]{zha05a}
 Zhang F., Han, Z., Li, L. \& J. R. Hurley, 2005a, MNRAS, 357, 1088

 \bibitem[\protect\citeauthoryear{Zhang et al.}{2005b}]{zha05b}
 Zhang F., Li, L., Han, Z., 2005b, MNRAS, 364, 503

 \bibitem[\protect\citeauthoryear{Zhang \& Li}{2006}]{zha06}
 Zhang F. \& Li, L. 2006, MNRAS, 370, 1181



\end{thebibliography}
\end{document}